\documentclass[a4paper,11pt]{article}
\usepackage{jheppub} % for details on the use of the package, please see the JINST-author-manual
\usepackage{lineno}

\usepackage{graphicx}

\usepackage{float}

\usepackage{amsmath}
\usepackage{amsfonts}
\usepackage{bbm}
\usepackage{cancel}
\usepackage{slashed}
\usepackage{amssymb}
\usepackage{feynman}

\usepackage{gensymb}
\usepackage{textcomp}
\usepackage{bm}
\usepackage{tikz}
\usepackage[compat=1.1.0]{tikz-feynman}
\usepackage{tgheros}
\usepackage{varwidth}
\usepackage{setspace}
\usepackage{float}
\usepackage{wrapfig}
\usepackage{url}
\usepackage{xcolor}
\usepackage{braket}

\usepackage{multirow}
\usepackage{tabularray}

\usepackage{xfrac}

\usepackage{caption,subcaption}
\title{\boldmath An Improved Precision Calculation of the $0\nu\beta\beta$ Contact Term within Chiral Effective Field Theory}

\author{Graham Van Goffrier}
\affiliation{University of Southampton,\\
Southampton, United Kingdom}
\affiliation{University College London,\\
London, United Kingdom}

\emailAdd{gwvg1e23@soton.ac.uk}

\abstract{Neutrinoless double-beta ($0\nu\beta\beta$) decay is an as-yet unobserved nuclear process, which stands to provide crucial insights for model-building beyond the Standard Model of particle physics. Its detection would simultaneously confirm the hypothesis that neutrinos are Majorana fermions, thus violating lepton-number conservation, and provide the first measurement of the absolute neutrino mass scale. This work aims to improve the estimation within chiral effective field theory of the so-called “contact term” for $0\nu\beta\beta$-decay, a short-range two-nucleon effect which is unaccounted for in traditional nuclear approaches to the process. We conduct a thorough review of the justifications for this contact term and the most precise computation of its size to date ($g_\nu^{NN} = 1.3(6)$ at renormalisation point $\mu=m_\pi$), whose precision is limited by a truncation to elastic intermediate hadronic states. We then perform an extension of this analysis to a subleading class of inelastic intermediate states which we characterise, delivering an updated figure for the contact coefficient ($g_\nu^{NN} = 1.4(3)$ at $\mu=m_\pi$) with uncertainty reduced by half. Such ab initio nuclear results, especially with enhanced precision, show promise for the resolution of disagreements between estimates of $0\nu\nu\beta\beta$ from different many-body methods.}

\begin{document}
\maketitle
\flushbottom

\section{Introduction}

From a theory perspective, the usefulness of neutrinoless double-beta  ($0\nu\beta\beta$) decay as a probe of neutrino mass and of lepton-number violation (LNV) is constrained by the precision with which its half-life may be related to physical model parameters. This precision is most severely limited by our ability to calculate many-body or even few-body nuclear structure, a challenge amplified by the inherent two-body nature of $0\nu\beta\beta$-decay. The factorisation of the $0\nu\beta\beta$ amplitude into a leptonic phase-space factor (PSF) and a hadronic nuclear matrix-element (NME) \cite{Doi1983}, although inexact, has become a standard approximation. This is in part because the imprecision introduced by this factorisation falls well below other theoretical uncertainties \cite{Dolinski2019}, and in part because the PSFs are calculable with relative ease compared to the NMEs.

No exact method to solve the nuclear many-body problem from first-principles has yet been developed; instead, a collection of efficient, truncated techniques are in common usage, which are reviewed extensively e.g. in \cite{Engel2017}. Estimates of the $0\nu\beta\beta$ decay NMEs are known to be correlated across isotopes, both within and across many-body methods, due to shared (and only partially-understood) systematics of the many-body procedures \cite{Agostini2023}. These correlations are sometimes interpreted through the lens of `quenching', a generic overestimation of NMEs for both $\beta$- and $2\nu\beta\beta$-decay, owing to the Gamow-Teller term. As Ref.~\cite{Engel2017} reviews in great detail, numerous underlying mechanisms for this systematic offset have been proposed but lack any conclusive evidence. Any or all of nuclear-structure correlations, multi-nucleon weak currents, heavier isobars such as the $\Delta$, pion-range properties of the nuclear medium and more could be responsible for the effect, and as such it is unclear whether a similarly-strong effect occurs for $0\nu\beta\beta$-decay.

Between 2018 and 2021, Cirigliano et al. surprised the $0\nu\beta\beta$ and nuclear theory communities with a series of works~\cite{Cirigliano2018,Cirigliano2019,Cirigliano2021short,Cirigliano2021long} purporting to show that even the two-body picture of $0\nu\beta\beta$-decay within conventional nuclear theory was, in fact, incomplete, containing its own intrinsic systematic offset. Employing an \textit{ab initio} framework derived from chiral effective field theory ($\chi EFT$), these works claimed that a short-range, two-nucleon LNV interaction necessarily induces a contribution to the decay similar in magnitude to the contribution from the usual light-neutrino-exchange. There has been debate in the community about the implications of this result for NME computations performed in standard phenomenological nuclear frameworks, and in particular whether the proposed ``contact counterterm" is already accounted for by the nucleon-nucleon correlations included within modern many-body nuclear models. However,~\cite{Cirigliano2021long} argued concisely that the EFT-based argument still proceeds even if the involved nucleons are maximally correlated, and so must be a distinct effect. At the time of writing, there is no published rebuttal.

Beyond such generic considerations, the $0\nu\beta\beta$-decay theorist is of course most interested in the quantitative impact of the contact counterterm on estimated NMEs. This question was partially answered in the most recent works by Cirigliano et al.~\cite{Cirigliano2021short,Cirigliano2021long}, which implemented a matching analysis between low-momentum $\chi EFT$ and high-momentum perturbative QCD in order to estimate the coefficient of the contact counterterm. Their final result, at renormalisation point $\mu = m_\pi$, was:
\begin{equation}
    g_\nu^{NN}|_{NN} \simeq 1.32(50)_{\text{inelastic}}(20)_{\text{V}_\text{S}}(5)_{\text{parameters}} = 1.3(6),
\end{equation}
where the largest error, about $38\%$ relative error, arises from the choice to only account for elastic $NN$ states in the intermediate stage of the process. The other acknowledged sources of error are the choice of a short-range nuclear potential $V_S$ and other model-dependent parameters used to extend the $\chi EFT$ result up to higher momenta, closer to the perturbative QCD regime so that a matching analysis is possible. The renormalisation point is a kinematic scale at which one chooses to match theory parameters to physical observables; it is not unexpected that an unobservable coupling such at $g_\nu^{NN}$ should vary with this choice of scale, although observables should not.

This work aims to improve upon the $38\%$ error from truncation to elastic states, by explicitly accounting for the lowest-lying collection of inelastic states, namely $NN\pi$ states. Our result will therefore include both an adjusted coefficient for the contact counterterm, and an improved error bar on the inelastic state truncation, and at the same renormalisation point will have numerical value:
\begin{equation}
    g_\nu^{NN}|_{NN + NN\pi} \simeq 1.40(20)_{V_S}(5)_{\text{parameters}}(3)_{\text{inelastic}} = 1.4(3).
\end{equation}
To contextualise this result, and our methods, we will first give an overviews of the mathematical and physical foundations of Cirigliano et al.'s work in Section 2: chiral symmetry, $\chi EFT$, and nucleon-nucleon interactions withn that framework. Section 3 will review the derivation of the contact counterterm from \cite{Cirigliano2021long}, including the aforementioned matching procedure for determining its coefficient. Finally, in Section 4, we will present our novel derivation of the contact counterterm coefficient accounting for both $NN$ and $NN\pi$ intermediate states, ultimately delivering an updated estimate of $g_\nu^{NN}$ with improved uncertainty, and comment on the uncertainties which then dominate.

\section{Chiral Effective Field Theory and Nuclear Forces}

%\begin{itemize}
%    \item Global $SU(N)_L \times SU(N)_R \times U(1)_V$ symmetry of massless QCD
%    \item Quick review of spontaneous symmetry breaking, Goldstone bosons
%    \item $SU(N)_V$ approximate hadronic symmetry, mesons as Goldstone bosons
%    \item Nonvanishing scalar singlet quark condensate as sufficient condition
%    \item Pion group realisation, lowest-order effective Lagrangian
%    \item General principles of effective theories, Weinberg's power-counting
%\end{itemize}

The experimental properties of both bound quark-gluon matter and hadronic scattering processes have constrained the gauge group of the strong interactions to be $SU(3)_{\text{colour}}$. This fact has numerous non-trivial implications, not the least of which is that quarks and gluons should be confined into hadronic matter in systems below the Hagedorn temperature \cite{Wilson1974,tHooft1982}. In studying QCD at low energies (MeV to GeV) relevant to \textit{nuclear} processes, typically MeV- to GeV-scale as opposed to TeV-scale collisions, our essential goal will be to make accurate predictions of hadronic properties, in particular the interactions of stable baryonic matter (protons and neutrons) via short-lived mesonic mediators. The running of the strong coupling leads to the breakdown of perturbation theory for low energy QCD; i.e. we cannot hope to describe these hadronic interactions through some convergent series of diagrams involving their constituent quark and gluon field quanta. 

However, nature has blessed QCD with a natural (chiral) symmetry, which is broken lightly enough that direct calculation of hadronic interactions is possible -- with quantifiable perturbative uncertainties. The lightest mesons, pions, emerge as the Goldstone bosons of this approximate spontaneously symmetry-breaking (SSB), and their interactions are severely constrained. The resultant $\chi EFT$, whose formulation is outlined in this section, will prove powerful enough to explain properties of nucleon-nucleon collisions from what are effectively QCD first principles. By including electroweak source terms in the $\chi EFT$ Lagrangian, $0\nu\beta\beta$-decay can also be described at the nucleon level.

\subsection{Chiral symmetry, SSB, and an effective lagrangian}

Chiral symmetry is most simply stated as the separation between the mass scale of the light quarks and that of the hadrons; with these light quark masses taken to be $0$, QCD classically enjoys an exact, global $U(3)_{\text{flavour}}$ symmetry. Furthermore, examining chirality eigenstates of the quark field reveals a copy of this flavour symmetry in each chirality sector, i.e.$U(3)_L \times U(3)_R \simeq SU(3)_L \times SU(3)_R \times U(1)_L \times U(1)_R$, where the latter factorisation emerges from extracting a global phase from each $U(3)$ transformation. 

This approximation of QCD still holds when we promote the $SU(3)_{\text{colour}}$ symmetry of the quark fields to a local symmetry, introducing non-Abelian gauge self-interactions and a $\theta$-term. However, the gauge interactions do constrain the collection of global flavour symmetries \textit{after quantisation}. The famous axial anomaly of $U(1)_A$~\cite{Adler1969,Bell1969}, with a violation inversely proportional to the number of colours, implies that $U(1)_A$ plays a role neither in chiral symmetry-breaking nor in the eventual $\chi EFT$. The remaining Abelian symmetry $U(1)_V$, on the other hand, is preserved after quantisation and corresponds to baryon-number conservation.

The action of chiral SSB is therefore restricted to $SU(3)_L \times SU(3)_R$, and its subgroup $SU(3)_V$ under which the QCD vacuum is invariant. Before exploring the consequences of this breaking, it is valuable to physically motivate its occurrence. Speaking phenomenologically, a $SU(3)_A$ symmetry on the baryon spectrum allows one to construct from any positive parity state a mass-degenerate negative parity state; however, such opposite-parity pairs are firmly absent from the observed baryon spectrum. Speaking theoretically, if the operator $\overline{q}q$ has a non-vanishing expectation value, which must be flavour-independent in the chiral limit, the QCD vacuum is no longer invariant under axial symmetry $SU(3)_A$~\cite{Scherer2011}. Thus the scalar singlet quark condensate $\overline{q}q$ provides not only an explanation, but a mechanism by which chiral symmetry can be spontaneously broken at zero temperature, yet be preserved/restored at higher energies. The pions, an isotriplet of pseudoscalars with masses $135.0$~MeV ($\pi^0$) and $139.6$~MeV ($\pi^{\pm}$), are interpreted as the pseudo-Goldstone bosons resulting from this SSB, and their significant mass hierarchy to the next-lightest mesons, kaons $\sim 490$~MeV and rho mesons $\sim 770$~MeV, may be taken as evidence of the chiral symmetry of QCD in action.

Some general comments about EFTs are in order. Although we have selected a symmetry-breaking pattern based on the theoretical and experimental properties of QCD, numerous quantum field theories could be constructed which obey this symmetry-breaking pattern. The natural question is then: which approximation is the best choice? A key result by Weinberg~\cite{Weinberg1979} is that the most general Lagrangian obeying the chosen symmetry will generate the most general complex-analytic S-matrix which also obeys that symmetry, as well as being consistent with unitary time evolution and ``cluster decomposition", where the latter refers to a conjected asymptotic independence between distant experiments. The conventional wisdom is that one cannot generically improve on this choice, as all specific physical content is encoded in the symmetry-breaking. One is free to select the constituent fields most appropriate at the observed scale; in $\chi EFT$, these degrees of freedoms are hadrons rather than the more fundamental quarks and gluons. The resultant Lagrangian will also in general be non-renormalisable, but this is in line with the expectation that the effective theory will only make accurate predictions within some range of scales. Even still, an infinite quantity of interaction terms appear in the Lagrangian, and there is no \textit{a priori} method for determining which will contribute dominantly in all situations. As a result, part of constructing the effective theory is the choice of ``power-counting" scheme, usually in the form of a small expansion parameter. In $\chi EFT$, the standard and physically well-motivated choice is an expansion in $\frac{Q}{\Lambda_\chi} \simeq \frac{m_\pi}{\Lambda_\chi} \sim \frac{1}{7.5}$ for characteristic momentum $Q$ and chiral symmetry-breaking scale $\Lambda_\chi \sim 1$~GeV~\cite{Manohar1984}.

We will not review the construction of the $\chi EFT$ Lagrangian from the above principles in detail, but instead direct the reader to standard resources such as \cite{Scherer2011}. In brief, one constructs the simplest non-trivial realisation of spontaneously broken chiral symmetry, building invariant interactions out of $SU(N)$-valued fields $U(x) = \exp \left( i \frac{\phi(x)}{F_0} \right)$ for pion fields $\phi$ and constant $F_0$. $\phi$ can be expanded $\phi = \phi_a \lambda_a$ over some basis $\lambda_{a=1,...,n^2}$ of $SU(N)$, with standard normalisation $Tr(\lambda_a \lambda_a) = 2$; for $N=2$, $\lambda_a$ are the (isospin) Pauli matrices, and for $N=3$ they are the Gell-Mann matrices. The chiral symmetry is promoted to a local symmetry by introducing two gauge fields $l_\mu$, $r_\mu$ and an appropriate gauge-covariant spacetime derivative. Explicit chiral symmetry-breaking is encoded in the light-quark mass matrix $\mathcal{M}$, and an argument due to Georgi~\cite{Manohar1984} shows that it is consistent to treat $\mathcal{M}$ as an auxiliary field and build the most general possible chiral-invariant Lagrangian accordingly, although it will always take on a constant value at evaluation.

If we are to apply $\chi EFT$ to nuclear problems, baryonic fields must be introduced into the formalism. It is not necessary, for low-energy applications, for these baryons to be full-fledged dynamical fields, as processes in which baryons are created and destroyed are already beyond the range of validity of $\chi EFT$ due to the heavy masses of baryons. Practically, this implies that our baryon realisations of the chiral symmetry need not be matrix-valued; instead they can be isospin multiplet states $N$ which transform under $SU(N)_V$ as $N \mapsto V N$. While previously we could only evaluate pion-diagrams between QCD-vacuum initial and final states, now these states can be (tensor products of) $N$ states, providing a language for n-nucleon interactions.

Again, invariance under chiral symmetry specifies the most general set of interactions possible between baryons $N$ and pion exponentials $U$. In nuclear applications, it is highly beneficial to take a ``static approximation" in which the nucleon mass is thought of as very large, and in which Dirac algebra reduces to Pauli algebra. Within the regime of our results, we implicitly take to be equivalent such formalisms as the heavy-baryon chiral perturbation theory (HBChPT) of~\cite{Jenkins1991}, or the simple neglect of a mass term for nucleons employed by \cite{Weinberg1990,Kaplan1996}.

Altogether, we have leading-order chiral Lagrangian:
\begin{align}
    \mathcal{L}_{2} = &\frac{F_0^2}{4} Tr \left( D_\mu U D^\mu U^\dagger \right) + \frac{F_0^2 B_0}{2} Tr \left( \mathcal{M} U^\dagger + U \mathcal{M}^\dagger \right) \\ \nonumber
    &+ \overline{N} \left( i \gamma^\mu \left( \partial_\mu + \Gamma_\mu \right) - m  + \frac{g_A}{2} \gamma^\mu \gamma_5 u_\mu \right) N,
\end{align}
where $B_0$ is an independent low-energy constant not determined by chiral symmetry, and where $\Gamma_\mu$ and $u_\mu$ are particular vector combinations of the pion fields and their derivatives, to be defined as needed. $F_0$ may be related to the expectation value for the decay of a pion to the vacuum via an axial current~\cite{Scherer2011}:
\begin{equation}
    \braket{0 | A^\mu_a(0) | \phi(b)(p)} = i p^\mu F_0 \delta_{ab}
\end{equation}
and is thus known as the ``pion decay constant". Normalisation varies by factors of $\sqrt{2}$ depending on the choice of pion field normalisation, but we follow~\cite{Scherer2011,Cirigliano2021long} and the bulk of recent literature in taking the relation above to define $F_0$. Corrections from $\mathcal{L}_4$ and beyond distinguish between $F_\pi$ for pions and $F_K$ for kaons, and the experimental value $F_\pi = 92.1(8)$~MeV~\cite{PDG2022} includes these corrections. Therefore in the following, we will follow~\cite{Cirigliano2021long} in writing $F_\pi$ with an awareness that its value includes higher-order chiral corrections even though the rest of our calculation does not.

Since even the $l=2$ Lagrangian contains infinitely many interaction terms, it is necessary to be able to truncate this series such that any resultant approximation error is bounded. This is the ethos of power-counting: to estimate the magnitudes of interaction vertices, and therefore the magnitudes of the diagrams they compose, before needing to evaluate the high-dimensional integrals which those diagrams encode. Of course these integrals must eventually be evaluated, and divergences cured through appropriate regularisation and the introduction of counterterms, but power-counting allows for the choice of truncation to be made without the evaluation of any \textit{unnecessary} integrals.

The more exact aspect of power-counting is the bookkeeping of both dimensionless and dimensionful constants which arise at interaction vertices. In $\chi EFT$, this is quite straightforward. All terms in the $U$-field Lagrangian come with fixed constants, e.g. $\frac{F_\pi^2}{4}$ for all terms of $\mathcal{L}_2$, in combination with $B_0$ for the scalar/pseudoscalar symmetry-breaking terms; external fields of course may come with their own dimensionless factors which are defined during matching, but are fixed for every appearance of the field. Where unknown, is it standard to take such dimensionless constants to be of order $\mathcal{O}(1)$, or vary them over $\sim 2$ orders of magnitude around $1$. In the exponential realisation of the spontaneously-broken axial symmetry, each pion field $\phi$ also comes with a factor of $\frac{1}{F_\pi}$. 

The less exact aspect of power-counting is the choice of representative values for the spatial momenta and energies which emerge from derivatives of field operators. Physically one can imagine that these virtual quanta have some characteristic scale for all processes within the range of validity of $\chi EFT$. Of course no choice can be truly universal, and indeed we encounter in $0\nu\beta\beta$ a process which in part astounds expectations of a nice hierarchy of scales for nuclear processes. For now, we will proceed by letting the characteristic four-momentum scale of pions be $Q \sim m_\pi$, which is considered in ratio to the chiral symmetry-breaking scale $\Lambda_\chi$; this is plausible since $r \sim \frac{1}{m_\pi}$ is the typical distance between nucleons. Any interaction vertex may then be assigned a label $k$ denoting approximate magnitude $\left( \frac{Q}{\Lambda_\chi} \right)^k$. Since our interactions will permit only low-energy pions and nucleons, we can write $Q \sim m_\pi \ll \Lambda_\chi \sim M_N$~\cite{Weinberg1991} such that the ratio $Q / \Lambda_\chi \sim \frac{1}{7.5}$ is an appropriate small parameter~\cite{Manohar1984}.

We will revisit the chiral power-counting in Section 4.4, when we define the specific rules necessary for estimating the magnitudes of our $NN\pi$ contributions.

\subsection{\texorpdfstring{The problematic $NN$ sector}{The problematic NN sector}}

Studies of two-nucleon ($NN$) interactions and bound-systems in $\chi EFT$, as distinct from the very broad literature on high-energy nucleon scattering e.g. in colliders, have been motivated by the desire to perform further analysis of many-nucleon systems. The first steps in this direction were phenomenological treatments of two-nucleon potentials, which were derived as fits to measured scattering data and bound-state (deuteron) characteristics~\cite{Stoks1994,Machleidt2001} and dominated the discipline until recently. The reasons are twofold for the broad acceptance~\cite{Epelbaum2020} of effective chiral potentials over these phenomenological ones: they come with quantitative error bars owing to the truncation of the chiral ordering, and are therefore systematically-improvable; and they are explainable in terms of the broken chiral symmetry of QCD.

The key distinction between the $NN$-sector of $\chi EFT$ and the $\pi$ and $\pi N$ sectors is that $NN$-interactions are known to be strong at low energies, and perturbation theory breaks down. While interactions involving pions, as Goldstone bosons, are ``protected" by chiral symmetry, those between nucleons are not. Pion-range interactions such as one-pion exchange may obey the chiral power-counting, but by the same logic, shorter-range nucleon interactions should also -- this paradoxically leads to a description of nuclear physics where the interactions are not strong, where nuclear bound states are not captured, and indeed where nuclei do not appear at all~\cite{Phillips2022}. More concretely, we will consider the effective $NN$-Lagrangian of~\cite{Kaplan1996}:
\begin{equation}
    \mathcal{L}_{NN} = N^\dagger i \partial_t N - N^\dagger \frac{\nabla^2}{2M_N} N - \frac{1}{2} C_S (N^\dagger N)^2 - \frac{1}{2}C_T (N^\dagger \vec{\sigma} N)^2 ...
    \label{NNLagrangian}
\end{equation}
which includes all interactions invariant under chiral symmetry and defines two undetermined four-nucleon low-energy constants (LECs), $C_S$ and $C_T$.

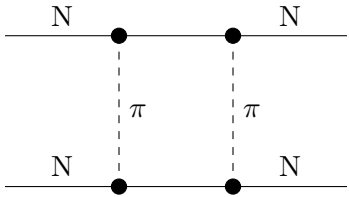
\begin{figure}[t]
\centering
    \begin{tikzpicture}
          \begin{feynman}
            \vertex (l);
            \vertex [right=1.5cm of l] (ma);
            \vertex [right=1.5cm of ma] (mb);
            \vertex [right=1.5cm of mb] (mc);
            \vertex [right=1.5cm of mc] (md);   
            \vertex [right=1.5cm of md] (r);   
            \vertex [above=1cm of ma] (ua);
            \vertex [above=1cm of mb] (ub);
            \vertex [above=1cm of mc] (uc);
            \vertex [above=1cm of md] (ud);
            \vertex [below=1cm of ma] (da);
            \vertex [below=1cm of mb] (db);
            \vertex [below=1cm of mc] (dc);
            \vertex [below=1cm of md] (dd);
            \vertex [above right=0.5cm of ub] (e1);
            \vertex [below right=0.5cm of db] (e2);         
            \diagram*{
                (ua) -- [solid, edge label=N] (ub) -- [solid] (uc) -- [solid, edge label=N] (ud);
                (da) -- [solid, edge label=N] (db) -- [solid] (dc) -- [solid, edge label=N] (dd);
                (ub) -- [dashed, edge label=$\pi$] (db);
                (uc) -- [dashed, edge label=$\pi$] (dc);
                %(ub) -- [fermion] (e1);
                %(db) -- [fermion] (e2);
            };
            \draw[fill=black] (ub) circle(1mm);
            \draw[fill=black] (db) circle(1mm);
            \draw[fill=black] (uc) circle(1mm);
            \draw[fill=black] (dc) circle(1mm);
          \end{feynman}
    \end{tikzpicture}
    \caption{The box diagram for two-pion exchange, shown in the text to be of similar order to the diagram for one-pion exchange.}
    \label{fig:boxdiagram}
\end{figure}

Evaluating all $LO$ diagrams between incoming and outgoing $NN$ states $\ket{\vec{p}}$ and $\ket{\vec{p}'}$ gives the $NN$-potential first elucidated by Weinberg~\cite{Weinberg1991}:
\begin{equation}
    \braket{\vec{p}' | V_{NN} | \vec{p}} = C - \tau_1 \cdot \tau_2 \frac{g_A^2}{2 F_\pi^2} \frac{ \left( \vec{p} - \vec{p'} \right) \cdot \sigma_1 \left( \vec{p} - \vec{p'} \right) \cdot \sigma_2}{(p-p')^2 + m_\pi^2}
\end{equation}
where $C = C_S - 3 C_T$ for a $^1S_0 \rightarrow ^1S_0$ transition. The derivation of the pion-exchange term is unimportant here, but will prove key to our argument in Section 4. From the chiral power-counting and now manifestly from the evaluation between states, both terms of the $LO$ potential are of order $\mathcal{O}(Q^0)$. Consider then the two-pion exchange ``box diagram'' of Figure \ref{fig:boxdiagram}. While the heavy-baryon $NN$-propagator contributes $Q^{-1}$, the non-relativistic loop integration contributes $Q^3$, and we find that the diagram is suppressed by $Q^2$ compared to one-pion exchange; the same result would hold with any or all of the pion-exchanges replaced by $NN$-contact interactions. 

This power-counting does not survive integration using the static nucleon propagator $\frac{i}{p_0}$, which diverges uncontrollably for the reason that the static baryon approximation cannot consistently hold while $p_0 \rightarrow 0$. To see this, it is enough to observe that there is no such thing as a soft nucleon in our chiral limit; all components of the on-shell nucleon four-momentum $p$ cannot at once be $0$.

$NN$-pairs of kinetic energy $E$ experience a pairwise Green's function
\begin{equation}
    \braket{\vec{p} | G_0(E) | \vec{p}} = \frac{i}{E - \frac{p^2}{M_N}},
\end{equation}
whose enhancement can be estimated to be $\frac{M_N}{m_\pi}$ over the naive estimate by letting $|\vec{p}| \sim m_\pi$. The chiral power-counting would have us take $M_N$ to be of order $Q^0$, as $M_N \sim $ chiral symmetry-breaking scale $\Lambda_\chi$, whereby the suppression of the doubly-iterated $NN$ interaction drops to $Q^1$, still allowing perturbative expansion. Weinberg argues~\cite{Weinberg1991} that large nuclear scattering lengths observed at low energies justify the treatment of $M_N$ as a ``free parameter'', in the sense that its value is determined more directly by the number of colours in QCD than by the chiral flavour symmetry. Then taking $M_N \sim Q^{-1}$, while not quite numerically satisfying~\cite{Phillips2022}, is a compact way to bring the iterated $NN$ interaction down to $\mathcal{O}(Q^0)$. In summary, the enhancement of pure-nucleon intermediate states by both the pinch singularity and the separation of scales $M_N \gg m_\pi$ implies that it is equally probable, roughly speaking, for a pair of low-energy nucleons to propagate freely or to interact via the contact/pion-range interactions of $\mathcal{L}_{NN}$.

The resultant procedure for generating nuclear forces from $\chi EFT$ has become ubiquitous in the literature. First, one produces an effective $n$-nucleon potential by evaluating only those interaction diagrams which \textit{do not} include $n$-nucleon intermediate states, corresponding to joint Green's functions like $G_0(E)$. Then, one iterates this potential, for example by numerically solving a complete Schr\"odinger equation with the potential as input~\cite{Weinberg1990}. This iteration may in some cases be treated as a formal geometric series, and indeed this is the approach that will prove efficient for the study of $0\nu\beta\beta$ decay. The great success of Weinberg's scheme is that it allows for the non-perturbative computation of $n$-nucleon forces within $\chi EFT$; the great drawbacks are that the power-counting is not wholly consistent with the single-nucleon sector, and that the extension to higher-order contact terms (with spacetime derivatives) is undefined~\cite{Phillips2022}.

%\subsection{\texorpdfstring{$NN$ scattering in pionless EFT}{NN scattering in pionless EFT}}

Perhaps inadvertently, a toy model Weinberg used to demonstrate the above procedure in~\cite{Weinberg1991} emerged as a significant tool for modern nuclear theory. This is the pionless EFT ($\cancel{\pi}EFT$) which contains only the interactions of Eq.~\eqref{NNLagrangian} at LO, and supports only nucleonic degrees of freedom. As such, it is the simplest nuclear EFT valid at short distance scales~\cite{Phillips2022}, and yet is robust enough to indicate the renormalization procedure for diagrams containing two (or more) nucleons. We note that because the $m_q \rightarrow 0$ limit is no longer meaningful, this theory cannot be said to obey even a broken chiral symmetry. 

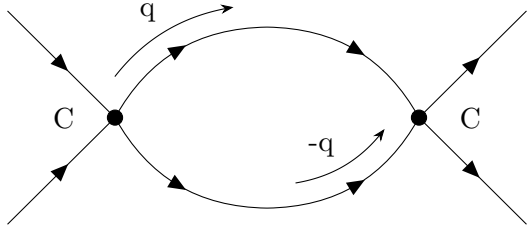
\begin{figure}[t]
\centering
    \begin{tikzpicture}
          \begin{feynman}
            \vertex (l);
            \vertex [above left=2cm of l] (ul);
            \vertex [below left=2cm of l] (dl);
            \vertex [right=2cm of l] (m);
            \vertex [above=12mm of m] (um);
            \vertex [below=12mm of m] (dm);
            \vertex [right=2cm of m] (r);
            \vertex [above right=2cm of r] (ur);
            \vertex [below right=2cm of r] (dr);
            \vertex [left=4mm of l] {C};
            \vertex [right=4mm of r] {C};
        
            \diagram* {
              (ul) -- [fermion] (l) -- [fermion, bend left, momentum=q] (um) -- [fermion, bend left] (r) -- [fermion] (ur),
              (dl) -- [fermion] (l) -- [fermion, bend right] (dm) -- [fermion, bend right, momentum=-q] (r) -- [fermion] (dr),
            };

            \draw[fill=black] (l) circle(1mm);
            \draw[fill=black] (r) circle(1mm);
          \end{feynman}
    \end{tikzpicture}
    \caption{The one-loop renormalisation diagram of four-nucleon contact interaction $C$ in pionless EFT ($\cancel{\pi}EFT$), as encoded in integral $T_2(E)$.}
    \label{fig:T2_diagram}
\end{figure}
First we recall the demonstration by~\cite{Weinberg1991} of the one-loop renormalisation of four-nucleon contact strength $C$, shown diagrammatically in Figure \ref{fig:T2_diagram}. The integral to be evaluated is
\begin{equation}
    T_2(E) = C^2 \int \frac{d^3q}{(2\pi)^3} \braket{\vec{q} | G_0(E) | \vec{q}} = C^2 \int \frac{d^3q}{(2\pi)^3} \frac{1}{E - \frac{q^2}{M_N} + i\epsilon},
\end{equation}
which in dimensional regularisation with $D = 3 - \varepsilon$ gives~\cite{Kaplan1996}
\begin{equation}
    T_2^{D=3-\varepsilon}(E) = -C^2 (4\pi)^{-\frac{3-\varepsilon}{2}} \mu^\epsilon M_N (-M_N E - i \epsilon)^{\frac{1-\varepsilon}{2}} \Gamma \left( \frac{\varepsilon-1}{2} \right).
    \label{dimregNN}
\end{equation}
Letting $\varepsilon \rightarrow 0$ gives a finite regularised result, with $E = \frac{p^2}{M}$ for the reduced-mass $NN$ system:
\begin{equation}
    T_2(E) = \lim_{\epsilon \rightarrow 0} T_2^{D=3-\varepsilon}(E) = \frac{-i C^2 M_N |\vec{p}|}{4\pi}.
\end{equation}
%
%How large is $C$? \textcolor{red}{Unsure why, but~\cite{Phillips2022} claims $C \sim \frac{4\pi}{M_N m_\pi}$ from ``naturalness"} Then, $T_2(E) \sim C \frac{|\vec{p}|}{m_\pi}$ -- suppression compared to hoped-for non-perturbativity?

However, Weinberg does not perform the renormalisation of $C$ order-by-order in~\cite{Weinberg1991}, nor even attempt it. Realizing that a chain of $k$ contact loops is in fact separable into $k+1$ interaction vertices $C$ and $k$ bubbles $\frac{T_2(E)}{C^2}$, and applying the infinite geometric series property $\sum_{k=0}^\infty a r^k = a (1-r)^{-1}$ with radius of convergence $|r| < 1$, he writes 
\begin{equation}
    T_\infty(E) = \left[ \frac{1}{C} - \int \frac{d^3q}{(2\pi)^3} \frac{1}{E - \frac{q^2}{M_N} + i\epsilon} \right]^{-1},
\end{equation}
which may be encoded diagrammatically, following~\cite{Kaplan1996}, as 
\begin{align}
T_\infty(E) &=~ 
\vcenter{\hbox{\begin{tikzpicture}
  \begin{feynman}
    \vertex (a);
    \vertex [above left=5mm of a] (ul);
    \vertex [above right=5mm of a] (ur);
    \vertex [below left=5mm of a] (dl);
    \vertex [below right=5mm of a] (dr);
    \diagram*{
      (a) -- [solid] (ul);
      (a) -- [solid] (ur);
      (a) -- [solid] (dl);
      (a) -- [solid] (dr);
    };
    \draw[fill=black] (a) circle(1mm);
  \end{feynman}
\end{tikzpicture}}}
~+~
\vcenter{\hbox{\begin{tikzpicture}
  \begin{feynman}
    \vertex (a);
    \vertex [above left=5mm of a] (ul);
    \vertex [below left=5mm of a] (dl);
    \vertex [right=8mm of a] (b);
    \vertex [above right=5mm of b] (ur);
    \vertex [below right=5mm of b] (dr);
    \diagram*{
      (a) -- [solid] (ul);
      (a) -- [solid] (dl);
      (a) -- [solid, bend left] (b);
      (a) -- [solid, bend right] (b);
      (b) -- [solid] (ur);
      (b) -- [solid] (dr);
    };
    \draw[fill=black] (a) circle(1mm);
    \draw[fill=black] (b) circle(1mm);
  \end{feynman}
\end{tikzpicture}}}
~+~
\vcenter{\hbox{\begin{tikzpicture}
  \begin{feynman}
    \vertex (a);
    \vertex [above left=5mm of a] (ul);
    \vertex [below left=5mm of a] (dl);
    \vertex [right=8mm of a] (b);
    \vertex [right=8mm of b] (c);
    \vertex [above right=5mm of c] (ur);
    \vertex [below right=5mm of c] (dr);
    \diagram*{
      (a) -- [solid] (ul);
      (a) -- [solid] (dl);
      (a) -- [solid, bend left] (b) -- [solid, bend left] (c);
      (a) -- [solid, bend right] (b) -- [solid, bend right] (c);
      (c) -- [solid] (ur);
      (c) -- [solid] (dr);
    };
    \draw[fill=black] (a) circle(1mm);
    \draw[fill=black] (b) circle(1mm);
    \draw[fill=black] (c) circle(1mm);
  \end{feynman}
\end{tikzpicture}}}
~+ \vcenter{\hbox{...}}\nonumber\\
&=~ 
\vcenter{\hbox{\begin{tikzpicture}
  \begin{feynman}
    \vertex (a);
    \vertex [above left=5mm of a] (ul);
    \vertex [above right=5mm of a] (ur);
    \vertex [below left=5mm of a] (dl);
    \vertex [below right=5mm of a] (dr);
    \diagram*{
      (a) -- [solid] (ul);
      (a) -- [solid] (ur);
      (a) -- [solid] (dl);
      (a) -- [solid] (dr);
    };
    \draw[fill=black] (a) circle(1mm);
  \end{feynman}
\end{tikzpicture}}}
~\left( 1 -~
\vcenter{\hbox{\begin{tikzpicture}
  \begin{feynman}
    \vertex (a);
    \vertex [above left=5mm of a] (ul);
    \vertex [below left=5mm of a] (dl);
    \vertex [right=8mm of a] (b);
    \vertex [above right=5mm of b] (ur);
    \vertex [below right=5mm of b] (dr);
    \diagram*{
      (a) -- [solid, bend left] (b);
      (a) -- [solid, bend right] (b);
    };
    \draw[fill=black] (a) circle(1mm);
  \end{feynman}
\end{tikzpicture}}}
  ~\right)^{-1}.
  \label{Tinf_diagrams}
\end{align}
Then absorbing any infrared divergences in $T_\infty(E)$ by defining $C_R \equiv T_\infty(0)$, the geometric series simplifies to
\begin{equation}
    T_\infty(E) = \left[ \frac{1}{C_R} + \frac{i M_N |\vec{p}|}{4\pi} \right]^{-1}.
    \label{pionlessNN}
\end{equation}
This result already contains all the key physics of $NN$ scattering in $\cancel{\pi}EFT$, even if some loose ends need tying up. It can be matched directly to the so-called effective range expansion, an experimentally-determined~\cite{Noyes1972} fit to $NN$-scattering phase-shift data:
\begin{equation}
    i |\vec{p}| - \frac{4\pi}{M_N} \frac{1}{T(E)} = - \frac{1}{a} + \frac{1}{2} r_0 p^2 + \mathcal{O}(p^4),
\end{equation}
where $a$ is known as the scattering length, and $r_0$ the effective range. Ref.~\cite{Kaplan1996} quotes $^1S_0$ np scattering values $a = -23.714 \pm 0.013$~fm and $r_0 = 2.73 \pm 0.03$~fm from~\cite{Preston2018} as a point of reference. Evaluating with $T_\infty(E)$ gives:
\begin{equation}
    a = \frac{M_N}{4\pi} C_R, \quad r_0 = 0.
\end{equation}
Two important conclusions must be drawn. The first is that large scattering lengths, as are observed for shallow $NN$ bound states (those near the unitarity limit), necessitate large contact couplings $C_R$ in Weinberg's approach. The second is that the ``natural" proportionality $C \sim \frac{4\pi}{M_N m_\pi}$~\cite{Phillips2022} implies $a \sim 1/m_\pi \sim 1.5$~fm; by defying this expectation, the experimental scattering length suggests that this effective theory formalism is not in fact consistent up to the $m_\pi$ scale. Ref.~\cite{Kaplan1996} cleanly shows the converse by working out $T(E)$ to NLO in $\cancel{\pi}EFT$, and observes a breakdown of the chiral expansion around $\Lambda \sim 35$~MeV, a factor of five below $m_\pi$.

In landmark work, Kaplan et al.~\cite{Kaplan1996} showed that both of these limitations can be overcome by an alternative regularisation scheme, referred to as power-divergence subtraction (PDS) or alternatively as the Kaplan-Savage-Wise (KSW) scheme. First, one notes that the dimensionally-regularised expression of Eq.~\eqref{dimregNN}, while finite at $\varepsilon = 0$ and therefore requiring no counterterm in the minimal subtraction $MS$ scheme, does diverge linearly at $\varepsilon = -1$. The idea of PDS is to subtract this divergence with an additional counterterm, dependent on a regularisation scale $\mu$~\cite{Kaplan1998}. Then,
\begin{equation}
    T_\infty^{PDS}(E) = \left[ \frac{1}{C(p^2,\mu)} + \frac{M_N (\mu + i|\vec{p}|)}{4\pi} \right]^{-1},
    \label{pionlessNN_PDS}
\end{equation}
which clearly matches the MS result if $\mu=0$.

The above approach may seem arbitrary, and indeed surprising given the familiar result that physical results obtained from different perturbative renormalisation schemes are identical; only the sum of divergence and counterterm is observable. As a toy model, Ref.~\cite{Phillips1997} instead argues that for delta potentials the physical amplitude can be dependent on the non-perturbative renormalisation scheme, in particular with different results for $MS$ dimensional regularisation compared with cutoff regularisation. Since the $MS$ removal of divergences is a choice of renormalisation scheme beyond the regularisation of a loop integral, $PDS$ achieves success by making an alternative choice more compatible with cutoff renormalisation.

More precisely, bound states with large scattering lengths, also known as shallow bound states, are those where the kinetic and potential energies of the state nearly cancel~\cite{Kaplan1998}. But, neither the kinetic nor the potential energy is an independent observable, and so this cancellation is an unfortunate side effect of MS renormalisation. The additional scale $\mu$ provides a sliding cutoff which separates the kinetic and potential terms and can be chosen at a value which provides a consistent power-counting. Specifically, matching to the effective range expansion now gives
\begin{equation}
    C_R = \frac{4\pi}{M_N} \left( \frac{1}{a} - \mu \right)^{-1},
\end{equation}
so even for large scattering length $a$, the natural choice $\mu \sim m_\pi$ facilitates $C \sim \mathcal{O}(Q^{-1})$ as initially claimed, while $\mu = 0$ again reproduces the $MS$ result. With renormalisation complete, the power-counting of all diagrams in the theory is well-defined: the iteration of the lowest order contact term $C$ has been justified, while all higher-order vertices are counted perturbatively~\cite{Kaplan1998}.

$\cancel{\pi}EFT$ has proven useful to the nuclear community because large scattering lengths manifest in many physical systems beyond $NN$. For example, it has been applied to so-called halo nuclei whose valence nucleons largely have density outside the classical interaction radius~\cite{Hammer2017}. Since a large scattering length in comparison to the interaction radius implies that the bound-state properties are driven by tunnelling, an EFT approach need not probe many details of the internucleon potential to be descriptive of such systems~\cite{Phillips2022}. $\cancel{\pi}EFT$ has also successfully been applied to the Efimov effect~\cite{Efimov1970}, the observation that three particles with short-range attractions near threshold can induce long-range, specifically three-body interactions~\cite{Naidon2022}. A review of practical applications of the pionless framework to few-nucleon systems may be found in~\cite{Bedaque2002}; these include electromagnetic form factors for the deuteron, deuteron formation in the context of Big Bang nucleosynthesis, and weak-deuteron interactions. While our focus is on the $NN$ system as relevant to $0\nu\beta\beta$-decay, many of our $\chi EFT$ conclusions will first be motivated within $\cancel{\pi}EFT$. 
% Could also mention Phillips line, Tjon line

% Weinberg90: https://www.sciencedirect.com/science/article/abs/pii/0370269390909383?fr=RR-2&ref=pdf_download&rr=7c20d0431cbc23e8
% Weinberg91: https://www.sciencedirect.com/science/article/abs/pii/055032139190231L?fr=RR-2&ref=pdf_download&rr=7c20ccbd2a3e23e8
% Weinberg92: https://arxiv.org/pdf/hep-ph/9209257.pdf
% Luke/Manohar96: https://arxiv.org/pdf/hep-ph/9610534.pdf
% Kaplan1996: https://arxiv.org/pdf/nucl-th/9605002.pdf
% another Kaplan1996: https://arxiv.org/pdf/nucl-th/9610052.pdf
% Kaplan98: https://arxiv.org/pdf/nucl-th/9801034.pdf
% another Kaplan98: https://arxiv.org/pdf/nucl-th/9802075.pdf
% Birse06: https://arxiv.org/pdf/nucl-th/0507077.pdf
% Epelbaum20: https://www.frontiersin.org/articles/10.3389/fphy.2020.00098/full#note3
% Phillips21: https://arxiv.org/pdf/2107.03558.pdf
% Might want to cite van Kolck 3N forces somewhere in here too?

%\subsection{\texorpdfstring{NN scattering in $\chi EFT$}{NN scattering in chiral EFT}}

The value of including explicit pion fields and interactions in our model as opposed to the $\cancel{\pi}EFT$ approach (beyond the realisation of chiral symmetry) is that the EFT should be valid to higher energies -- naively, up to the mass scale of the next ``integrated-out'' degree of freedom, the rho meson.

Weinberg's power-counting implied that a one-pion exchange potential should be iterated to all orders, just as the four-nucleon contact potential was. The joint iteration of two potentials requires  some careful algebra, detailed in ~\cite{Kaplan1996}. Let us define the free, retarded two-nucleon Green's function as $\hat{G}^0_E = \left[ E - \hat{H^0} + i\epsilon \right]^{-1}$ and the two-nucleon potential as $\hat{V}$, both of which will have simple actions on the space of momentum eigenstates $\ket{\vec{p}}$. Writing the geometric series
\begin{equation}
    \hat{V} + \hat{V}\hat{G}^0_E \hat{V} + \hat{V}(\hat{G}^0_E \hat{V})^2 + ... = \sum_{k=0}^\infty \hat{V}(\hat{G}^0_E \hat{V})^k \equiv \hat{V} + \hat{V}\hat{G}_E \hat{V},
\end{equation}
allows us to define an interacting Green's function $\hat{G}_E$. Reducing the geometric series gives the simple relations
\begin{equation}
    \hat{G}_E = \frac{\hat{G}_E^0}{1 - \hat{G}_E^0 \hat{V}}, \quad \hat{G}_E^{-1} = \left( \hat{G}^0_E \right)^{-1} - \hat{V} = \left[ E - \hat{H}^0 - \hat{V} + i\epsilon \right]^{-1},
\end{equation}
which relate the `dressed' and `undressed' Green's functions. It is then straightforward to uniquely define the scattering state $\ket{\chi_{\vec{p}}}$ as the eigenvector of $\hat{H}$ with eigenvalue $E = \frac{p^2}{M_N}$, or equivalently such that 
\begin{equation}
    \hat{G}_E^{-1} \ket{\chi_{\vec{p}}} = 0, \quad \ket{\chi_{\vec{p}}} = (1 + \hat{G}_E\hat{V})\ket{\vec{p}},
\end{equation}
where the latter relation is consistent with factorising the iterated potential amplitude as 
\begin{equation}
    \braket{\vec{p} | \hat{V} + \hat{V}\hat{G}_E\hat{V} | \vec{p}'} = \braket{\vec{p} | \hat{V} | \chi_{\vec{p}'}} = \int d^3r e^{-i \vec{p} \cdot \vec{r}} V(\vec{r}) \chi_{\vec{p}'}(\vec{r}).
\end{equation}
With these definitions in place, the strategy of~\cite{Kaplan1996} is to absorb the four-nucleon contact interaction $C$ into $\hat{H_0}$ in the sense of $\cancel{\pi}EFT$, making use of the same explicit power series solution. $\hat{V}$ then consists only of the one-pion-exchange potential, and it is with (iterated) pion exchange that the scattering state $\ket{\chi_{\vec{p}}}$ and two-nucleon propagator $\hat{G}_E$ are therefore dressed. In $\cancel{\pi}EFT$, $\ket{\chi_{\vec{p}}}$ reduces to the plane-wave state $\ket{\vec{p}}$, $\hat{G}_E$ reduces to the free two-nucleon propagator $\hat{G}_E^0$, and the $NN$-scattering amplitude corresponds to Eq.~\eqref{pionlessNN} or Eq.~\eqref{pionlessNN_PDS} depending on subtraction scheme.

We can now precisely state how this framework changes when pions are included. First, the pion-exchange potential induces a correction in the $NN$ contact coupling, $C \mapsto \tilde{C} = C + \frac{g_A^2}{2 f_\pi^2}$. Iterating as before, the T-matrix is~\cite{Kaplan1996}
\begin{equation}
    T_\infty^{\chi EFT}(E) = T_\infty^\pi(E) + \left[ \chi_{\vec{p}}(0) \right]^2 \left[ \frac{1}{C} + \tilde{G}_E \right]^{-1},
    \label{pionlessNN_PDSkaplan}
\end{equation}
so long as one can evaluate scattering state $\ket{\chi_{\vec{p}}}$ at $\vec{r}=0$, and compute $\tilde{G}_E = \int\int \frac{d^3q}{(2\pi)^3} \frac{d^3q'}{(2\pi)^3} \braket{\vec{q}'|\hat{G}_E|\vec{q}}$, the dressed two-nucleon propagator from $\vec{r}=0$ to $\vec{r}'=0$. $T_\infty^\pi(E)$ is the iterated one-pion-exchange contribution, i.e. the collection of terms containing no contact interactions. This series can be represented diagrammatically as follows, adapted from~\cite{Kaplan1996}:
\begin{align}
T_\infty^{\chi EFT}(E) &=~ 
\vcenter{\hbox{\begin{tikzpicture}
  \begin{feynman}
    \vertex (a);
    \vertex [above left=5mm of a] (ul);
    \vertex [left=3mm of ul] (ull);
    \vertex [above right=5mm of a] (ur);
    \vertex [right=3mm of ur] (urr);
    \vertex [below left=5mm of a] (dl);
    \vertex [left=3mm of dl] (dll);
    \vertex [below right=5mm of a] (dr);
    \vertex [right=3mm of dr] (drr);
    \diagram*{
      (ull) -- [solid] (urr);
      (dll) -- [solid] (drr);
      (ul) -- [dashed] (dl);
    };
  \end{feynman}
  \draw[line width=1.5mm] (ur) -- (dr);
\end{tikzpicture}}}
~+~
\vcenter{\hbox{\begin{tikzpicture}
  \begin{feynman}
    \vertex (a);
    \vertex [above left=5mm of a] (ul);
    \vertex [left=3mm of ul] (ull);
    \vertex [above right=5mm of a] (ur);
    \vertex [right=3mm of ur] (urr);
    \vertex [below left=5mm of a] (dl);
    \vertex [left=3mm of dl] (dll);
    \vertex [below right=5mm of a] (dr);
    \vertex [right=3mm of dr] (drr);
    \diagram*{
      (ull) -- [solid] (ul) -- [solid, bend left] (a) -- [solid, bend left] (ur) -- [solid] (urr);
      (dll) -- [solid] (dl) -- [solid, bend right] (a) -- [solid, bend right] (dr) -- [solid] (drr);
    };
    \draw[fill=black] (a) circle(1mm);
  \end{feynman}
  \draw[line width=1.5mm] (ul) -- (dl);
  \draw[line width=1.5mm] (ur) -- (dr);
\end{tikzpicture}}}
~+~
\vcenter{\hbox{\begin{tikzpicture}
  \begin{feynman}
    \vertex (a);
    \vertex [above left=5mm of a] (ul);
    \vertex [left=3mm of ul] (ull);
    \vertex [below left=5mm of a] (dl);
    \vertex [left=3mm of dl] (dll);
    \vertex [right=4mm of a] (ab);
    \vertex [above=3mm of ab] (uab);
    \vertex [below=3mm of ab] (dab);
    \vertex [right=4mm of ab] (b);
    \vertex [above right=5mm of b] (ur);
    \vertex [right=3mm of ur] (urr);
    \vertex [below right=5mm of b] (dr);
    \vertex [right=3mm of dr] (drr);
    \diagram*{
      (ull) -- [solid] (ul) -- [solid, bend left] (a) -- [solid, bend left] (uab) -- [solid, bend left] (b) -- [solid, bend left] (ur) -- [solid] (urr);
      (dll) -- [solid] (dl) -- [solid, bend right] (a) -- [solid, bend right] (dab) -- [solid, bend right] (b) --[solid, bend right] (dr) -- [solid] (drr);
    };
    \draw[fill=black] (a) circle(1mm);
    \draw[fill=black] (b) circle(1mm);
  \end{feynman}
  \draw[line width=1.5mm] (ul) -- (dl);
  \draw[line width=1.5mm] (uab) -- (dab);
  \draw[line width=1.5mm] (ur) -- (dr);
\end{tikzpicture}}}
~+ \vcenter{\hbox{...}}~\nonumber\\
&=~ 
\vcenter{\hbox{\begin{tikzpicture}
  \begin{feynman}
    \vertex (a);
    \vertex [above left=5mm of a] (ul);
    \vertex [left=3mm of ul] (ull);
    \vertex [above right=5mm of a] (ur);
    \vertex [right=3mm of ur] (urr);
    \vertex [below left=5mm of a] (dl);
    \vertex [left=3mm of dl] (dll);
    \vertex [below right=5mm of a] (dr);
    \vertex [right=3mm of dr] (drr);
    \diagram*{
      (ull) -- [solid] (urr);
      (dll) -- [solid] (drr);
      (ul) -- [dashed] (dl);
    };
  \end{feynman}
  \draw[line width=1.5mm] (ur) -- (dr);
\end{tikzpicture}}}
~+~
\vcenter{\hbox{\begin{tikzpicture}
  \begin{feynman}
    \vertex (a);
    \vertex [above left=5mm of a] (ul);
    \vertex [left=3mm of ul] (ull);
    \vertex [above right=5mm of a] (ur);
    \vertex [right=3mm of ur] (urr);
    \vertex [below left=5mm of a] (dl);
    \vertex [left=3mm of dl] (dll);
    \vertex [below right=5mm of a] (dr);
    \vertex [right=3mm of dr] (drr);
    \diagram*{
      (ull) -- [solid] (ul) -- [solid, bend left] (a) -- [solid, bend left] (ur) -- [solid] (urr);
      (dll) -- [solid] (dl) -- [solid, bend right] (a) -- [solid, bend right] (dr) -- [solid] (drr);
    };
    \draw[fill=black] (a) circle(1mm);
  \end{feynman}
  \draw[line width=1.5mm] (ul) -- (dl);
  \draw[line width=1.5mm] (ur) -- (dr);
\end{tikzpicture}}}
~\left( 1 -~
\vcenter{\hbox{\begin{tikzpicture}
  \begin{feynman}
    \vertex (a);
    \vertex [right=4mm of a] (ab);
    \vertex [above=3mm of ab] (uab);
    \vertex [below=3mm of ab] (dab);
    \vertex [right=4mm of ab] (b);
    \diagram*{
      (a) -- [solid, bend left] (uab) -- [solid, bend left] (b);
      (a) -- [solid, bend right] (dab) -- [solid, bend right] (b);
    };
    \draw[fill=black] (a) circle(1mm);
  \end{feynman}
  \draw[line width=1.5mm] (uab) -- (dab);
\end{tikzpicture}}}
  ~\right)^{-1},
  \label{Tinf_chieft_diagrams}
\end{align}
where
\begin{align}
    \vcenter{\hbox{\begin{tikzpicture}
      \begin{feynman}
        \vertex (a);
        \vertex [above left=5mm of a] (ul);
        \vertex [left=3mm of ul] (ull);
        \vertex [above right=5mm of a] (ur);
        \vertex [right=7mm of ur] (urr);
        \vertex [below left=5mm of a] (dl);
        \vertex [left=3mm of dl] (dll);
        \vertex [below right=5mm of a] (dr);
        \vertex [right=7mm of dr] (drr);
        \diagram*{
          (ul) -- [solid] (urr);
          (dl) -- [solid] (drr);
        };
      \end{feynman}
      \draw[line width=1.5mm] (ur) -- (dr);
    \end{tikzpicture}}}
    ~&=~
    \vcenter{\hbox{\begin{tikzpicture}
      \begin{feynman}
        \vertex (a);
        \vertex [above left=5mm of a] (ul);
        \vertex [left=3mm of ul] (ull);
        \vertex [above right=5mm of a] (ur);
        \vertex [right=7mm of ur] (urr);
        \vertex [below left=5mm of a] (dl);
        \vertex [left=3mm of dl] (dll);
        \vertex [below right=5mm of a] (dr);
        \vertex [right=7mm of dr] (drr);
        \diagram*{
          (ul) -- [solid] (urr);
          (dl) -- [solid] (drr);
        };
      \end{feynman}
    \end{tikzpicture}}}
    ~+~
    \vcenter{\hbox{\begin{tikzpicture}
      \begin{feynman}
        \vertex (a);
        \vertex [above left=5mm of a] (ul);
        \vertex [left=3mm of ul] (ull);
        \vertex [above right=5mm of a] (ur);
        \vertex [right=7mm of ur] (urr);
        \vertex [below left=5mm of a] (dl);
        \vertex [left=3mm of dl] (dll);
        \vertex [below right=5mm of a] (dr);
        \vertex [right=7mm of dr] (drr);
        \diagram*{
          (ul) -- [solid] (urr);
          (dl) -- [solid] (drr);
          (ur) -- [dashed] (dr);
        };
      \end{feynman}
    \end{tikzpicture}}}
    ~+~
    \vcenter{\hbox{\begin{tikzpicture}
      \begin{feynman}
        \vertex (a);
        \vertex [above left=5mm of a] (ul);
        \vertex [left=3mm of ul] (ull);
        \vertex [left=7mm of ull] (ulll);
        \vertex [above right=5mm of a] (ur);
        \vertex [right=3mm of ur] (urr);
        \vertex [below left=5mm of a] (dl);
        \vertex [left=3mm of dl] (dll);
        \vertex [left=7mm of dll] (dlll);
        \vertex [below right=5mm of a] (dr);
        \vertex [right=3mm of dr] (drr);
        \diagram*{
          (ull) -- [solid] (urr);
          (dll) -- [solid] (drr);
          (ur) -- [dashed] (dr);
          (ul) -- [dashed] (dl);
        };
      \end{feynman}
    \end{tikzpicture}}}
    ~+~
    \vcenter{\hbox{\begin{tikzpicture}
      \begin{feynman}
        \vertex (a);
        \vertex [above left=5mm of a] (ul);
        \vertex [left=7mm of ul] (ull);
        \vertex [left=3mm of ull] (ulll);
        \vertex [above right=5mm of a] (ur);
        \vertex [right=3mm of ur] (urr);
        \vertex [below left=5mm of a] (dl);
        \vertex [left=7mm of dl] (dll);
        \vertex [left=3mm of dll] (dlll);
        \vertex [below right=5mm of a] (dr);
        \vertex [right=3mm of dr] (drr);
        \diagram*{
          (ulll) -- [solid] (urr);
          (dlll) -- [solid] (drr);
          (ur) -- [dashed] (dr);
          (ul) -- [dashed] (dl);
          (ull) -- [dashed] (dll);
        };
      \end{feynman}
    \end{tikzpicture}}}
    ~+ \vcenter{\hbox{...}},\nonumber\\
    \tilde{G}_E &=~
    \vcenter{\hbox{\begin{tikzpicture}
      \begin{feynman}
        \vertex (a);
        \vertex [right=4mm of a] (ab);
        \vertex [above=3mm of ab] (uab);
        \vertex [below=3mm of ab] (dab);
        \vertex [right=4mm of ab] (b);
        \diagram*{
          (a) -- [solid, bend left] (uab) -- [solid, bend left] (b);
          (a) -- [solid, bend right] (dab) -- [solid, bend right] (b);
        };
      \end{feynman}
      \draw[line width=1.5mm] (uab) -- (dab);
    \end{tikzpicture}}}
    ~,\hspace{1cm}~
    \chi_{\vec{p}}(0) =~
    \vcenter{\hbox{\begin{tikzpicture}
      \begin{feynman}
        \vertex (a);
        \vertex [above left=5mm of a] (ul);
        \vertex [left=3mm of ul] (ull);
        \vertex [below left=5mm of a] (dl);
        \vertex [left=3mm of dl] (dll);
        \diagram*{
          (ull) -- [solid] (ul) -- [solid, bend left] (a);
          (dll) -- [solid] (dl) -- [solid, bend right] (a);
        };
      \end{feynman}
      \draw[line width=1.5mm] (ul) -- (dl);
    \end{tikzpicture}}}
    ~.
    \label{pioniteration_diagrams}
\end{align}
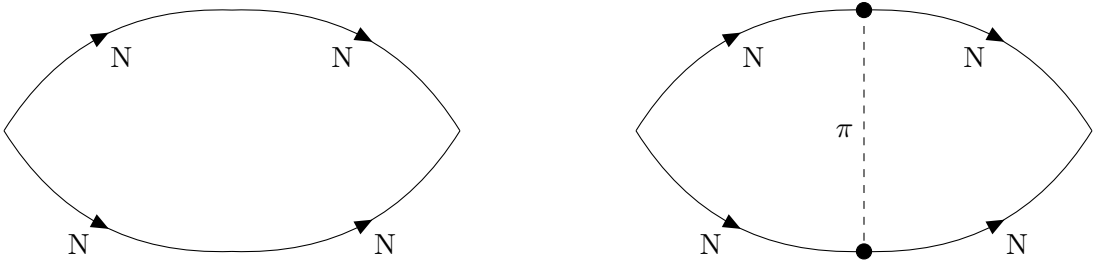
\begin{figure}[t]
    \centering % <-- added
    \begin{subfigure}{0.45\textwidth}
         \begin{tikzpicture}
          \begin{feynman}
            \vertex (l);
            \vertex [right=of l] (ma);
            \vertex [right=of ma] (mb);
            \vertex [right=of mb] (mc);
            \vertex [right=of mc] (r);

            \vertex [above=0.8cm of ma] (na);
            \vertex [above=0.8cm of mb] (nb);
            \vertex [above=0.8cm of mc] (nc);
            \vertex [above=0.8cm of na] (a);
            \vertex [above=0.8cm of nb] (b);
            \vertex [above=0.8cm of nc] (c);
            \vertex [below=0.8cm of ma] (ka);
            \vertex [below=0.8cm of mb] (kb);
            \vertex [below=0.8cm of mc] (kc);
            \vertex [below=0.8cm of ka] (d);
            \vertex [below=0.8cm of kb] (e);
            \vertex [below=0.8cm of kc] (f);
        
            \diagram* {
              (l) -- [fermion, bend left, edge label'=N] (b) -- [fermion, bend left, edge label'=N] (r),
              (l) -- [fermion, bend right, edge label'=N] (e) -- [fermion, bend right, edge label'=N] (r),
            };

          \end{feynman}
    \end{tikzpicture}
    \end{subfigure}\hfill % <-- added
    \begin{subfigure}{0.45\textwidth}
              \begin{tikzpicture}
          \begin{feynman}
            \vertex (l);
            \vertex [right=of l] (ma);
            \vertex [right=of ma] (mb);
            \vertex [right=of mb] (mc);
            \vertex [right=of mc] (r);

            \vertex [above=0.8cm of ma] (na);
            \vertex [above=0.8cm of mb] (nb);
            \vertex [above=0.8cm of mc] (nc);
            \vertex [above=0.8cm of na] (a);
            \vertex [above=0.8cm of nb] (b);
            \vertex [above=0.8cm of nc] (c);
            \vertex [below=0.8cm of ma] (ka);
            \vertex [below=0.8cm of mb] (kb);
            \vertex [below=0.8cm of mc] (kc);
            \vertex [below=0.8cm of ka] (d);
            \vertex [below=0.8cm of kb] (e);
            \vertex [below=0.8cm of kc] (f);
        
            \diagram* {
              (l) -- [fermion, bend left, edge label'=N] (b) -- [fermion, bend left, edge label'=N] (r),
              (l) -- [fermion, bend right, edge label'=N] (e) -- [fermion, bend right, edge label'=N] (r),
              (b) -- [dashed, edge label'=\(\pi\)] (e),
            };

            \draw[fill=black] (b) circle(1mm);
            \draw[fill=black] (e) circle(1mm);
          \end{feynman}
    \end{tikzpicture}
    \end{subfigure}
    \caption{Leading perturbative diagrams of chiral two-nucleon bubble $\tilde{G}(E)$, both of which contain regularisable divergences.}
    \label{fig:GEtwodiagrams}
\end{figure}
Green's function $\tilde{G}_E$ contains divergences from the first two diagrams in its perturbative series, shown in Figure \ref{fig:GEtwodiagrams}. These divergences must be regularised and absorbed into the renormalised interaction strength $\tilde{C}(\mu)$. It will be instructive to review the $\overline{MS}$ regularisation of these diagrams by~\cite{Kaplan1996}, as an analogous computation for divergent diagrams containing a neutrino-exchange operator will later prove relevant to $0\nu\beta\beta$-decay. The goal is to examine the divergent parts of these diagrams only; the finite parts, while calculable through numerical solution of the relevant Schr\"odinger equation, will not be directly relevant to us. Because the S-wave solutions to this Schr\"odinger equation are independent of $E$ as $r \rightarrow 0$,~\cite{Kaplan1996} sets $E = 0$ for convenience in these calculations.

As an example, we can explicitly evaluate the integral corresponding to the left of Figure \ref{fig:GEtwodiagrams}, the amplitude with no pion-exchange:
\begin{align}
    & \braket{\vec{r}'|\hat{G}_0^0|\vec{r}=0} \nonumber\\
    &= \iiint \frac{dr^3}{(2\pi)^3} \frac{dq^3}{(2\pi)^3} \frac{dq'^3}{(2\pi)^3} \braket{\vec{r}'|\vec{q}'} \frac{(2\pi)^3 \delta^{(3)}(\vec{q}' - \vec{q})}{-\frac{q^2}{M_N} + i\epsilon} \nonumber\\
    &= 2\pi \int_0^\infty q^2 dq \int_0^\pi d\theta \cos\theta e^{-i q r \cos\theta} \frac{M_N}{q^2 - i\epsilon} \nonumber\\
    &= 2\pi M_N \int_0^\infty dq \int_0^\pi d\theta \cos\theta e^{- q r \cos\theta} \nonumber\\
    &= \frac{2\pi M_N}{(2\pi)^3} \int_0^\pi d\theta \left[ -\frac{1}{r} e^{-q r \cos\theta} \right]_{q=\infty}^{q=0}\nonumber\\
    &= -\frac{M_N}{4\pi r'},
\end{align}
where in the fourth line, we take $q \mapsto -i q$ by the Cauchy theorem in the absence of any poles off the imaginary $q^2$-axis. The limit $r' \rightarrow 0$ is then singular. Ref.~\cite{Kaplan1996} shows that the irregular s-wave solution takes the following asymptotic form at small distances:
\begin{equation}
    \lim_{r' \rightarrow 0} K_E^\lambda(r' )  \rightarrow \frac{M}{4\pi r'} - \frac{\alpha_\pi M_N^2}{4\pi} \ln (\lambda r) + \mathcal{O}(r \ln r),
\end{equation}
where $\lambda$ is a choice of separation scale between the regular and irregular Schr\"odinger solutions. Crucially, the $1/r'$-divergence here exactly cancels that from our first divergence perturbative diagram. Since $\overline{MS}$ dimensional-regularisation reduces this diagram to $0$~\cite{Kaplan1996}, the conclusion is that this asymptotic Schr\"odinger regularisation agrees with dimensional regularisation to $1/r'$-order. 

What of the $\ln r'$-divergence? Ignoring finite contributions, direct integration of the amplitude $\braket{\vec{r}'|\hat{G}_0^0 \hat{V}_\pi \hat{G}_0^0 |\vec{r}=0}$ shown in the right of Figure \ref{fig:GEtwodiagrams} gives a divergent term $\frac{\alpha_\pi M_N^2}{4\pi} \ln r'$, which again precisely cancels the divergence of $K_E^\lambda(r')$. $\overline{MS}$ dimensional-regularisation at scale $\mu$ gives the divergent contribution $\frac{\alpha_\pi M_N^2}{4\pi} \ln \left( \frac{m_\pi}{\mu} \right)$ identical to the Schr\"odinger divergence with $\lambda \mapsto \mu$~\cite{Kaplan1996}, albeit with a distinct finite part. Therefore we learn two things: the renormalisation of $\tilde{C}$ is scheme-independent, and the kinematic point $\mu \sim m_\pi$ provides the most precise perturbative series, as expected from dimensional analysis.  

Ref.~\cite{Kaplan1996} promotes the numerical Schr\"odinger procedure for practical calculations in part because it is easily extended to higher-order chiral potentials. Indeed, this approach will also extend to the inclusion of neutrino potentials, but backed up with rigorous agreement to dimensional regularisation. A clever way to see the manifest symmetry-preservation of dimensional regularisation is to observe~\cite{Long2016} that the $D\pi DN DN^\dagger$ path-integral measure transforms non-trivially under chiral symmetry, requiring the introduction of a counterterm; this counterterm vanishes with dimensional regularisation. Therefore momentum cutoffs must be used with some caution, either by checking for symmetry violations post-calculation~\cite{Bernard2004}, or by including counterterms in the Lagrangian which cancel the effects of cutoff regularisation at each chiral order~\cite{Long2016}.

In effect, the choice of PDS subtraction during regularisation of the leading divergences has implied a new power-counting for $NN$ interactions in $\chi EFT$, in which pion exchanges are treated perturbatively with convergence guaranteed for momentum scales less than $\Lambda_{NN} \simeq 300$~MeV~\cite{Kaplan1998} -- the Kaplan-Savage-Wise (KSW) power-counting, in contrast to the Weinberg power-counting. This approach has not been without controversy, shown in~\cite{Fleming2000} to not converge for spin-triplet channels such as $^3S_1$, where two-pion exchange contributions are large.  One resolution to this quandary has been to apply KSW counting only in the $^1S_0$ partial-wave, where the distinctive hard-core interaction between nucleons facilitates a perturbative treatment of pion-exchanges~\cite{Beane2002}. Because the singular tensor component of the pion potential in higher partial-waves seeded the failure of the KSW counting in these cases, a direct regularisation of the pion propagator~\cite{Beane2009} is another route to perturbative pion physics beyond $^1S_0$. A more modern approach which has been shown to be scaleable through $N^4LO$ in the chiral expansion is a for-purpose regularisation of contact interactions by nonlocal Gaussian cutoffs, while pion-range interactions are regularised at the level of the propagator -- thus, semilocal momentum-space regularisation~\cite{Epelbaum2022}. Although the renormalisation of pionic $NN$-forces has been thoroughly computed in higher partial-waves~\cite{Nogga2005,Birse2006}, this will play only a small role in our analysis -- the enhanced contact term for $0\nu\beta\beta$ to be discussed will only appear in the $^1S_0$ partial-wave.

%\subsection{\texorpdfstring{Few- and many-nucleon forces from $\chi EFT$}{Few- and many-nucleon forces from chiral EFT}}

We have now introduced all the primary tools and methodologies which facilitate~\cite{Cirigliano2018} in treating the leading-order divergent diagrams of $0\nu\beta\beta$ in $\chi EFT$ --  these will prove essential to our refined estimate of the contact term. The unifying feature which binds all these EFT-driven nuclear computations together, whether in the symmetry-constrained framework of $\chi EFT$ or the satisfying simplicity of $\cancel{\pi} EFT$, is that each and every observable quantity comes with a quantitative estimate of error. As in any EFT with a consistent power-counting, this bound emerges from the truncation in the chiral expansion, which can be either treated somewhat approximately~\cite{Epelbaum2015} or formalised as a Bayesian inference procedure~\cite{Furnstahl2015,Melendez2017,Wesolowski2019} such that any prior-assumption dependence, regarding the coefficients of the chiral expansion of the observable, is controlled. Without such tools, nuclear theorists have historically had to rely upon variations of cutoffs and fit parameters to indirectly estimate the theoretical uncertainty of a computation. Although informative, such estimates are not statistically rigorous, and unlike an EFT power-counting, do not allow the practitioner to anticipate at what scale higher-order mechanisms may become relevant to a process

%To quote one advocate of the ab initio nuclear programme, ``real theorists have error bars" (Timmermanns, 1999)~\cite{Phillips2022}. 

\section{\texorpdfstring{Renormalisation Enhancement and the $0\nu\beta\beta$ ``Contact Term"}{Renormalisation Enhancement and the 0vbb ``Contact Term"}}

%\begin{itemize}
%    \item \todo{Neutrino potential in chiral EFT, with and without nucleon form factors}
%    \item \todo{Resummation of NN-scattering, pion-exchange, and neutrino potential}
%    \item \todo{Leading divergence in bubble diagrams, [Cirigliano2019]}
%    \item \todo{Contact counterterm unnecessary in other partial waves, dimensional analysis}
%    \item \todo{Intermediate momentum exchange: nucleon and half-off-shell form factors}
%    \item \todo{High momentum exchange: pQCD and operator product expansion}
%    \item \todo{Matching estimate for contact counterterm, [Cirigliano2021]}
%\end{itemize}

We have now overviewed the development of $\chi EFT$-derived nuclear potentials, and in particular their application to shallow two-nucleon bound states. Now we will present the state of the field in applying this ab initio technology to the $0\nu\beta\beta$ transition in the two-nucleon system, following the seminal works by Cirigliano et al.~\cite{Cirigliano2018,Cirigliano2019,Cirigliano2021short,Cirigliano2021long}. To our knowledge, this is the first thorough review of the Cirigliano group's approach in print outside of their own publications, and as such we will aim to present the core results simply, but with an emphasis on how they represent a general viewpoint towards calculations of nuclear amplitudes (both SM and BSM) which is very much in the spirit of systematically-improveable effective theories. We will also highlight the explicit and implicit assumptions which are manifest in this approach, the most numerically significant of which we will attempt to expand upon in Section 4.

The procedure for realizing the two-nucleon contribution to $0\nu\beta\beta$-decay at LO in the $\chi EFT$ expansion has three stages. First, a ``neutrino potential" corresponding to the $\Delta L = 2$ $NN \rightarrow NN$ operator of choice must be defined between any relevant $NN$ partial waves, as first performed in~\cite{Cirigliano2018}. Second, in the style of the $NN$ scattering computations presented in Section 2.2~\cite{Kaplan1996}, this potential must be dressed with resummed contact and pion-exchange potential operators, and renormalised where divergences appear to produce a consistent non-perturbative amplitude. The key result of~\cite{Cirigliano2019} is that this renormalisation enhances a doubly-LNV dimension-9 contact counterterm to LO in the chiral power-counting, which is not accounted for in standard phenomenological nuclear calculations of the decay. Third and finally, although the contact-term contribution to the full two-nucleon amplitude depends on an unknown LEC which will ultimately require input from LQCD, its size may be estimated by a matching analysis as illustrated in~\cite{Cirigliano2021long}. 

\subsection{\texorpdfstring{Neutrino potential in $\chi EFT$}{Neutrino potential in chiral EFT}}

In order to evaluate the T-matrix for an $NN$ neutrino exchange, the neutrino-lepton vertex on a single nucleon must first be specified. Given that Lorentz-invariance severely constrains the form of this vertex, an efficient approach is to first decompose the vertex into all permitted Lorentz structures, whose unknown kinematic dependence are encoded in form factors. In purely phenomenological approaches to nuclear theory, these form factors are derived from experiment using simple parameterisations such as dipoles; in an ab initio approach, each form factor is computable from $\chi EFT$. 
%There is of course no obstacle to hybridising these two approaches, for example in using a lesser degree of phenomenological data to inform a form factor computation on the lattice. 

Following~\cite{Simkovic1999}, the effective single-$\beta$ decay Hamiltonian is
\begin{equation}
    \mathcal{H}_\beta = \frac{G_F}{\sqrt{2}} \left[ \overline{e} \gamma_\mu (1 - \gamma_5) \nu_{e,L} \right] J_L^{\mu\dagger} + h.c.,
\end{equation}
where the expression in square brackets is the usual leptonic weak current, which is left-handed ($V-A$) -- note that full details of neutrino potentials for ``leptonic'' BSM operators of all Lorentz structures may be found in~\cite{Cirigliano2017}. Taking the impulse approximation for nucleon-level calculations, the hadronic current consists of six independent Lorentz structures~\cite{Towner1995}:
\begin{align}
   J^\mu_L = J^\mu_V - J^\mu_A = \overline{N} \tau^+ \bigg[ &g_V(q^2) \gamma^\mu + g_M(q^2) \frac{\sigma^{\mu\nu}}{2M_N}q_\nu  + i g_S(q^2) q^\mu \\ \nonumber
   &- g_A(q^2) \gamma^\mu \gamma_5 - g_T(q^2) \frac{\sigma^{\mu\nu}}{2M_N}q_\nu \gamma_5 - g_P(q^2) q^\mu \gamma_5 \bigg] N,
\end{align}
where $\sigma^{\mu\nu} = \frac{i}{2}[\gamma^\mu,\gamma^\nu]$. Imposing hermiticity and time-reversal invariance, scalar and tensor terms $g_S(q^2) = g_T(q^2) = 0$, and the other form factors are guaranteed to be real~\cite{Towner1995}.  It is convenient to write this current in the non-relativistic limit~\cite{Simkovic1999}:
\begin{equation}
   J_{L,NR}^\mu = \overline{N} \tau^+ \left[ g^{\mu,0} g_V(q^2) + g^{\mu,i}  \left( i g_M(q^2) \frac{\epsilon_{ijk} \sigma^j q^k }{2 M_N} + g_A(q^2) \sigma^i - g_P(q^2) \frac{q^i \sigma \cdot q}{2 M_N} \right) \right] N.
   \label{piNNcurrents}
\end{equation}
While the above separation is guided by phenomenological simplicity, symmetry properties also support the conserved vector-current (CVC) hypothesis and partially-conserved axial-current (PCAC) hypothesis. CVC allows one to set $g_V(0) = 1$ regardless of host isotope, and to assign dipole approximations to both $g_V$ and $g_M$~\cite{Simkovic1999}:
\begin{equation}
   g_V(q^2) = g_V(0) \frac{m_V^2}{q^2 + m_V^2},
\end{equation}
\begin{equation}
   g_M(q^2) = (1 + \kappa_1) g_V(q^2).
\end{equation}
PCAC connects the divergence of the axial-current to measurable pion decay, allowing both $g_P$ and $g_A$ to be expressed in terms of the same phenomenological vertex form factor $F_{ANN}(q^2)$ accurate up to $q^2 \sim 1$~GeV:
\begin{equation}
   g_A(q^2) = g_A(0) F_{ANN}(q^2) \frac{m_A^2}{q^2 + m_A^2},
   \label{eq:dipolega}
\end{equation}
\begin{equation}
   g_P(q^2) = \frac{2 M_N g_A(q^2)}{q^2 + m_\pi^2} \left( 1 - \frac{m_\pi^2}{m_A^2} \right),
\end{equation}
which take the functional form of a dipole because they emerge from a single graph mediated by an axial-meson. Sample values for the constants in this model are~\cite{Simkovic1999}: $g_V = 1$, $g_A = 1.27$, $\kappa_1 = 3.7$, $m_V = 0.84$~GeV, $m_A = 1.09$~GeV. Although the CVC/PCAC hypotheses pre-date the modern EFT approach and indeed QCD itself, their predictions generically hold at the first nontrivial leading-orders of $\chi EFT$~\cite{Scherer2011} with small corrections at higher chiral orders.

\renewcommand{\arraystretch}{1.3}
\begin{table}[t!]
\centering
\begin{tabular}{cll}
    \hline
    $h_{(\cdot,\cdot)}$ & $\overline{N}_1 \overline{N}_2 \tau^{1+} \tau^{2+} ( \quad ... \quad ) N_2 N_1$ & LO $\chi EFT$ \\
    \hline
    (V,V)   &  $g_V^2(q^2)$ & 1 \\
    (A,A)   &  $g_A^2(q^2) \vec{\sigma}_1 \cdot \vec{\sigma}_2$ & $g_A^2$\\
    (P,P)   &  $g_P^2(q^2) \frac{1} {4 M_N^2} \left[ \vec{\sigma}_1 \cdot \vec{q} \vec{\sigma}_2 \cdot \vec{q} \right] q^2$ & $\frac{g_A^2}{3} \left[ S_{12} + \sigma_{12} \right] \frac{q^4}{(q^2 + m_\pi^2)^2} $ \\
    (M,M)   &  $g_M(q^2) \frac{-1}{M_N^2} \left[ \vec{\sigma}_1 \cdot \vec{\sigma}_2 q^2 - \vec{\sigma}_1 \cdot \vec{q} \vec{\sigma}_2 \cdot \vec{q} \right] $ & $\frac{(1+\kappa_1)^2}{3} \left[ S_{12} - 2 \sigma_{12} \right] \frac{q^2}{M_N^2}$ \\
    (A,P)   &  $g_A(q^2) g_P(q^2) \frac{1}{2 M_N} \vec{\sigma}_1 \cdot \vec{q} \vec{\sigma}_2 \cdot \vec{q}$ &  $\frac{g_A^2}{3} \left[ S_{12} + \sigma_{12} \right] \frac{q^2}{q^2 + m_\pi^2} $ \\
    \hline
\end{tabular}
\caption{Allowed pairs of Lorentz structures, their contributions to the $NN$ neutrino potential, and $LO$ $\chi EFT$ evaluation. }
\label{tab:nupotterms}
\end{table}

Separately from the determination of form factors, we can construct the neutrino potential by taking a product of two copies of the non-relativistic one-body current $J_{L,NR}^\mu$, contracted over the index of exchange momentum $q$. Those product terms which do not evaluate to zero are collected in Table \ref{tab:nupotterms}, where notably the vector term decouples in the non-relativistic limit, and $h_{(M,A)} = h_{(M,P)} = 0$ by symmetrisation. Of these, only $h_{(V,V)}$ is a Fermi contribution and is purely so; $h_{(A,A)}$ is purely Gamow-Teller; and all other terms include both a Gamow-Teller and tensor contributions. We define the Gamow-Teller and tensor contributions as the respective coefficients of Pauli inner product $\sigma_{12} \equiv \vec{\sigma}_1 \cdot \vec{\sigma}_2$ and Pauli quadrupole moment $S_{12} \equiv 3(\vec{\sigma}_1 \cdot \hat{q} \vec{\sigma}_2 \cdot \hat{q}) - \sigma_{12}$. 

Finally, to make contact with $\chi EFT$ predictions for the couplings of nucleons to external currents,~\cite{Cirigliano2017} computes the single-nucleon form factors at LO in the chiral expansion, obtaining
\begin{align}
    g_V(q^2) = 1, \quad &g_M(q^2) = 1  + \kappa_1 = 4.7, \nonumber\\
    g_A(q^2) = g_A = 1.27, \quad &g_P(q^2) = - g_A \frac{2 M_N}{q^2 + m_\pi^2}
\end{align}
where $\kappa_1$ encodes the difference between the proton and neutron magnetic moments~\cite{Towner1995}. Form factors have also been derived at higher order in $\chi EFT$, for example to NNLO in~\cite{Bernard1998}, but these will be beyond the requirements of our precision. It is noteworthy that LO $\chi EFT$ reproduces the pseudoscalar form factor ``induced" by the axial coupling in CPAC. Then the resultant expression for each Lorentz product is given in the right-hand column of Table \ref{tab:nupotterms}. Factoring out global constants and the lepton bilinear as $ V_\nu = (4 G_F^2 V_{ud}^2) \overline{u}(k_1) P_R C \overline{u}^T(k_2) \hat{V}_\nu$, we can now write the neutrino potential induced by light Majorana-neutrino exchange~\cite{Simkovic1999,Cirigliano2017},
\begin{align}
    \hat{V}_\nu(q) &= \frac{\tau^{(1)+} \tau^{(2)+}}{q^2} \left[ h_F(q^2) - \sigma_{12} h_{GT}(q^2) - S_{12} h_T(q^2) \right] \nonumber\\
    &\simeq \frac{\tau^{(1)+} \tau^{(2)+}}{q^2} \left[ 1 - \frac{2 g_A^2}{3} \sigma_{12} \left( 1 + \frac{m_\pi^4}{2(q^2 + m_\pi^2)^2}\right) - \frac{g_A^2}{3} S_{12} \left( 1 - \frac{m_\pi^4}{(q^2 + m_\pi^2)^2}\right) \right].
\end{align}
Coordinate-space realisations of the above are presented in~\cite{Cirigliano2019}. Our primary interest will be the insertion of the neutrino potential between $^1S_0$ $NN$ states, where $S_{12} = 0$ and $\sigma_{12} = -3$, the above expression simplifying to
\begin{equation}
    \hat{V}_\nu^{^1S_0}(q) = \frac{\tau^{(1)+} \tau^{(2)+}}{q^2} \left[ 1 + 2 g_A^2 + \frac{g_A^4 m_\pi^4}{(q^2 + m_\pi^2)^2} \right].
\end{equation}
When $q \gg m_\pi$, the quantity in brackets has a limiting behavior of $1 + 2 g_A^2$, while it is enhanced to $1 + 3 g_A^2$ if $m_\pi \rightarrow \infty$ as in $\cancel{\pi}EFT$.

% Consider adding a plot of the above behavior.

\subsection{Resummation, the leading divergence, and its counterterm}

With the $NN$ potential corresponding to doubly-LNV light neutrino exchange in hand, 
the naive expectation might be that the amplitude between well-defined $NN$ states should be directly calculable. More precisely, if we write~\cite{Cirigliano2019}
\begin{equation}
    \mathcal{A}_\nu(E,E') = -_{^1S_0}\braket{\Psi_{pp}(E')|V_\nu^{^1S_0}(q)|\Psi_{nn}(E)}_{^1S_0}
\end{equation}
for center-of-mass energies $E = p^2 / m_n$ and $E' = p'^2 / m_p$, with momentum transfer $\vec{q} = \vec{p} - \vec{p}'$, the naive expectation is that $\mathcal{A}_\nu$ should be finite for all physical kinematics and represent the LO perturbation theory of this weak operator, naturally regularised by the leading $q^{-2}$-dependence of the neutrino potential. This expectation fails as a consequence of the enhanced contact counterterms required to renormalise $\mathcal{A}_\nu$, after the strong interactions intrinsic to $\Psi_{NN}$ are accounted for.

In agreement with the analysis of shallow bound-state $NN$-scattering reviewed in Section 2.2, the $\chi EFT$ $\ket{\Psi_{NN}}$ differs from a free nucleon pair state $\ket{\vec{p}}$ by the iteration to all orders of both contact and pion-exchange terms in $V_{NN}$. In the $^1S_0$ channel, we therefore have three potentials to be combined:
\begin{align}
     _{pp,^1S_0}\braket{\vec{p}' | V_\nu | \vec{p}}_{nn,^1S_0} &= \frac{1}{(\vec{p} - \vec{p}')^2} \left[ 1 + 2 g_A^2 + \frac{g_A^4 m_\pi^4}{((\vec{p} - \vec{p}')^2 + m_\pi^2)^2} \right], \nonumber \\
     _{^1S_0}\braket{\vec{p}' | V_C | \vec{p}}_{^1S_0} &= C, \nonumber \\
     _{^1S_0}\braket{\vec{p}' | V_\pi | \vec{p}}_{^1S_0} &= -\frac{g_A^2}{4 F_\pi^2} \frac{m_\pi^2}{(\vec{p} - \vec{p}')^2 + m_\pi^2}.
\end{align}
Although they appear two orders suppressed in the chiral expansion, we should have in mind the leading local $\Delta L = 2$ structures in the $\chi EFT$ Lagrangian~\cite{Cirigliano2018LNE},
\begin{align}
    &\mathcal{L}^{(2)}_{\Delta L=2} = \\ \label{eq:local_gNN}
    &\quad \frac{2G_F^2 V_{ud}^2 m_{\beta\beta}}{(4\pi F_\pi)^2} \left[ \frac{5}{6}F_\pi^2 g_\nu^{\pi\pi} \partial_\mu \pi^- \partial^\mu \pi^- + \sqrt{2} g_A F_\pi g_\nu^{\pi N} \overline{p} S_\mu n \partial^\mu \pi^- + g_\nu^{NN} \overline{p} n \overline{p} n \right] \overline{e}_L C \overline{e}^T_L, \nonumber
\end{align}
where $g_\nu^{\pi\pi}$, $g_\nu^{\pi N}$, and $g_\nu^{NN}$ are $\mathcal{O}(1)$ LECs corresponding to respective tree-level amplitudes $\pi^- \pi^- \rightarrow e^- e^-$, $n \pi^- \rightarrow p^+ e^- e^-$, and $n n \rightarrow p^+ p^+ e^- e^-$. Because the latter is the trivial counterterm topology for our neutrino potential amplitude, we can make the ansatz
\begin{equation}
    _{pp,^1S_0}\braket{\vec{p}' | V_\nu | \vec{p}}_{nn,^1S_0} \quad \mapsto \quad _{pp,^1S_0}\braket{\vec{p}' | V_\nu - 2 g_\nu^{NN}(\mu_S) | \vec{p}}_{nn,^1S_0}
\end{equation}
to adjust for any regulator-dependence of $\mathcal{A}_\nu$, where $\mu_S$ is the regulator scale which separates any short-range $\Delta L=2$ physics.

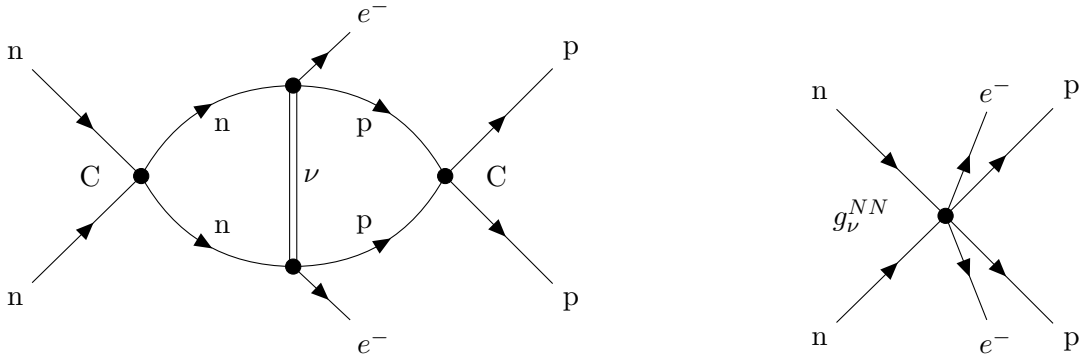
\begin{figure}[t]
\centering
    \begin{subfigure}{0.5\textwidth}
        \begin{tikzpicture}
              \begin{feynman}
                \vertex (l);
                \vertex [above left=2cm of l] (ul) {n};
                \vertex [below left=2cm of l] (dl) {n};
                \vertex [right=2cm of l] (m);
                \vertex [above=12mm of m] (um);
                \vertex [above right=1cm of um] (e1) {$e^-$};
                \vertex [below=12mm of m] (dm);
                \vertex [below right=1cm of dm] (e2) {$e^-$};
                \vertex [right=2cm of m] (r);
                \vertex [above right=2cm of r] (ur) {p};
                \vertex [below right=2cm of r] (dr) {p};
                \vertex [left=4mm of l] {C};
                \vertex [right=4mm of r] {C};
            
                \diagram* {
                  (ul) -- [fermion] (l) -- [fermion, bend left, edge label'=n] (um) -- [fermion, bend left, edge label'=p] (r) -- [fermion] (ur),
                  (dl) -- [fermion] (l) -- [fermion, bend right, edge label=n] (dm) -- [fermion, bend right, edge label=p] (r) -- [fermion] (dr),
                  (um) -- [double,double distance=0.8mm, edge label=$\nu$] (dm),
                  (um) -- [fermion] (e1),
                  (dm) -- [fermion] (e2),
                };
    
                \draw[fill=black] (l) circle(1mm);
                \draw[fill=black] (r) circle(1mm);
                \draw[fill=black] (um) circle(1mm);
                \draw[fill=black] (dm) circle(1mm);
              \end{feynman}
        \end{tikzpicture}
    \end{subfigure}\hfill % <-- added
    \begin{subfigure}{0.3\textwidth}
        \begin{tikzpicture}
              \begin{feynman}
                \vertex (l);
                \vertex [above left=2cm of l] (ul) {n};
                \vertex [below left=2cm of l] (dl) {n};
                \vertex [above right=2cm of l] (ur) {p};
                \vertex [left=1cm of ur] (e1) {$e^-$};
                \vertex [below right=2cm of l] (dr) {p};
                \vertex [left=1cm of dr] (e2) {$e^-$};
                \vertex [left=6mm of l] {$g_\nu^{NN}$};
            
                \diagram* {
                  (ul) -- [fermion] (l) -- [fermion] (ur),
                  (dl) -- [fermion] (l) -- [fermion] (dr),
                  (l) -- [fermion] (e1),
                  (l) -- [fermion] (e2),
                };
    
                \draw[fill=black] (l) circle(1mm);
              \end{feynman}
        \end{tikzpicture}
    \end{subfigure}
    \caption{Leading $\Delta L = 2$ $NN$-scattering amplitude (left) and its counterterm diagram (right) in pionless EFT ($\cancel{\pi}EFT$).}
    \label{fig:NNscattering_diagrams}
\end{figure}

The earliest evidence that the size of this counterterm is larger than anticipated by the chiral power-counting comes from a $\cancel{\pi}EFT$ analysis in~\cite{Cirigliano2018LNE}. The argument ultimately parallels that of~\cite{Kaplan1996} in the pure $NN$-scattering case; the large scattering length requires that any contact $NN$ interaction be enhanced if the iteration of the interaction is to reflect non-perturbative nuclear physics. The $g_\nu^{NN}$ interaction itself cannot be iterated because it includes leptonic final states, but intuitively it can induce the iteration of both incoming and outgoing $C$ interactions, resulting in two orders of enhancement. To see this explicitly,~\cite{Cirigliano2018LNE} employs dimensional regularisation with the PDS scheme, with the result
\begin{equation}
    g_\nu^{NN}(\mu_S) = (4\pi F_\pi)^2 \left[ \frac{M_N C_R(\mu_S)}{4\pi}\right]^2 \tilde{g}_\nu^{NN}(\mu_S),
\end{equation}
where $\tilde{g}_\nu^{NN}(\mu_S)$ is $\mathcal{O}(1)$ and the quantity in square brackets is the inverse scattering length $a^{-1}$. Thus $g_\nu^{NN} \sim \Lambda_\chi^2 a^2 \sim \left(\frac{Q}{\Lambda_\chi}\right)^{-2}$. This amplitude and its counterterm diagram are shown in Figure \ref{fig:NNscattering_diagrams}.
\begin{figure}[t]
\centering
    \begin{align*}
        \scalebox{1.5}{$\mathcal{A}_A =$}&~
        \vcenter{\hbox{\begin{tikzpicture}
          \begin{feynman}
            \vertex (l);
            \vertex [right=1.5cm of l] (ma);
            \vertex [right=1.5cm of ma] (mb);
            \vertex [right=1.5cm of mb] (mc);
            \vertex [right=1.5cm of mc] (r);   
            \vertex [above=1cm of ma] (ua);
            \vertex [above=1cm of mb] (ub);
            \vertex [above=1cm of mc] (uc);
            \vertex [below=1cm of ma] (da);
            \vertex [below=1cm of mb] (db);
            \vertex [below=1cm of mc] (dc);
            \vertex [above right=0.5cm of ub] (e1);
            \vertex [below right=0.5cm of db] (e2);         
            \diagram*{
                (ua) -- [solid, edge label=n] (ub) -- [solid, edge label=p] (uc);
                (da) -- [solid, edge label'=n] (db) -- [solid, edge label'=p] (dc);
                (ub) -- [double,double distance=0.8mm, edge label=$\nu$] (db),
                %(ub) -- [fermion] (e1);
                %(db) -- [fermion] (e2);
            };
            \draw[fill=black] (ub) circle(1mm);
            \draw[fill=black] (db) circle(1mm);
            \draw [->] (ub) -- (e1);
            \draw [->] (db) -- (e2);
          \end{feynman}
        \end{tikzpicture}}}
        ~,\nonumber\\
        \scalebox{1.5}{$\mathcal{A}_B =$}&~
        \vcenter{\hbox{\begin{tikzpicture}
          \begin{feynman}
            \vertex (l);
            \vertex [right=1.5cm of l] (ma);
            \vertex [right=1.5cm of ma] (mb);
            \vertex [right=1.5cm of mb] (mc);
            \vertex [right=1.5cm of mc] (md);   
            \vertex [right=1.5cm of md] (r);   
            \vertex [above=1cm of ma] (ua);
            \vertex [above=1cm of mb] (ub);
            \vertex [above=1cm of mc] (uc);
            \vertex [above=1cm of md] (ud);
            \vertex [below=1cm of ma] (da);
            \vertex [below=1cm of mb] (db);
            \vertex [below=1cm of mc] (dc);
            \vertex [below=1cm of md] (dd);
            \vertex [above right=0.5cm of ub] (e1);
            \vertex [below right=0.5cm of db] (e2);         
            \diagram*{
                (ua) -- [solid, edge label=n] (ub) -- [solid, bend left, edge label=p] (mc)  -- [solid, bend left, edge label=p] (ud);
                (da) -- [solid, edge label'=n] (db) -- [solid, bend right, edge label'=p] (mc) -- [solid, bend right, edge label'=p] (dd);
                (ub) -- [double,double distance=0.8mm, edge label=$\nu$] (db),
                %(ub) -- [fermion] (e1);
                %(db) -- [fermion] (e2);
            };
            \draw[fill=black] (ub) circle(1mm);
            \draw[fill=black] (db) circle(1mm);
            \draw[fill=black] (mc) circle(1mm);
            \draw [->] (ub) -- (e1);
            \draw [->] (db) -- (e2);
          \end{feynman}
        \end{tikzpicture}}}
        ~+~
        \vcenter{\hbox{\begin{tikzpicture}
          \begin{feynman}
            \vertex (l);
            \vertex [right=1.5cm of l] (ma);
            \vertex [right=1.5cm of ma] (mb);
            \vertex [right=1.5cm of mb] (mc);
            \vertex [right=1.5cm of mc] (md);   
            \vertex [right=1.5cm of md] (me);
            \vertex [right=1.5cm of me] (mf);
            \vertex [right=1.5cm of mf] (r);
            \vertex [above=1cm of ma] (ua);
            \vertex [above=1cm of mb] (ub);
            \vertex [above=1cm of mc] (uc);
            \vertex [above=1cm of md] (ud);
            \vertex [above=1cm of me] (ue);
            \vertex [above=1cm of mf] (uf);
            \vertex [below=1cm of ma] (da);
            \vertex [below=1cm of mb] (db);
            \vertex [below=1cm of mc] (dc);
            \vertex [below=1cm of md] (dd);
            \vertex [below=1cm of me] (de);
            \vertex [below=1cm of mf] (df);
            \vertex [above right=0.5cm of ub] (e1);
            \vertex [below right=0.5cm of db] (e2);         
            \diagram*{
                (ua) -- [solid, edge label=n] (ub) -- [solid, bend left, edge label=p] (mc)  -- [solid, bend left, edge label=p] (ud) -- [solid, bend left, edge label=p] (me) -- [solid, bend left, edge label=p] (uf);
                (da) -- [solid, edge label'=n] (db) -- [solid, bend right, edge label'=p] (mc) -- [solid, bend right, edge label'=p] (dd) -- [solid, bend right, edge label=p] (me) -- [solid, bend right, edge label=p] (df);
                (ub) -- [double,double distance=0.8mm, edge label=$\nu$] (db),
                %(ub) -- [fermion] (e1);
                %(db) -- [fermion] (e2);
            };
            \draw[fill=black] (ub) circle(1mm);
            \draw[fill=black] (db) circle(1mm);
            \draw[fill=black] (mc) circle(1mm);
            \draw[fill=black] (me) circle(1mm);
            \draw [->] (ub) -- (e1);
            \draw [->] (db) -- (e2);
          \end{feynman}
        \end{tikzpicture}}}
        ~+ \vcenter{\hbox{...}}~,\nonumber\\
        \scalebox{1.5}{$\mathcal{A}_{\overline{B}} =$}&~
        \vcenter{\hbox{\begin{tikzpicture}
          \begin{feynman}
            \vertex (l);
            \vertex [right=1.5cm of l] (ma);
            \vertex [right=1.5cm of ma] (mb);
            \vertex [right=1.5cm of mb] (mc);
            \vertex [right=1.5cm of mc] (md);   
            \vertex [right=1.5cm of md] (r);   
            \vertex [above=1cm of ma] (ua);
            \vertex [above=1cm of mb] (ub);
            \vertex [above=1cm of mc] (uc);
            \vertex [above=1cm of md] (ud);
            \vertex [below=1cm of ma] (da);
            \vertex [below=1cm of mb] (db);
            \vertex [below=1cm of mc] (dc);
            \vertex [below=1cm of md] (dd);
            \vertex [above right=0.5cm of ue] (e1);
            \vertex [below right=0.5cm of de] (e2);         
            \diagram*{
                (uc) -- [solid, bend left, edge label=n] (md) -- [solid, bend left, edge label=n] (ue) -- [solid, edge label=p] (uf) ;
                (dc) -- [solid, bend right, edge label'=n] (md) -- [solid, bend right, edge label'=n] (de) -- [solid, edge label'=p] (df) ;
                (ue) -- [double,double distance=0.8mm, edge label=$\nu$] (de),
                %(ub) -- [fermion] (e1);
                %(db) -- [fermion] (e2);
            };
            \draw[fill=black] (ue) circle(1mm);
            \draw[fill=black] (de) circle(1mm);
            \draw[fill=black] (md) circle(1mm);
            \draw [->] (ue) -- (e1);
            \draw [->] (de) -- (e2);
          \end{feynman}
        \end{tikzpicture}}}
        ~+~
        \vcenter{\hbox{\begin{tikzpicture}
          \begin{feynman}
            \vertex (l);
            \vertex [right=1.5cm of l] (ma);
            \vertex [right=1.5cm of ma] (mb);
            \vertex [right=1.5cm of mb] (mc);
            \vertex [right=1.5cm of mc] (md);   
            \vertex [right=1.5cm of md] (me);
            \vertex [right=1.5cm of me] (mf);
            \vertex [right=1.5cm of mf] (r);
            \vertex [above=1cm of ma] (ua);
            \vertex [above=1cm of mb] (ub);
            \vertex [above=1cm of mc] (uc);
            \vertex [above=1cm of md] (ud);
            \vertex [above=1cm of me] (ue);
            \vertex [above=1cm of mf] (uf);
            \vertex [below=1cm of ma] (da);
            \vertex [below=1cm of mb] (db);
            \vertex [below=1cm of mc] (dc);
            \vertex [below=1cm of md] (dd);
            \vertex [below=1cm of me] (de);
            \vertex [below=1cm of mf] (df);
            \vertex [above right=0.5cm of ue] (e1);
            \vertex [below right=0.5cm of de] (e2);         
            \diagram*{
                (ua) -- [solid, bend left, edge label=n] (mb)  -- [solid, bend left, edge label=n] (uc) -- [solid, bend left, edge label=n] (md) -- [solid, bend left, edge label=n] (ue) -- [solid, edge label=p] (uf) ;
                (da) -- [solid, bend right, edge label'=n] (mb)  -- [solid, bend right, edge label'=n] (dc) -- [solid, bend right, edge label'=n] (md) -- [solid, bend right, edge label'=n] (de) -- [solid, edge label'=p] (df) ;
                (ue) -- [double,double distance=0.8mm, edge label=$\nu$] (de),
                %(ub) -- [fermion] (e1);
                %(db) -- [fermion] (e2);
            };
            \draw[fill=black] (ue) circle(1mm);
            \draw[fill=black] (de) circle(1mm);
            \draw[fill=black] (md) circle(1mm);
            \draw[fill=black] (mb) circle(1mm);
            \draw [->] (ue) -- (e1);
            \draw [->] (de) -- (e2);
          \end{feynman}
        \end{tikzpicture}}}
        ~+ \vcenter{\hbox{...}}~,\nonumber\\
        \scalebox{1.5}{$\mathcal{A}_{C} =$}&~
        \vcenter{\hbox{\begin{tikzpicture}
          \begin{feynman}
            \vertex (l);
            \vertex [right=1.5cm of l] (ma);
            \vertex [right=1.5cm of ma] (mb);
            \vertex [right=1.5cm of mb] (mc);
            \vertex [right=1.5cm of mc] (md);   
            \vertex [right=1.5cm of md] (me);
            \vertex [right=1.5cm of me] (mf);
            \vertex [right=1.5cm of mf] (r);
            \vertex [above=1cm of ma] (ua);
            \vertex [above=1cm of mb] (ub);
            \vertex [above=1cm of mc] (uc);
            \vertex [above=1cm of md] (ud);
            \vertex [above=1cm of me] (ue);
            \vertex [above=1cm of mf] (uf);
            \vertex [below=1cm of ma] (da);
            \vertex [below=1cm of mb] (db);
            \vertex [below=1cm of mc] (dc);
            \vertex [below=1cm of md] (dd);
            \vertex [below=1cm of me] (de);
            \vertex [below=1cm of mf] (df);
            \vertex [above right=0.5cm of uc] (e1);
            \vertex [below right=0.5cm of dc] (e2);         
            \diagram*{
                (ua) -- [solid, bend left, edge label=n] (mb)  -- [solid, bend left, edge label=n] (uc) -- [solid, bend left, edge label=p] (md) -- [solid, bend left, edge label=p] (ue);
                (da) -- [solid, bend right, edge label'=n] (mb)  -- [solid, bend right, edge label'=n] (dc) -- [solid, bend right, edge label'=p] (md) -- [solid, bend right, edge label'=p] (de);
                (uc) -- [double,double distance=0.8mm, edge label=$\nu$] (dc),
                %(ub) -- [fermion] (e1);
                %(db) -- [fermion] (e2);
            };
            \draw[fill=black] (uc) circle(1mm);
            \draw[fill=black] (dc) circle(1mm);
            \draw[fill=black] (md) circle(1mm);
            \draw[fill=black] (mb) circle(1mm);
            \draw [->] (uc) -- (e1);
            \draw [->] (dc) -- (e2);
          \end{feynman}
        \end{tikzpicture}}}
        ~+ \vcenter{\hbox{...}}~.
    \end{align*}
    \caption[Decomposition of $NN$ $0\nu\beta\beta$ diagrammatic series in pionless EFT ($\cancel{\pi}EFT$). The leading (leftmost) $\mathcal{A}_A$ and $\mathcal{A}_B \simeq \mathcal{A}_{\overline{B}}$ diagrams are evaluated in the text and shown to be finite to all orders of contact iteration, while the leading diagram $\mathcal{A}_C$ is divergent and requires regularisation through contact $\Delta L=2$ diagram $g_\nu^{NN}$.]{Decomposition of $NN$ $0\nu\beta\beta$ diagrammatic series in pionless EFT ($\cancel{\pi}EFT$), adapted from~\cite{Cirigliano2019}. The leading (leftmost) $\mathcal{A}_A$ and $\mathcal{A}_B \simeq \mathcal{A}_{\overline{B}}$ diagrams are evaluated in the text and shown to be finite to all orders of contact iteration, while the leading diagram $\mathcal{A}_C$ is divergent and requires regularisation through contact $\Delta L=2$ diagram $g_\nu^{NN}$.}
    \label{fig:pionless0vbb_diagrams}
\end{figure}
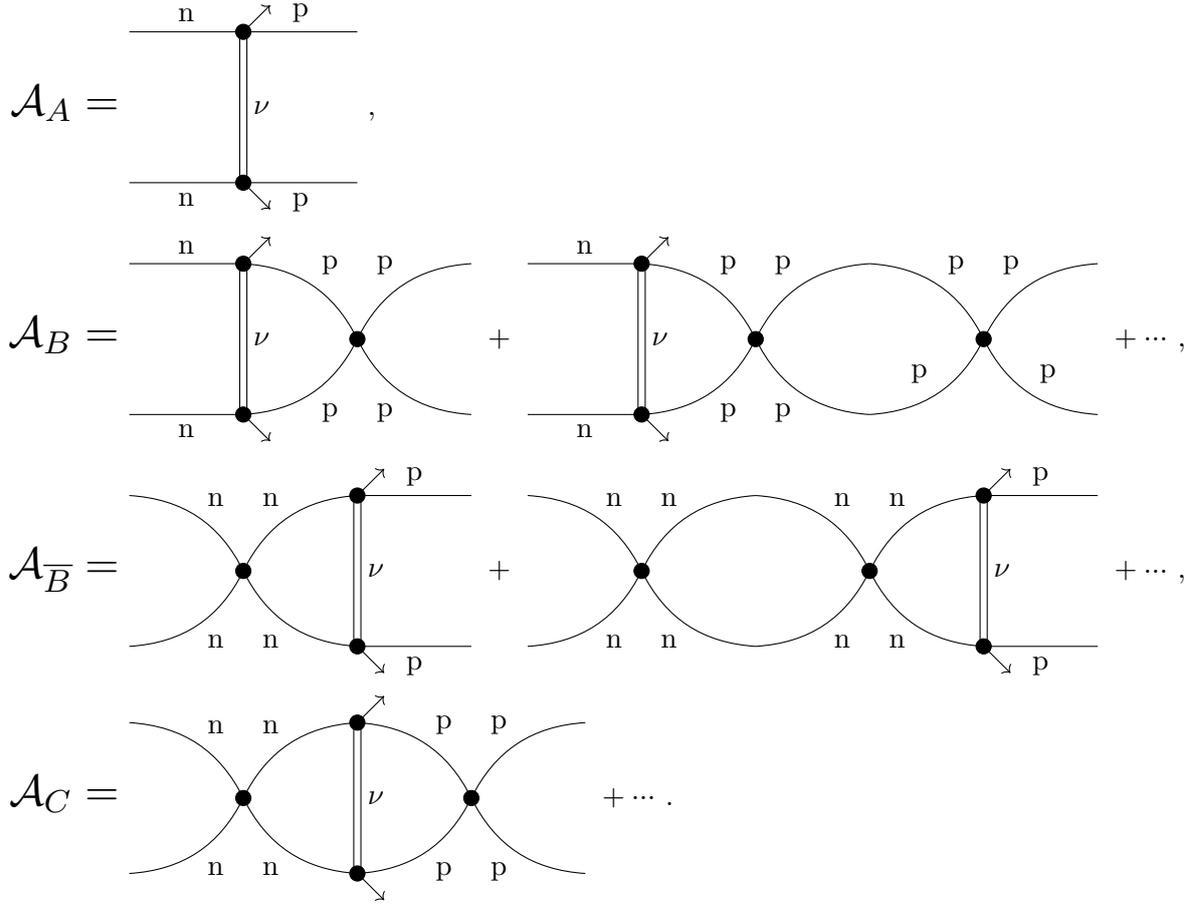
To see how this back-of-the-envelope calculation generalises to $\chi EFT$,~\cite{Cirigliano2018,Cirigliano2019} constructs a geometric series in the style of~\cite{Kaplan1996}. Recall that in the pure strong-interaction case, it was possible to decompose the T-matrix as $T_\infty^{\chi EFT}(E) = T_\infty^\pi(E) + \chi_{\vec{p}}(0) K_E \chi_{\vec{p}}(0) $, where $K_E \equiv \left[ \frac{1}{C} + \tilde{G}_E \right]^{-1}$ is the iteration of the $NN$ contact interaction to all orders, with each $\vec{r}=0$ to $\vec{r}'=0$ bubble dressed by iterated one-pion exchange. Analogously, an insertion of the LO $V_\nu$ can be dressed by contact strong-interactions on either or both side, which are then summed coherently:
\begin{equation}
    \mathcal{A}_\nu = \mathcal{A}_A + \chi_{\vec{p}'}(0) K_{E'} \mathcal{A}_B + \overline{\mathcal{A}}_B K_E \chi_{\vec{p}}(0) + \chi_{\vec{p}'}(0) K_{E'} \left( \mathcal{A}_C + \frac{2 g_\nu^{NN}}{C^2} \right) K_E \chi_{\vec{p}}(0),
    \label{nuampseries}
\end{equation}
where all $\mathcal{A}$ implicitly have incoming $E$ and outgoing $E'$ two-nucleon kinematics, and the constant factors on $g_\nu^{NN}$ account for the two $C$ interactions which are absorbed compared to the bubble diagram, as well as the symmetry of the counterterm. Because both the scattering states $\chi_{\vec{p}}(0)$ and $K_E$ can be regularised in a scheme-independent manner~\cite{Kaplan1996}, it is sufficient to separately evaluate and regulate $\mathcal{A}_{A,B,C}$ to obtain $\mathcal{A}_\nu$. As it happens, $\mathcal{A}_A$ and $\mathcal{A}_B$ are UV-finite and so do not require regularisation. This can be seen exactly in $\cancel{\pi}EFT$~\cite{Cirigliano2019}:
\begin{align}
    \mathcal{A}_A &= \frac{1 + 3 g_A^2}{2 p p'} \tanh^{-1} \left( \frac{-2 p p'}{p^2 + p'^2} \right), \nonumber\\
    \mathcal{A}_B &= \overline{\mathcal{A}}_B = \frac{i M_N}{4\pi} \frac{1 + 3 g_A^2}{2 p'} \log \frac{p + p'}{p - p' + i m_\nu} + \mathcal{O}(m_\nu),
\end{align}
where dimensional regularisation and the Feynman integral trick have been applied, analogous to Coulomb corrections to nucleon scattering in~\cite{Kong2000}. The apparent IR-divergence in both cases is controlled by including some small neutrino mass $m_\nu \sim 0^+$ in the propagator. $\mathcal{A}_C$, on the other hand, exhibits a logarithmic divergence at $d=4$, for example in the $\overline{MS}$ regularisation scheme~\cite{Cirigliano2019,Kong2000},
\begin{equation}
    \mathcal{A}_C = \frac{- M_N^2 \left( 1 + 3 g_A^2 \right)}{2 \cdot (4\pi)^2} \left( \frac{1}{\varepsilon} - \gamma_E + \log 4\pi + 1 + \log \frac{\mu^2}{-(p + p')^2} \right),
\end{equation}
where $\gamma_E \sim 0.577$ is the Euler-Mascheroni constant and $\mu$ is the chosen regularisation scale. In alignment with the PDS subtraction scheme proposed in~\cite{Kaplan1996} and defined in Section 2.2, one can also observe that $\mathcal{A}_C$ is finite at $d=3$, and so should be consistent with cutoff regularisation -- in other words, the calculation performed here should not fall victim to the same energetic cancellations which plagued the $^1S_0$ $NN$-scattering state.

Recalling that the scale-dependent amplitude was defined with normalisation $\mathcal{A}_C(\mu) + \frac{2 g_\nu^{NN}(\mu)}{C^2}$, the counterterm has beta-function
\begin{equation}
    \beta_{g_\nu^{NN}} \equiv \frac{d g_\nu^{NN}}{d \ln \mu} = \frac{1 + 3 g_A^2}{2},
\end{equation}
which can be satisfied, e.g., by coupling $g_\nu^{NN}(\mu) = \frac{1 + 3 g_A^2}{2} \ln(\mu/\mu_0) + g_\nu^{NN}(\mu_0)$.

All of the above calculations have been performed in $\cancel{\pi}EFT$, and the critical reader might reasonably expect that proceeding to $\chi EFT$ could substantively modify the structure of divergences. Diagrammatically, this corresponds to inserting an iterated pion exchange of the form in Eq.~\eqref{pioniteration_diagrams} into every two-nucleon reducible component of each diagram in Figure \ref{fig:pionless0vbb_diagrams}. However, the difference is minor, for the convenient reason that each pion-exchange potential inserted between the two nucleons contributes $\int \frac{d^3k}{(2\pi)^3} \frac{1}{k^4}$~\cite{Cirigliano2019} and therefore improves the convergence of a diagram by one order. Therefore only the leading diagram of type C with no pion-exchanges could require an enhanced counterterm. Following~\cite{Cirigliano2021long}, the divergent contribution is denoted $\mathcal{A}_C^{sing}$, while the convergent part of this diagrams and any pion-iterated convergent diagrams are collected in $\delta \mathcal{A}_C$.

$\mathcal{A}_C^{sing}$ itself is modified in passing from $\cancel{\pi}EFT$ to $\chi EFT$ in two ways. First, as before, the pion-exchange potential induces a correction in the $NN$ contact coupling, $C \mapsto \tilde{C} = C + \frac{g_A^2}{2 f_\pi^2}$, wherever it appears in a diagram. Second, the coefficient of the neutrino potential recovers its induced-pseudoscalar component: $1 + 3 g_A^2 \mapsto 1 + 2 g_A^2 + \frac{g_A^2 + m_\pi^2}{k^2 + m_\pi^2}$. For the same reason as the pion-exchange insertions above, this additional term can only improve the convergence of diagrams, and its effects are absorbed into $\delta \mathcal{A}_C$. $\mathcal{A}_C^{sing}$ is then recovered in $\chi EFT$ by means of the simple substitution $1 + 3 g_A^2 \mapsto 1 + 2 g_A^2$, and the beta function of $g_\nu^{NN}$ is similarly adjusted.

Overall, we see that the leading-order $\chi EFT$ renormalisation requirements on neutrino exchange in an $NN$ system can be extracted from a $\cancel{\pi}EFT$ analysis. We could interpret this to mean that the separation of nuclear forces into short- and pion-range interactions remains appropriate for a process such as $0\nu\beta\beta$-decay -- but that some short-range physics specific to neutrino exchange is not captured by the naive truncation of the short-range diagrams, and so must be encoded in $g_\nu^{NN}$. Our more precise investigation of the extent to which short- and pion-range physics can be separated will play a central role in Section 4.

%could explain parallel method by Cirigliano which employs numerical Schr\"odinger solutions and cutoff regularisation to arrive at an equivalent result

%\textcolor{red}{(should probably mention that there is no renormalisation enhancement in other partial waves)}

%\subsection{Extending the EFT to intermediate momenta: form factor approach}
%\subsection{Constraining the EFT at high momenta: operator product expansion approach}
% for now, merging these topics into the following section

\subsection{Quantitative estimates for the contact contribution}

Having established that a sizable two-nucleon contact term enters into the $\chi EFT$ description of $0\nu\beta\beta$-decay, it remains to quantify its size. A preliminary attempt at this estimation was made in~\cite{Cirigliano2018}, based on a comparison to the $I=2$ charge-independence-breaking (CIB) $NN$ process, with a more thorough and systematically improvable estimate made in~\cite{Cirigliano2021short,Cirigliano2021long}, based on a generalisation of the Cottingham approach to electromagnetic hadron-mass corrections~\cite{Cottingham1963}. In this section we will summarise the former approach, and detail the latter, which we will then expand upon in Section 4.5.

\subsubsection{\texorpdfstring{Approximating $g_\nu^{NN}$ from CIB scattering data}{Approximating gnuNN from CIB scattering data}}

In the mid-term, precise LQCD computations of $NN$ systems are expected to become possible to the effect that quantities like $g_\nu^{NN}$ may be extracted from a matching between $\cancel{\pi}EFT / \chi EFT$ and LQCD~\cite{Barnea2015}. At present, a substitute is to consider nuclear processes similar in topology to $0\nu\beta\beta$-decay for which experimental scattering data is available. Ref.~\cite{Cirigliano2018} makes the first such comparison with CIB, or isospin-breaking, $NN$-scattering. Just as in the $\Delta L=2$ case discussed above, the leading diagrams of this process require renormalization by a contact counterterm. Two such four-nucleon contact operators are available~\cite{Cirigliano2018}, which correspond the insertion of currents with handedness $LL/RR$ or $LR$, associated with respective LECs $C_1$ and $C_2$. In the electromagnetic case, the $LH$ and $RH$ currents are identical and so the process can only probe the sum $C_1 + C_2$, which~\cite{Cirigliano2018} reports to have regularisation scale-dependence:
\begin{equation}
    \frac{d \left( C_1 + C_2 \right)}{d \log \mu} = 1 + g_A^2 \frac{m_{\pi^+}^2 - m_{\pi^0}^2}{e^2 F_\pi^2}.
\end{equation}
Choosing $\mu = m_\pi$ as the natural regularisation scale within $\chi EFT$, experimental scattering lengths imply the dimensionless value $\tilde{C}_1 + \tilde{C}_2 = \left( C_1 + C_2 \right) \left[\frac{M_N \tilde{C}}{4\pi} \right]^2 \simeq 5.0$. For $0\nu\beta\beta$-decay, however, the weak-current insertion can only occur with $LL$ handedness, and so $g_\nu^{NN} = C_1$. Thus Ref.~\cite{Cirigliano2018} makes the assumption that $C_1 = C_2$ to arrive at estimate $\tilde{g}_\nu^{NN} = \frac{\tilde{C}_1 + \tilde{C}_2}{2} \simeq 2.5$. More conservatively assuming that $|C_1|$ and $|C_2|$ differ at most by an order of magnitude, one can conclude that $-0.5 < \tilde{g}_\nu^{NN} < 5.5$.

Once $g_\nu^{NN}$ is determined, the magnitude of its contribution to the amplitude can be assessed by comparing the divergent amplitude $\mathcal{A}_{\Delta L=2}^\nu(\mu)$ and counterterm $\mathcal{A}_{\Delta L=2}^{CT}(\mu)$, the sum of which is seen to be scale-invariant. Ref.~\cite{Cirigliano2018} finds that $R_{CT} \equiv \mathcal{A}_{\Delta L=2}^{CT}(\mu)/\mathcal{A}_{\Delta L=2}^\nu(\mu) \sim 10\%$ for $\mu \sim 350$~MeV, increasing to $R_{CT} \sim 30\%$ for $\mu \sim 2$~GeV, although beyond the chiral symmetry-breaking scale $\Lambda_\chi$ one can expect these results to come with significant uncertainties. Further computational estimates on $A = 6,12$ nuclei affirm this substantial impact, with $R_{CT} \sim 25-60\%$ depending on the choice of CIB contact potential and the fit of $C_1 + C_2$.

\subsubsection{Aside: the neutron-proton mass difference a la Cottingham}

The second, more thorough estimate of the contact contribution by~\cite{Cirigliano2021short,Cirigliano2021long} is conceptually based on a classic dispersive calculation of the electromagnetic nucleon mass correction~\cite{Cottingham1963}, which we will now briefly review. 

The forward Compton scattering amplitude for a virtual photon with momentum $\vec{q}$ and polarisation $\vec{\varepsilon}$ may be written
\begin{equation}
    \varepsilon^\mu \varepsilon^\nu T_{\mu\nu}(\vec{q}) = i\pi \varepsilon^\mu \varepsilon^\nu \int d^4x e^{i q^\mu x_\mu} \braket{N|\mathcal{T}\{j_\mu(\vec{x}),j_\nu(0)\} + \mathcal{T}\{j_\nu(\vec{x}),j_\mu(0)\}|N},
\end{equation}
where $\mathcal{T}$ is the time-ordering operator and $j_\mu$ is the electromagnetic current. By inserting a complete set of intermediate nuclear states and performing the Fourier integrals in the nucleon rest frame, one obtains
\begin{align}
    T_{\mu\nu}(\vec{q}) = \frac{(2\pi)^4}{2} \sum_\kappa &\braket{N|j_\mu(0)|\kappa} \braket{\kappa|j_\nu(0)|N} \left[ \frac{\delta^{(3)}(k - q)}{k_0 - q_0 - M - i\epsilon} + \frac{\delta^{(3)}(k + q)}{k_0 + q_0 - M - i\epsilon} \right] \nonumber\\
    &+ (\mu \leftrightarrow \nu).
\end{align}
Dispersive analysis relies on the examination of analyticity constraints when a real function $f(q_0)$ or $f(q^2)$ is continued to complex-valued $q_0$ or $q^2$. In general, two-point functions in QFTs may be written in the K\"all\'en-Lehmann spectral representation~\cite{Zwicky2016}:
\begin{equation}
    f(q^2) = \int_0^\infty ds \frac{\rho(s)}{s - p^2 - i0^+},
\end{equation}
where the analogous representation for $f(q_0)$ may be obtained by taking $\mathbf{q}^2$ as a fixed parameter. Unitarity requires that the spectral function $\rho(s)$ is positive definite, and it frequently may be decomposed as 
\begin{equation}
    \rho(s) = Z \delta(s - m^2) + \theta(s-s_0) \sigma(s)
\end{equation}
where the pole term (first) represents a one-particle state of total energy $m^2$, and the branch-cut term (second) represents a continuum of multi-particle states beginning at threshold energy $s_0$. For a single nucleon in $\chi EFT$, $m^2 = M_N^2$ and $s_0 = (M_N + m_\pi)^2$ imply the following poles and branch-cuts in $q_0$~\cite{Cottingham1963}: 
%\textcolor{red}{(could adapt Figure 2 of~\cite{Zwicky2016} here)}.
%
\begin{equation}
    \pm q_0 = \sqrt{M_N^2 + \mathbf{q}^2} - M_N - i\epsilon , \quad \pm q_0 \geq \sqrt{(M_N + m_\pi)^2 + \mathbf{q}^2} - M_N - i\epsilon.
\end{equation}
%
%\begin{figure}[t]
%    \centering
%    \begin{tikzpicture}
%    % Axes:
%    \draw [-latex] (-7,0) -- (2,0) node [above left]  {$\Re$};
%    \draw [-latex] (0,-5) -- (0,5) node [below right] {$\Im$};
%    \node[cross out,draw=black] at (0,0) {};
%    \draw[dashed] (0,0) -- node[pos=0.8, above right] {$\omega_p$}(145:6) node[solid, cross out,draw=black] {};
%    \draw[red, -stealth] (0,2) arc (90:145:2);
%    
%    \draw[dashed] (0,0) -- node[pos=0.8, above right] {$\omega_z$}(125:3.5) node[solid, fill=white, circle,draw=black] {};
%    \draw[blue, -stealth] (0,1) arc (90:125:1);
%    
%    \draw[dashed]  (-5,0) node[below left] {$-\xi_p\omega_p$} --  (-5,-3) node[solid, cross out,draw=black] {};
%    \draw[dashed]  (-2,0) node[below left] {$-\xi_z\omega_z$} --  (-2,-3) node[solid, fill=white, circle,draw=black] {};
%    
%    \end{tikzpicture}
%\end{figure}
%
Cottingham's essential observation is that the one-loop electromagnetic correction to the nucleon mass is simply the forward Compton scattering amplitude with the photon line contracted. This is not immediately useful. However, the absence of any poles or branch cuts from the $++$ quadrant of the complex plane (including the new photon pole $\pm q_0 = \mathbf{q}^2 - i \epsilon$) allows for a rotation to Euclidean four-momentum, i.e.
\begin{equation}
    \Delta M = \frac{i}{8 \pi^2} \int d^4q \frac{\eta^{\mu\nu}}{q^\mu q_\mu - i \epsilon} T_{\mu\nu}(\vec{q}) = - \frac{1}{8 \pi^2} \int d^4q \frac{\eta^{\mu\nu}}{\mathbf{q}^2 + q_0^2} T_{\mu\nu}(i q_0,\mathbf{q})
\end{equation}
This result is very consequential, because the now-spacelike virtual ``photons" of energy $i q_0$ are equivalent to those which mediate measurable electron-nucleon scattering. Ref.~\cite{Cottingham1963} decomposes $T_{\mu\nu}(i q_0,\mathbf{q})$ into two independent Lorentz structures $t_i$, which can be further decomposed into elastic pole contributions $f_i$ and inelastic cut contributions $h_i$ (which must be integrated along the cut).

In modern incarnations of the Cottingham approach, elastic photon contributions are able to quantitatively capture the final result for the electromagnetic nucleon mass difference within $30\%$ error~\cite{Gasser1975,Gasser2021}. Determinations of the inelastic contribution between dispersive and lattice methods disagree: in contrast to the elastic term $m_{p-n}^{el} = 0.72(2)$~MeV approaching percent-level precision~\cite{Gasser2021,Cirigliano2021long}, the dispersive evaluation gives $m_{p-n}^{inel} = -0.17(16)$~MeV~\cite{Gasser2021} while the lattice gives $m_{p-n}^{inel} = 0.28(11)$~MeV~\cite{Horlsey2019,Cirigliano2021long}. A resolution of this tension at some intermediate value would thus only improve the $30\%$ error of an elastic-only calculation. It is noteworthy that calculations for electromagnetic meson-mass corrections do not show the same elastic dominance, with the vector and axial-vector resonances first applied by~\cite{Das1967} delivering over $93\%$ of the experimental pion mass difference in a modern calculation~\cite{Donoghue1997} while elastic diagrams give only $10\%$, the excess being negated by other inelastic corrections.

%\textcolor{red}{(need to re-locate Hoferichter citation for this section)}
% dispersive and lattice methods disagree ... "at the $2.3\sigma$ level~\cite{Hoferichter2022}" not sure where I got this figure from, I can't find the citation anywhere now so removing it

\subsubsection{Extracting contact contribution from matching procedure}

On the above grounds,~\cite{Cirigliano2021short,Cirigliano2021long} argues that there is no conceptual reason why a loop correction to a two-nucleon system cannot be treated according to Cottingham's approach, and furthermore that the nominal error from only including elastic contributions should be taken to be $\sim 30\%$. A more rigorous dispersive derivation would likely verify and clarify this claim, but to date has not been attempted. In the following, we will see how the Cottingham representation of the two-nucleon $0\nu\beta\beta$-decay amplitude allows for a controlled estimate of the impact of the contact counterterm.

The construction begins by defining the factorised amplitude (Eq. (2.9) in~\cite{Cirigliano2021long}):
\begin{equation}
    \mathcal{A} = (2\pi)^4 \delta^{(4)}(p_{e_1} + p_{e_2} + p_f - p_i) \left( 4 G_F^2 V_{ud}^2 m_{\beta\beta} \overline{u}_L(p_1) u_L^c(p_2) \right) \times \mathcal{A}_\nu,
\end{equation}
where the hadronic amplitude $\mathcal{A}_\nu$ encodes the nuclear evaluation which is our focus. The neutrino propagator becomes that of a massless boson while mass effects are absorbed into $m_{\beta\beta}$. With the further assumption that the outgoing electron momenta are sufficiently similar that their Dirac algebra may be approximated as $\gamma^\mu \gamma^\nu \rightarrow \eta^{\mu\nu}$, $\mathcal{A}_\nu$ is reduced to the form necessary for the Cottingham approach,
\begin{align}
    \mathcal{A}_\nu &= 2 \int \frac{d^4k}{(2\pi)^4} \frac{\eta^{\mu\nu}}{k^2 + i\epsilon} T_{\mu\nu}^{(\nu)}(k,p_i,p_f) \nonumber\\
    &= 2 \int \frac{d^4k}{(2\pi)^4} \frac{\eta^{\mu\nu}}{k^2 + i\epsilon} \braket{NN_f(p_f)| \int d^4r e^{i k\cdot r} \mathcal{T}\{ J_\mu^L(r/2) J_\nu^L(-r/2) \} |NN_i(p_i)},
    \label{anu_timeordered}
\end{align}
where $T_{\mu\nu}$ is now a forward two-nucleon neutrino scattering amplitude. Just as for forward Compton scattering, one can insert a complete set of intermediate nuclear states and perform not just the Fourier integrals, but the $k_0$-integral as well, obtaining (Eq. (2.22) in~\cite{Cirigliano2021long}):
\begin{align}
    T_{\mu\nu}^{(\nu)}(k,p_i,p_f) = - \sum_\kappa &\frac{\braket{NN_f(p_f)|J_\mu^L(0)|\kappa(q)} \braket{\kappa(q)|J_\nu^L(0)|NN_i(p_i)}} {q + E_\kappa(q) - \frac{1}{2}(p_i^0 + p_f^0) - i\epsilon} \nonumber\\
    &\cdot \frac{1}{|k|} \left[ \delta^{(3)}(k - q) + \delta^{(3)}(k + q)  \right].
    \label{NNkappasum}
\end{align}
$\ket{NN_i}$ and $\ket{NN_f}$ are scattering states, i.e. these two-nucleon wavefunctions must take into account the strong interactions, both short-range and pion-range, which bind the system together. They can be described by making contact with the $NN$-scattering results discussed of~\cite{Kaplan1996,Kaplan1998} reviewed in Section 2.2, which imply
\begin{equation}
    \ket{NN_i}(p_i) = (1 + \hat{G}(E) \hat{V})\ket{\vec{p_i}} = (1 + \hat{G}^0(E) \hat{T}^{(S)}(E))\ket{\vec{p_i}},
\end{equation}
for plane-wave state $\ket{\vec{p_i}}$, where the scattering operator is defined $\hat{T}^{(S)}(E) = \hat{V} \left( \mathbb{I} +  \hat{G}^0(E) \hat{V} \right)^{-1}$ in the notation of~\cite{Cirigliano2021long}. While both representations are convenient in their separation of the free and interacting components, the latter is perhaps more intuitive, because $\hat{T}^{(S)}(E)$ contains all iterated strong interactions while $\hat{G}^0(E)$ simply encodes the free two-nucleon propagator at energy $E$.

This transition to a Euclidean spatial-momentum picture, and in particular the smoothness of the resultant single-current amplitudes and energy denominators, is the practical benefit of having treated $0\nu\beta\beta$-decay through the Cottingham lens. Again inserting complete sets of nuclear states between all defined operators,~\cite{Cirigliano2021long} decomposes the amplitude into four parts:
\vspace{-0.2cm}
\begin{align}
    \mathcal{A}_\nu = \int &\frac{d^3k}{(2\pi)^3} \Big[ \braket{p_f| T^{(\nu)}(k) | p_i} \\
    &+ \sum_{\kappa'} \braket{p_f| \hat{T}^{(S)}(E') |\kappa'} \braket{\hat{G}^0(E')}_{\kappa'} \braket{\kappa'| \hat{T}^{(\nu)}(k) |p_i} \nonumber\\
    &+ \sum_\kappa \braket{p_f| \hat{T}^{(\nu)}(k) |\kappa} \braket{\hat{G}^0(E)}_\kappa \braket{\kappa| \hat{T}^{(S)}(E) |p_i} \nonumber\\
    &+ \sum_{\kappa'} \sum_\kappa \braket{p_f| \hat{T}^{(S)}(E') |\kappa'} \braket{\hat{G}^0(E')}_{\kappa'} \braket{\kappa'| \hat{T}^{(\nu)}(k) |\kappa} \braket{\hat{G}^0(E)}_\kappa \braket{\kappa| \hat{T}^{(S)}(E) |p_i} \Big] \nonumber
    \label{nuampfourterms}
\end{align}

where the Green's function expectation is $\braket{\hat{G}^0(E)}_\kappa = \frac{1}{E - E_\kappa^0 + i\epsilon}$. This expression is directly useful, giving the full two-nucleon $0\nu\beta\beta$-decay amplitude so long as one can calculate matrix elements of the strong T-matrix and the neutrino-exchange T-matrix between relevant nuclear states. However, obtaining even those single-operator expectation values for a large selection of nuclear states is highly non-trivial. One simplification arises from the fortunate circumstance that only the neutrino-exchange potential in the $^1S_0$ partial wave requires a contact counterterm at $LO$ in the chiral power-counting~\cite{Cirigliano2019}, i.e. this entire procedure need only be performed with $\ket{\vec{p_{i,f}}}$ in that partial wave. However, this still leaves an unbounded nuclear state space from which intermediate states $\kappa$ and $\kappa'$ are to be selected. For this reason, ~\cite{Cirigliano2021short,Cirigliano2021long} restrict to elastic intermediate states, i.e. pure $\ket{NN}$ states. Conservation of angular momentum then restricts both intermediate states purely to the $^1S_0$ partial wave.

If we take the spatial momentum to have magnitude $|\vec{k}| < m_\pi$ such that $LO$ $\chi EFT$ is certainly valid, then the above construction is exactly equivalent to the result Eq.~\eqref{nuampseries} computed by~\cite{Cirigliano2018,Cirigliano2019}. Indeed, since $\mathcal{A}_{A,B,\overline{B},C}$ were distinguished from one another by the absence or presence of $NN$ rescattering before and after the insertion of the neutrino-exchange potential, the four terms in Eq.~\eqref{nuampseries} exactly correspond to the decomposition of Eq.~\eqref{nuampfourterms}. As a consequence, the target of this matching-based analysis need only be the divergent part of $\mathcal{A}_C = \mathcal{A}_C^{div} + \delta \mathcal{A}_C$.

After much preamble, we can now define the matching procedure conceived by~\cite{Cirigliano2021short,Cirigliano2021long}. As presented earlier and in~\cite{Cirigliano2019}, $\mathcal{A}_C^{div}$ was calculated in $LO$ $\chi EFT$ to have real component 
\begin{align}
    \mathcal{A}_C^{div}(\mu_\chi) &= \left( \frac{4\pi}{M_N} \right)^2 \left[ - \frac{1 + 2 g_A^2}{2}  + \int_0^{\mu_\chi} dk a_\chi(k) \right] \nonumber\\
    &= \left( \frac{4\pi}{M_N} \right)^2 \left[ - \frac{1 + 2 g_A^2}{2}  - \int_0^{\mu_\chi} dk (1 + 2 g_A^2) \frac{1}{k} \theta(k - 2p) \right],
\end{align}
where the role of $\mu_\chi$ is apparent as a cutoff on the validity of this EFT amplitude. It is precisely the ignorance of short-range physics beyond $\mu_\chi$ which our enhanced contact term parameterises. As such,~\cite{Cirigliano2021long} conceives of a hypothetical $\mathcal{A}_C^{exact}(\mu_\chi)$ which integrates accounts for all $0 < k < \infty$ and is finite. $\chi EFT$ of course cannot provide such an amplitude; however, QCD may be treated perturbatively at scales above some $\Lambda \sim 1.5$~GeV, allowing us to compute some partial amplitude $\mathcal{A}_C^>$ such that 
\begin{equation}
    \mathcal{A}_C^{exact} = \mathcal{A}_C^< + \mathcal{A}_C^> = \int_0^\Lambda dk a_<(k) + \int_\Lambda^\infty dk a_>(k).
    \label{ampc_exact}
\end{equation}
Perturbative methods are valid for QCD in the $|k| > \Lambda$ regime because asymptotic freedom ensures that the effective strong-coupling is small, i.e. $\alpha_s \ll 1$~\cite{Colangelo2001}. If we were modelling the interactions of free, high-energy quarks and gluons, direct perturbative computation would be largely sufficient. However, our quarks and gluons are quanta of a nuclear (ground) state, as opposed to the QCD vacuum, and so our model clearly requires input from non-perturbative nuclear scale around $\Lambda_{QCD}$. In fact, even in the free vacuum case, QCD vacuum condensates of both quarks and gluons are present due to the nonlinearities of the QCD Lagrangian~\cite{Colangelo2001}, and some non-perturbative input is eventually necessary.

A clean way to decompose these perturbative and non-perturbative QCD inputs in the case of a two-point correlator is by constructing a Wilsonian operator product expansion (OPE)~\cite{Wilson1964,Wilson1972}, first performed in this context by~\cite{Shifman1979}. Formally, this expansion may be written
\begin{equation}
    \lim_{x \rightarrow 0} \braket{0|\mathcal{T}\{\mathcal{O}_1(x) \mathcal{O}_2(0)\}|0} = \sum_d C_d(x) \braket{0|\mathcal{O}_{(d)}(0)|0},
    \label{OPEdef}
\end{equation}
where $\mathcal{T}$ is the time-ordering symbol, $\mathcal{O}_{1,2}$ are two particular local operators, $\mathcal{O}_{(d)}$ are (typically other) local operators of dimension $d$, and $C_d$ are their respective coefficients. Note that there may well be distinct operators $\mathcal{O}_{(d)}$ of the same dimension requiring independent coefficients, although this will not occur in our case. There is typically a unit operator $\mathcal{O}_0$ with vacuum expectation value $1$ (in our case it will instead have expectation $\braket{NN_f | \overline{u}_L \gamma_\mu d_L \overline{u}_L \gamma^\mu d_L | NN_i}$), such that $C_0(x)$ encodes the usual perturbative expansion of QCD diagrams. This perturbative series is only the first term in the summation over vacuum condensates $\braket{0|\mathcal{O}_{(d)}(0)|0}$ seen in Eq.~\eqref{OPEdef}, whose lowest-order non-trivial terms include the quark condensate $\overline{\psi}\psi$ and the gluon condensate $G_{\mu\nu}^a G^{a,\mu\nu}$. We will follow~\cite{Cirigliano2021long} in restricting ourselves to $d=0$, but note that the subleading corrections would arise from the gluon condensate. Implicitly, the separation between perturbative and non-perturbative QCD has occurred at a chosen scale $\mu$, where physics at scales greater than $\mu$ is explicitly calculated in coefficients $C_d(x)$, and physics at scales less than $\mu$ is absorbed into the QCD vacuum condensates.

A full derivation of $C_0(k^2)$ leads to the following amplitude truncated at the first term of the OPE~\cite{Cirigliano2021long}:
\begin{equation}
    \mathcal{A}^> = \frac{3 \alpha_S}{2\pi} \braket{NN_f | \overline{u}_L \gamma_\mu d_L \overline{u}_L \gamma^\mu d_L | NN_i} \int_\Lambda^\infty dk \frac{1}{k^3},
\end{equation}
where the condensate contains all contributions softer than $\Lambda$. From Eq.~\eqref{eq:local_gNN}, those contributions are encoded in the local LECs $g_\nu^{\pi\pi},g_\nu^{\pi N},g_\nu^{N} \sim O(1)$, which together contribute a factor $\frac{2}{C^2} \left( g_\nu^{NN} + \frac{g_A^2}{2} \left( \frac{5}{6}g_\nu^{\pi\pi} - g_\nu^{\pi N}\right) \right)$. For simplicity and since $g_\nu^{NN}$ is poorly constrained,~\cite{Cirigliano2021long} elects to set the latter term in parentheses to zero, absorbing any inaccuracy into the uncertainty assessment of $g_\nu^{NN}$. Because the share of $\mathcal{A}_C$ from the $k > \Lambda$ regime is vanishingly small compared with uncertainties of this approach, only a very unnaturally large $g_\nu^{NN}$ or other LECs would confound this assumption. Altogether, the integrand for the matching analysis is therefore written:
\begin{equation}
a_>(k) = \frac{3 \alpha_S}{\pi} \frac{g_\nu^{NN}}{C^2} \frac{1}{k^3}.
\end{equation}
Model-independent information about $a_<(k)$ is not so readily available as it is about $a_>(k)$ in the perturbative QCD limit. A multitude of hadronic states not included in $\chi EFT$ are accessible between $\mu_\chi$ and $\Lambda$, including strange pseudoscalar mesons $\eta,\eta'$ and $K^{+,0,-}$, vector mesons $\rho^{+,0,-}$, $\omega$, $\phi$, and baryonic $\Delta^{++,+,0,-}$ resonances. Rather than account for these effects directly,~\cite{Cirigliano2021long} treats the intermediate momentum regime as a perturbation on the low-momentum regime already computed in $\chi EFT$. To the extent that nucleons are still appropriate degrees of freedom for the interaction, the topology of the bubble diagrams and therefore the decomposition Eq.~\eqref{nuampfourterms} is retained. Thus $\mathcal{A}_C^<$ can be obtained from $\mathcal{A}_C^{\chi EFT}$ by some appropriate generalisations of $\hat{T}^{(S)}(E)$ and $\hat{T}^{(\nu)}(E)$, i.e. by introducing form factors, which are accurate up to $\Lambda \sim 1.5$~GeV.

The modification of $\hat{T}^{(\nu)}(E)$ can be expected to proceed smoothly, since for larger momentum exchange the $NN$ state should be at least as separable as in the $\chi EFT$ regime. Therefore nucleon-level form factors at the neutrino vertices are an appropriate generalisation;~\cite{Cirigliano2021long} chooses to include the form factors for all pairs of Lorentz structures compatible with the product of two left-handed currents: $AA$, $PP$, $AP$, and $MM$. However, the induced pseudoscalar current only contributes to the convergent part of the amplitude, so we need not include it in $\mathcal{A}_C^<$. Therefore this form factor contribution amounts to a substitution:
\begin{equation}
    (1 + 2 g_A^2) \mapsto g_V^2(k^2) + 2 g_A^2(k^2) + \frac{k^2}{2 M_N^2} g_M^2(k^2),
\end{equation}
where the form factors were defined in Section 3.1 and depend on scales $\Lambda_V = 0.84$~GeV and $\Lambda_A = 1.0$~GeV. Ref.~\cite{Cirigliano2021long} demonstrates that the dependence on the amplitude on these precise scales is a ``by far subdominant uncertainty" in contrast to, for example, the truncation of nuclear states.

The modification of $\hat{T}^{(S)}(E)$ is somewhat more subtle, because it directly relies upon a choice of phenomenological short range potential between nucleons. This data is encoded in a so-called ``half-off-shell" (HOS) form factor on the iterated strong interaction vertex represented by $\hat{T}^(S)$, i.e. $T^{(S)}_<(q,p) \equiv T^{(S)}(p,p) \times f_S(q,p)$. Part of the appeal of this parameterisation is that $f_S(q,p) = 1$ in $LO$ $\chi EFT$~\cite{Cirigliano2021long}, so the form factor will specifically encode deviations from the low-momentum amplitude. Simply extending $\chi EFT$ to $NLO$ gives 
\begin{equation}
    f_S^{\chi EFT} = 1 - \frac{C_2}{2C}(p^2 - q^2),
\end{equation}
%
%\begin{figure}[t!]
%    \centering
%    \includegraphics[width=.49\textwidth]{Figures/C4_ContactTerm/rs_paper1.png}
%    \includegraphics[width=0.49\textwidth]{Figures/C4_ContactTerm/rs_paper2.png}
%    \caption[Left: $r(k)$ computed using LO $\chi EFT$ (black), NLO $\cancel{\pi}EFT$ (blue), NLO $\chi EFT$ (red) and the Kaplan-Steele potential (green). Right: $r(k)$ computed using the Kaplan-Steele potential (blue, solid), the Reid potential (green, dotted) and the AV18 potential (red, dashed).]{Left: $r(k)$ computed using LO $\chi EFT$ (black), NLO $\cancel{\pi}EFT$ (blue), NLO $\chi EFT$ (red) and the Kaplan-Steele potential~\cite{Kaplan1999} (green). Right: $r(k)$ computed using the Kaplan-Steele potential~\cite{Kaplan1999} (blue, solid), the Reid potential~\cite{Reid1968} (green, dotted) and the AV18 potential~\cite{Wiringa1995} (red, dashed); both taken from \cite{Cirigliano2021long}.}
%    \label{fig:rk_compare}
%\end{figure}
%
where the two form factors together deliver a real-valued suppression factor $r^{NLO}(k) = 1 - \frac{8 C_2}{M_N C^2} k$ to  $\mathcal{A}_C^<$. The ratio of constants $C_2/C^2$ evaluates to $0.52$ in $\cancel{\pi}EFT$ and $0.38$ on $\chi EFT$. To obtain more precision at momenta up to $\Lambda \sim 1.5$~GeV, Ref.~\cite{Cirigliano2021long} calculates the form factor $f_S(q,p)$ using three popular phenomenological $NN$ potentials, the Kaplan-Steele potential~\cite{Kaplan1999}, the Reid potential~\cite{Reid1968} and the AV18 potential~\cite{Wiringa1995}, which match very well to $NLO$ $\chi EFT$ below $k=200$~MeV, and all show a suppression $r(k) \sim 0.1$ before $k=400$~MeV. Model-dependencies for momenta larger than this are thus highly suppressed in their contribution to the overall amplitude, whose integrand is written:
\begin{equation}
    a_<(k) = - \left( g_V^2(k^2) + 2 g_A^2(k^2) + \frac{k^2}{2 M_N^2} g_M^2(k^2) \right) \frac{r(k)}{k} \theta (k - 2p).
\end{equation}
$a_\chi(k)$, $a_<(k)$, and $a_>(k)$ are plotted in Figure \ref{fig:matchingboth}, adapted from Figure 6 in~\cite{Cirigliano2021long}, alongside the corrected integrands from the $NN\pi$ analysis in Section 4.4. These pieces together are sufficient to estimate the size of the enhanced contact counterterm $g_\nu^{NN}(\mu)$, since the scale-invariant amplitude $\mathcal{A}_C(\mu) + \frac{2 g_\nu^{NN}(\mu)}{C^2}$ should have equal value to our constructed $\mathcal{A}_C^{exact} = \mathcal{A}_C^< + \mathcal{A}_C^>$. Here we note some ambiguity in the notation; the constant $g_\nu^{NN}$ which appears in the OPE is more precisely the $\mu \rightarrow \infty$ limit of $g_\nu^{NN}(\mu)$, and performing the matching analysis is equivalent to determining the action of the renormalisation group on this constant. In terms of the integrands~\cite{Cirigliano2021long}:
\begin{equation}
    2 g_{\nu}^{NN}(\mu) = \frac{1 + 2 g_A^2}{2} - \int_0^\mu dk a_\chi(k) + \int_0^\Lambda dk a_<(k) + \int_\Lambda^\infty dk a_<(k),
    \label{integrandmatching}
\end{equation}
which is plotted as the pure $NN$ comparison point for our later $NN\pi$ analysis in Figures \ref{fig:allgnn} and \ref{fig:allgnn_nu_both} for a range of matching scales $\Lambda$, where the dependence on the latter is comfortingly weak. Ref.~\cite{Cirigliano2021long} selects the value $g_{\nu}^{NN}(\mu=m_\pi) = 1.32$ as a representative value, which is essentially independent of $\Lambda$ between $1-4$~GeV. However this number by itself is not especially meaningful, because it is specific to a decomposition of the amplitude into convergent and divergent parts, i.e. the $MS/PDS$ regularisation scheme. Section 7 of~\cite{Cirigliano2021long} performs the analogous matching calculation for several choices of momentum and coordinate-space cutoff-regulators, The material point is that the relative importance of light neutrino exchange and the contact counterterm is scheme-dependent and therefore non-physical; however one can expect that the total $\mathcal{A}_\nu$ will be scheme-independent, and in general have been modified with a significance similar to the $\sim 15 \%$ adjustment seen in the $MS$ scheme. To facilitate other groups in evaluating the contact contribution within their own chosen regularisation scheme,~\cite{Cirigliano2021short,Cirigliano2021long} provide ``synthetic data" by computing $\mathcal{A}_\nu$ in full at a kinematic point relevant to the many-body nuclear community, $p = 25$~MeV and $p' = 30$~MeV. 

However, our primary interest is in the two-nucleon matrix element, and in particular the improvement of the leading uncertainty on the matching calculation which arises from truncating the sum over intermediate nuclear states to elastic $NN$ states only. Ref.~\cite{Cirigliano2021long} demonstrates that this uncertainty is dominant primarily by constraining all other sources of uncertainty. The impact of model-dependent form factors for the strong interactions in the intermediate momentum range is assessed by parameter variation, and contributes an uncertainty of $\pm 0.2$ to $\delta g_\nu^{NN}$, only $\pm 0.05$ of which arises from the less well-described but suppressed region $0.4 < k < 1.5$~GeV. A further $\pm 0.05$ (at most) enters $\delta g_\nu^{NN}$ from numerical parameter selections, especially the scales $\Lambda_{A,V}$ entering the single-nucleon form factors. However, a rough evaluation of one divergent diagram with a $NN\pi$ intermediate state, which can be interpreted as a leading inelastic correction, contributes between $-0.1$ and $0.35$ to $\delta g_\nu^{NN}$; this of course is just one of numerous possible diagrams containing the first accessible inelastic state, and~\cite{Cirigliano2021long} consequently assigns an uncertainty of $\pm 0.5$ in the absence of further information. This source of uncertainty, its more precise quantification, and steps towards its improvement will be the focus of the following section.

\section{\texorpdfstring{Quantifying the $NN\pi$ Intermediate State}{Quantifying the NNpi Intermediate State}}

%\begin{itemize}
%    \item Sum of states and statement of aims
%    \item \todo{Spin/isospin/A.M. selection rules for $NN \rightarrow NN\pi$ transitions}
%    \item \todo{Pion production from nucleon pairs, review of threshold analysis}
%    \item \todo{Derivation of $\pi \pi$ and $\pi N$ neutrino potentials.}
%    \item \todo{Full diagrammatic structure of $NN\pi$ contribution}
%    \item \todo{Example integrals and numerical estimates \textcolor{red}{(in progress)}}
%\end{itemize}
In the Cottingham procedure for studying loop corrections from nuclear operators, the superposition principle directly requires the completeness of the intermediate set of nuclear states. Taking a perturbative viewpoint of the nuclear state space, these intermediate states need not have the same particle content as the external $NN$ states -- such states are called inelastic. At the scales well-modelled by $\chi EFT$, this will mean that dynamical fluctuations of the nucleon field induce fluctuation quanta in ``nearby" strongly-interacting fields, most accessibly the pseudo-Goldstone pion fields. Of course, all this is a convenient perturbative realisation of systems that are more precisely non-perturbative bound configurations of QCD fields, where, as the particle number of a nucleon state is a distribution rather than a single value, the line between elastic and inelastic nuclear states is blurred. We are able to take advantage of a truncation between elastic and inelastic nuclear states in the case of $0\nu\beta\beta$ for two fortunate reasons: the bulk of the neutrino momentum transfer occurs well below energy scales that probe the internal structure of nucleons; and chiral symmetry is sufficiently broken such that, roughly speaking, momentum at the order of the transferred neutrino is required to induce even a single (virtual) excitation of the pion field.

For these same reasons, we take a further truncation of inelastic nuclear states between $\ket{NN_\pi}$ and $\ket{NN_\pi+}$ (all other inelastic states) to be well-defined. As~\cite{Cirigliano2021long} is careful to note, the two-particle kinematics of this process greatly complicate a direct relation between elastic states of the forward scattering amplitude and the Cottingham-style loop correction, and we accept the same limitation in framing our $\ket{NN_\pi}$ intermediate states. We retain the state decomposition of~\cite{Cirigliano2021long} from Eq.~\eqref{NNkappasum} and Eq.~\eqref{nuampfourterms}, where $\ket{\kappa},\ket{\kappa'}$ now are summed over both elastic and $NN\pi$ states. We will also refer to the leading elastic correction as the difference $\Delta \mathcal{A}_\nu^{NN\pi} = \mathcal{A}_\nu - \mathcal{A}_\nu^{NN}$. $\Delta \mathcal{A}_\nu^{NN\pi}$ receives contributions from diagrams (1) where only one of $\ket{\kappa},\ket{\kappa'}$ is inelastic (distinguished by isospin-breaking) and (2) where both $\ket{\kappa}$ and $\ket{\kappa'}$ are inelastic. 

A significant complication regarding $NN\pi$ states is that both angular momentum and isospin may now be distributed between the component particles in multiple ways, and strong scattering $\hat{T}^{(S)}$ will transfer these quantum numbers. We cannot therefore retain the $^1S_0$ partial wave or some $NN\pi$ equivalent throughout our analysis as~\cite{Cirigliano2021long} does, and the principles of angular momentum, isospin, and parity conservation will constrain our result via selection rules.

We will proceed as follows. First the role of selection rules in constraining the angular momentum and isospin properties of our intermediate states will be reviewed. Then $\braket{NN|\hat{T}^{(S)}|NN\pi}$ will be characterised within $\chi EFT$ and compared to the broader literature on threshold pion production from nucleon pairs. $\braket{NN|\hat{T}^{(\nu)}|NN\pi}$ and $\braket{NN\pi|\hat{T}^{(\nu)}|NN\pi}$ will also be characterised with the derivation of $\pi\pi$ and $\pi N$ neutrino potentials within $\chi EFT$. Combining these two building blocks will lead to the complete diagrammatic structure of the divergent contributions to $\Delta \mathcal{A}_\nu^{NN\pi}$. Naive dimensional analysis a la Weinberg will be used to estimate the size of each diagram in comparison to $\mathcal{A}_\nu^{NN}$, and finally the leading divergences will be evaluated in the dimensional regularisation scheme such that a numerical adjustment to the elastic computation of contact term $g^{NN}$ can be claimed.

\subsection{\texorpdfstring{Selection rules for $NN \rightarrow NN\pi$ transitions}{Selection rules for NN to NNpi transitions}}

An excellent description of the selection rules for $NN \rightarrow NNx$ transitions is given in the review~\cite{Hanhart2004} and the thesis~\cite{Malafaia2006}. First, let us state the relevant quantum numbers. Two-nucleon systems may have total spin $S = 0,1$, total isospin $T = 0,1$, and total orbital angular momentum $L = 0,1,2,...$, while a pion in an $NN\pi$ system has spin $S_\pi = 0$, total isospin $T_\pi = 1$, and angular momentum $l_\pi = 0,1,2,...$ about the centre of mass of the $NN$ component. Total isospin $T$ and total angular momentum $J = |\vec{L} + \vec{S}|$ are conserved by strong interactions, which may be thought of as the zeroth selection rule. It is crucial that $L$ and $S$ combine vectorially, so a given $J$ can arise from any $L$ and $S$ satisfying $|L - S| \leq J \leq |L + S|$.

Nucleons, as fermions, obey Pauli statistics, so the wavefunction of an $NN$ system (including as a component of an $NN\pi$ system) must be antisymmetric under particle exchange. Orbital angular momentum contributes $(-1)^L$ to this antisymmetry, while spin and isospin contribute $(-1)^{S+1}$ and $(-1)^{T+1}$ respectively. As a result, we have the first selection rule $(-1) = (-1)^{L+S+T}$ on both initial and final $NN$ states, satisfied for example by $^1S_0$ $nn$ and $pp$ states ($S=0, L=0, T=1$). Here we make use of the $NN$-spectroscopic notation $^{2S+1}X(L)_J$, where $X(L)$ is the conventional mapping $0 \rightarrow S, 1 \rightarrow P, 2 \rightarrow S, 3 \rightarrow F$ (lowercase will be used for pions).

Parity, the eigenvalue of a physical system under reversal of all spatial coordinates, is another quantum number which is at least approximately conserved under strong interactions~\cite{Wilkinson1958}. Spin/isospin do not influence parity as they do not contribute to the spatial wavefunction, but each angular momentum quantum contributes $(-1)$ to parity, which together with the intrinsic parities of field quanta implies the second selection rule~\cite{Hanhart2004}:
\begin{equation}
    (-1)^{L_i} = \pi_\pi \cdot (-1)^{L_f + l_\pi} \implies (-1)^{\Delta S + \Delta T} = \pi_\pi (-1)^{l_\pi},
\end{equation}
where $\pi_\pi = -1$ is the intrinsic parity of the pion (relative to the proton, by convention), and the implication holds only in the presence of Pauli statistics.

As an example application of these rules, consider the $NN \rightarrow NN\pi$ processes of an initial $^1S_0$ state. Since $J_i = 0 = J_f$, conservation implies $J_f = l_\pi = {0,1,2,...}$, with each choice allowing for a multiplet of $(L_f,S_f)$ selections. $NN$ isospin $T = {0,1}$, and Pauli statistics require that $T_i = 1$ ($nn$ or $pp$) as well as that $T_f = S_f + L_f + 1 (\text{mod}~2)$.

If $T_f = 1$ ($\Delta T = 0$), Pauli statistics require $S_f + L_f$ to be even. Meanwhile, parity conservation requires $S_f + l_\pi$ to be odd. For example, we could take $S_f = 1$ and $l_\pi = 0$. Then $J_f = 0$, so $L_f = 1$ and the process is $^1S_0 \rightarrow ^3P_0 + s$. We might also try $S_f = 0$ and $l_\pi = 1$, so $J_f = 1$ and again $L_f = 1$; but now $S_f + L_f$ is odd, so this combination is deselected. The next lowest-order candidates are: $S_f = 2$ and $l_\pi = 1$, which gives $J_f = 1$ and $L_f = 2$ ($^1S_0 \rightarrow {^5}D_1 + p$); and $S_f = 1$ and $l_\pi = 2$, which gives $J_f = 2$ and $L_f = 1,3$ ($^1S_0 \rightarrow {^3}P_2 + d$, $^1S_0 \rightarrow {^3}F_2 + d$).

If $T_f = 0$ ($\Delta T = -1$), Pauli statistics require $S_f + L_f$ to be odd, and parity conservation requires $S_f + l_\pi$ to be even. For example, we might try $S_f = 0$ and $l_\pi = 0$, so $J_f = 0$ and $L_f = 0$; but now $S_f + L_f$ is even, so this combination is deselected. Next we could take $S_f = 1$ and $l_\pi = 1$, so $J_f = 1$ and $L_F = 1$, selecting the process $^1S_0 \rightarrow ^3S_1 + p$. The next lowest-order candidates are: $S_f = 3$ and $l_\pi = 1$, which gives $J_f = 1$ and $L_f = 2,4$ ($^1S_0 \rightarrow {^7}D_1 + p$,$^1S_0 \rightarrow {^7}G_1 + p$); and $S_f = 1$ and $l_\pi = 3$, which gives $J_f = 3$ and $L_f = 2,4$ ($^1S_0 \rightarrow {^3}D_3 + f$, $^1S_0 \rightarrow {^3}G_3 + f$).

The above results are summarized in Table~\ref{tab:nnnnpiprocesses}. Overall, Pauli statistics and the conservation of angular momentum and parity impose quite strong restrictions on what $NN \rightarrow NN\pi$ channels can contribute (virtually) to our process. Presuming suppression of higher angular-momentum pion states, we will restrict ourselves to ($\Delta T = 0$) $^1S_0 \rightarrow ^3P_0 + s$ and ($\Delta T = -1$) $^1S_0 \rightarrow ^3S_1 + p$. Significantly, the neutrino-exchange insertion cannot occur in any $^1S_0 + x$ $NN\pi$ state for $^1S_0$ external $NN$ states. Note that only $\Delta T = 0$ can contribute to neutral-pion production, while both isospin channels contribute to charged-pion production.
%
%\textcolor{red}{(perhaps recreate subset of Table 5.4/5 of Malafaia thesis here?)}
%
\renewcommand{\arraystretch}{1.3}
\begin{table}[t!]
\centering
\begin{tabular}{ccccc}
    \hline
    $NN \rightarrow NN\pi$ process & $\Delta T$ & $J_f = l_\pi$ & $S_f$ & $L_f$\\
    \hline
    $^1S_0 \rightarrow {^3}P_0 + s$ & $0$ & $0$ & $1$ & $1$ \\
    $^1S_0 \rightarrow {^5}D_1 + p$ & $0$ & $1$ & $2$ & $2$ \\
    $^1S_0 \rightarrow {^3}P_2 + d$ & $0$ & $2$ & $1$ & $1$ \\
    $^1S_0 \rightarrow {^3}F_2 + d$ & $0$ & $2$ & $1$ & $3$ \\
    \hline
    $^1S_0 \rightarrow {^3}S_1 + p$ & $-1$ & $1$ & $1$ & $1$ \\
    $^1S_0 \rightarrow {^7}D_1 + p$ & $-1$ & $1$ & $3$ & $2$ \\
    $^1S_0 \rightarrow {^7}G_1 + p$ & $-1$ & $1$ & $3$ & $4$ \\
    $^1S_0 \rightarrow {^3}D_3 + f$ & $-1$ & $3$ & $1$ & $2$ \\
    $^1S_0 \rightarrow {^3}G_3 + f$ & $-1$ & $3$ & $1$ & $4$ \\
    \hline
\end{tabular}
\caption{$NN \rightarrow NN\pi$ processes permitted by Pauli statistics, conservation of angular momentum and conservation of parity for an initial $^1S_0$ state, derived in the text and truncated at $J_F = l_\pi \leq 3$.}
\label{tab:nnnnpiprocesses}
\end{table}

\subsection{Off-shell pion production from nucleon pairs}

There exists a long and thorough literature on \textit{threshold} pion production from nucleon pairs in connection to low-energy scattering experiments. This literature will provide us with the necessary tools to account for virtual $NN \rightarrow NN\pi$ processes in our $\hat{T}^{(S)}$. As we are only interested in the divergent part of our $0\nu\beta\beta$-decay amplitude, and will ultimately compute this through $\overline{MS}$ at the diagram level, it will not be necessary to explicitly compute $\hat{T}^{(S)}$ as a matrix operator. Instead, here we will identify the leading partial diagrams for both $s$- and $p$-wave pion production through dimensional analysis and bound the subleading uncertainties through a comparison to the threshold scenario.

Early works such as~\cite{Mandelstam1958} and~\cite{Ferrari1963} generally made use of of the distorted-wave Born approximation (DWBA), where the process is decomposed into initial- and final-state strong interactions around a fixed, perturbative pion-production core. The three-parameter model described in~\cite{Mandelstam1958}, based on the~\cite{Chew1956} effective-range approach to $p$-wave pion-nucleon resonances, was the first to achieve broad agreement with scattering data, ranging from near threshold (where $s$-wave pion production becomes relevant) to $\sim 600$~MeV (beyond which $d$-wave pion production becomes relevant). Ref.~\cite{Ferrari1963} attempts to model higher-energy production using a one-pion exchange model without final state interactions, appropriate since deuteron formation is suppressed above $\sim 800$~MeV. This approach is quite prescient of the later role one-pion exchange will play in $\chi EFT$-based $NN$ and $NN\pi$ theory. In conceptual agreement with the methods of~\cite{Cirigliano2021long}, we will also pursue a DWBA-like separation, but a review of early works which instead made use of fully-nonperturbative, coupled-channel computations can be found in~\cite{Hanhart2004}.

Refs.~\cite{Woodruff1960} and~\cite{Koltun1965} developed a formalism for $s$-wave pion production near threshold, based on a phenomenological pion-nucleon interaction Hamiltonian which is reminiscent of later $\chi EFT$ Lagrangians, and is accurate within $10\%$ for the process $pp \rightarrow d + \pi^+$ while agreeing with limited experimental data on $pp \rightarrow pp + \pi^0$. Crucially, $p$- and $s$-wave pion production arise from distinct interaction terms, of respective size $\frac{f_\pi}{m_\pi}$ and $\frac{1}{2M_N}$, as well as two rescattering coefficients fit from $\pi N$ scattering lengths. Ref.~\cite{Bertsch1978}, focusing on pion absorption, extends this methodology to include virtual $\rho$- and $\omega$-meson diagrams in an attempt to explain discrepancies with experimental pion absorption in heavier nuclei, ultimately concluding that the fault lies with irreducible three-nucleon contributions. Such additional meson diagrams did however account for discrepancies with threshold $pp \rightarrow pp \pi^0$ as better data became available~\cite{Lee1993}, albeit in conflict with similarly-sized corrections from off-shell $\pi N$ rescattering contributions~\cite{Hernandez1995}.

The first $\chi EFT$- and therefore QCD-based computations of $pp \rightarrow pp \pi^0$ near threshold were presented in~\cite{Cohen1996} and~\cite{Park1996}. As we have seen for $NN$-scattering, the systematic organisation of diagrams via a power-counting facilitates a concrete calculation with controlled errors, with typical expansion parameter $\varepsilon \sim m_\pi/M_N \sim 0.1$. This typical power-counting arises from treating both the typical three-momentum exchange and the typical nucleon three-momentum of a process as being of the order of the pion mass: $Q \sim p_N \sim m_\pi$. Ref.~\cite{Cohen1996} observes that at threshold, this approximation is not kinematically sound, since at minimum $p_N \sim \sqrt{M_N m_\pi}$ is required to produce a real pion excitation. Then one expects also $Q \sim \sqrt{M_N m_\pi}$, and the revised expansion parameter is $\varepsilon_{\text{thr}} \sim \sqrt{m_\pi/M_N} \sim 0.4$; a significantly slower perturbative expansion results, and we will see that in many cases the ordering of diagrams is changed substantially. This detail is relevant to us not because we will use $\varepsilon_{\text{thr}}$; in fact for off-shell pions, $Q \sim p_N \sim m_\pi$ is still a valid expectation, so we will use the standard $\varepsilon$ power-counting. However, most of the subsequent threshold literature has been expressed in $\varepsilon_{\text{thr}}$, and it will be necessary to carefully convert between the two.

At this point it is useful to define the dominant irreducible diagrams which in some order will fill out the LO through NNLO classes of contributions. The impulse or Born diagram is evidently the simplest, with a single $\pi NN$ vertex facilitating the process. At threshold, this diagram is kinematically forbidden, as one of the external $NN$ pairs must be off-shell by $m_\pi$; Ref.~\cite{Cohen1996} resolves this by defining connected impulse diagrams with a single (contact or pion-range) strong interaction, which together may be thought of as the impulse contribution. Equivalently, the connected impulse diagrams are irreducible because no kinematical two-nucleon unitary cut is permitted~\cite{Hanhart2004}. Of course in our application, both the produced pion and the final $NN$ state are expected to be off-shell, so we can treat these subsidiary strong interactions as part of the scattered $NN$ wavefunctions.
% These are tabulated in Figure \textcolor{red}{(make figure based on Cohen1996 and Hanhart2004)}.

\begin{figure}[t]
    \centering % <-- added
    \hspace{-1cm}
\begin{subfigure}{0.2\textwidth}
    \centering
        \begin{tikzpicture}
          \begin{feynman}
            \vertex (a) {N};
            \vertex [right=1.5cm of a] (b);
            \vertex [right=1.5cm of b] (c) {N};
            \vertex [above right=of b] (g) {\(\pi\)};;
            \vertex [below=1.5cm of b] (e);
            \vertex [below=1.5cm of a] (d) {N};
            \vertex [right=1.5cm of e] (f) {N};
        
            \diagram* {
              (a) -- [fermion] (b) -- [fermion] (c),
              (d) -- [fermion] (e) -- [fermion] (f),
              (b) -- [dashed] (g),
            };
        
            \draw[fill=black] (b) circle(1mm);
          \end{feynman}
        \end{tikzpicture}
\end{subfigure}\hspace{2cm} % <-- added
\begin{subfigure}{0.2\textwidth}
    \centering
        \begin{tikzpicture}
          \begin{feynman}
            \vertex (a) {N};
            \vertex [right=1.5cm of a] (b);
            \vertex [right=1.5cm of b] (c) {N};
            \vertex [above right=of b] (g) {\(\pi\)};;
            \vertex [below=1.5cm of b] (e);
            \vertex [below=1.5cm of a] (d) {N};
            \vertex [right=1.5cm of e] (f) {N};
        
            \diagram* {
              (a) -- [fermion] (b) -- [fermion] (c),
              (d) -- [fermion] (e) -- [fermion] (f),
              (b) -- [dashed] (g),
            };
        
            \draw[fill=white] (b) circle(1mm);
          \end{feynman}
        \end{tikzpicture}
\end{subfigure}\hspace{2cm}
\begin{subfigure}{0.2\textwidth}
    \centering
        \begin{tikzpicture}
          \begin{feynman}
            \vertex (a) {N};
            \vertex [right=1.5cm of a] (b);
            \vertex [right=1.5cm of b] (c) {N};
            \vertex [above right=of b] (g) {\(\pi\)};;
            \vertex [below=1.5cm of b] (e);
            \vertex [below=1.5cm of a] (d) {N};
            \vertex [right=1.5cm of e] (f) {N};
        
            \diagram* {
              (a) -- [fermion] (b) -- [fermion] (c),
              (d) -- [fermion] (e) -- [fermion] (f),
              (b) -- [dashed] (g),
              (b) -- [dashed] (e),
            };
        
            \draw[fill=black] (b) circle(1mm);
            \draw[fill=black] (e) circle(1mm);
          \end{feynman}
        \end{tikzpicture}
\end{subfigure} 
\caption{$NN \rightarrow NN\pi$ production via the $p$-wave $NN\pi$ vertex (left), the suppressed $s$-wave $NN\pi$ vertex (middle) and the $NN\pi\pi$ Weinberg-Tomozawa vertex (right).}
\label{fig:weinbergtomozawa}
\end{figure}
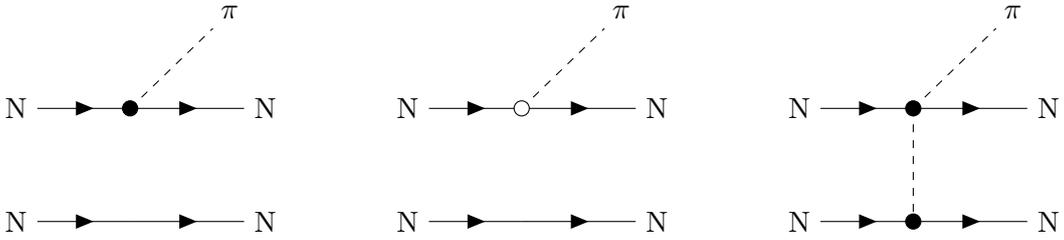

Several irreducible diagrams may be seen as modifications of the impulse diagram. The $\Delta$-excitation impulse diagrams provide for an excited nucleon state between the pion production vertex and the subsidiary strong interaction. The recoil diagrams have distinct kinematics to their one-pion-exchange impulse counterparts. Furthermore, as mentioned previously, there exist two distinct $NN\pi$ vertices corresponding to $p$- and $s$-wave pion production, the latter of which is suppressed; impulse diagrams with both vertices are shown in Figure \ref{fig:weinbergtomozawa}.
The seagull or pion-rescattering diagram arises from the inclusion of the $NN\pi\pi$ or Weinberg-Tomozawa (WT) interaction vertex, and is irreducible, as seen in Figure \ref{fig:weinbergtomozawa}. Irreducible diagrams containing higher-dimensional vertices of course can be constructed, but are surfeit to our needs. A collection of one-loop irreducible diagrams generically prove to be more relevant~\cite{Cohen1996,Hanhart2004}.

The power-counting of~\cite{Cohen1996} for $s$-wave pion production places impulse and $\Delta$-excitation contributions at leading order, with recoil corrections, the seagull diagram, and all one-loop diagrams subleading by a ratio $m_\pi/M_N$. However, numerical calculations with phenomenological $NN$ wavefunctions reveal that the seagull contribution is approximately equal (and opposite in sign) to the impulse contribution, which the authors attribute to some combination of spin/isospin factors and off-shell behavior. They also reveal a suppression of the $\Delta$-excitation contributions which is strongly dependent on the choice of $NN$-potential. Overall, the calculated result is much too small to agree with experiment, and so~\cite{Cohen1996} is taken as a methodological point of comparison rather than a generalisable result. Ref.~\cite{Park1996} similarly realises the cancellation between impulse contributions, attributing the mismatch with experiment to unaccounted-for heavy-meson exchange diagrams.

The fully relativistic approach of~\cite{Bernard1999} delivered near percent-precision cross sections for $s$-wave $pp \rightarrow pp \pi^0$ and $pp \rightarrow pn \pi^+$ at threshold, including a verification of the importance of heavy-meson exchange diagrams. The decomposition of an amplitude into perturbative diagrams is (always) model-dependent, in the sense that we cannot take this relativistic computation as a comparable breakdown for the power-counting of diagrams in $\chi EFT$ with non-relativistic nucleons. However,~\cite{Bernard1999} also presents an excellent spin-operator decomposition of both processes,
\begin{align}
    T^{\text{com}}_{\text{thr}}(pp \rightarrow pp \pi^0) &= \mathcal{A} ( i \vec{\sigma}_1 - i \vec{\sigma}_2 + \vec{\sigma}_1 \times \vec{\sigma}_2) \cdot \vec{p} \nonumber \\
    T^{\text{com}}_{\text{thr}}(pp \rightarrow pn \pi^+) &= \frac{\mathcal{A}}{\sqrt{2}} ( i \vec{\sigma}_1 - i \vec{\sigma}_2 + \vec{\sigma}_1 \times \vec{\sigma}_2) \cdot \vec{p} - \sqrt{2} \mathcal{B} i ( \vec{\sigma}_1 + \vec{\sigma}_2) \cdot \vec{p},
\end{align}
where two complex-valued coefficients $\mathcal{A},\mathcal{B}$ are then evaluated within one's chosen model, and correspond respectively to $^3P_0 \rightarrow ^1S_0 + s$ and $^3P_1 \rightarrow ^3S_1 + s$ transitions~\cite{Hanhart2002}. Refs.~\cite{Hanhart2002,Hanhart2004} employ the same representation in a demonstration that, with the modified power-counting introduced by~\cite{Cohen1996}, the perturbative expansion of $\chi EFT$ for threshold pion production does indeed converge. 

It is easier to show this for $p$-wave pion production because irreducible loop diagrams should only appear at $N^3LO$~\cite{Hanhart2000}; the impulse term alone gives agreement within experimental error bars for $p_\pi < 0.7 m_\pi$, with seagull and subleading $NN\pi$ vertex contributions entering at $NNLO$ and pushing agreement up to $p_\pi \sim 0.9 m_\pi$. For $s$-wave pion production, irreducible loop diagrams enter at $NLO$, but Ref.~\cite{Hanhart2002} shows an exact cancellation between divergences of these diagrams, which in the neutral-pion case has zero finite part. This is to be expected from consistency of the power-counting, Ref.~\cite{Hanhart2004} explains, because there is no $NNLO$ counterterm diagram to absorb $NLO$ divergences. To $NLO$, $\mathcal{A}$ is then built only from the impulse and $\Delta$-excitation terms, while $\mathcal{B}$ is built from the impulse, seagull, and finite loop terms.

$NNLO$ $p$-wave pion-production amplitudes were produced by~\cite{Baru2009} for three experimentally comparable channels ($pn \rightarrow pp \pi^-$, $pp \rightarrow d \pi^+$, and $pp \rightarrow pn \pi^+$), with a comfortable fit to data with $p_\pi < 0.4 m_\pi$. Later,~\cite{Filin2012,Filin2013} followed a similar formalism in computing the first complete $NNLO$ $s$-wave pion-production amplitude, employing a $\chi EFT$ Lagrangian with explicit $\Delta$ degrees of freedom. Order-of-magnitude expectations from the modified power-counting of~\cite{Cohen1996} were again verified. Later still,~\cite{Baru2016} performed the phenomenological $NN$ wavefunction convolution necessary to compare the $NNLO$ $s$-wave $pp \rightarrow d \pi^+$ amplitude to experimental data, finding a $10-20\%$ deficit for cutoffs between $600-1000$~MeV which is attributed to short-range processes e.g. $\rho$- and $\omega$-meson exchanges.

All this has been a rather roundabout way of justifying the application of $LO$ and $NLO$ $\chi EFT$ to the virtual pion production and absorption within our $T^{(S)}$ operators. In the absence of an on-shell requirement for our intermediate $NN\pi$ state, we will follow the Weinberg power-counting with small parameter $\varepsilon \sim m_\pi/M_N \sim 0.1$.

For $p$-wave charged-pion production, the leading terms~\cite{Hanhart2004} are the impulse and seagull diagrams of size $p_\pi/m_\pi \sim 1$ and $p_\pi m_\pi / p_N^2 \sim 1$, with error estimate $p_\pi / M_N \sim 0.1$ provided by the same topologies with subleading pion vertices.

For $s$-wave neutral- and charged-pion production, we need only consider partial amplitude $\mathcal{A}$ since we require one $^1S_0$ external $NN$ state. The the seagull term vanishes regardless of kinematics as the WT vertex is isovector, so the leading contributions are in fact $NLO$, arising from the subleading impulse and $\Delta$-excitation diagrams of size $p_N / M_N \sim 0.1$.

\subsection{\texorpdfstring{$\pi\pi$ and $\pi N$ neutrino potentials from $\chi EFT$}{pi-pi and pi-N neutrino potentials from chiral EFT}}

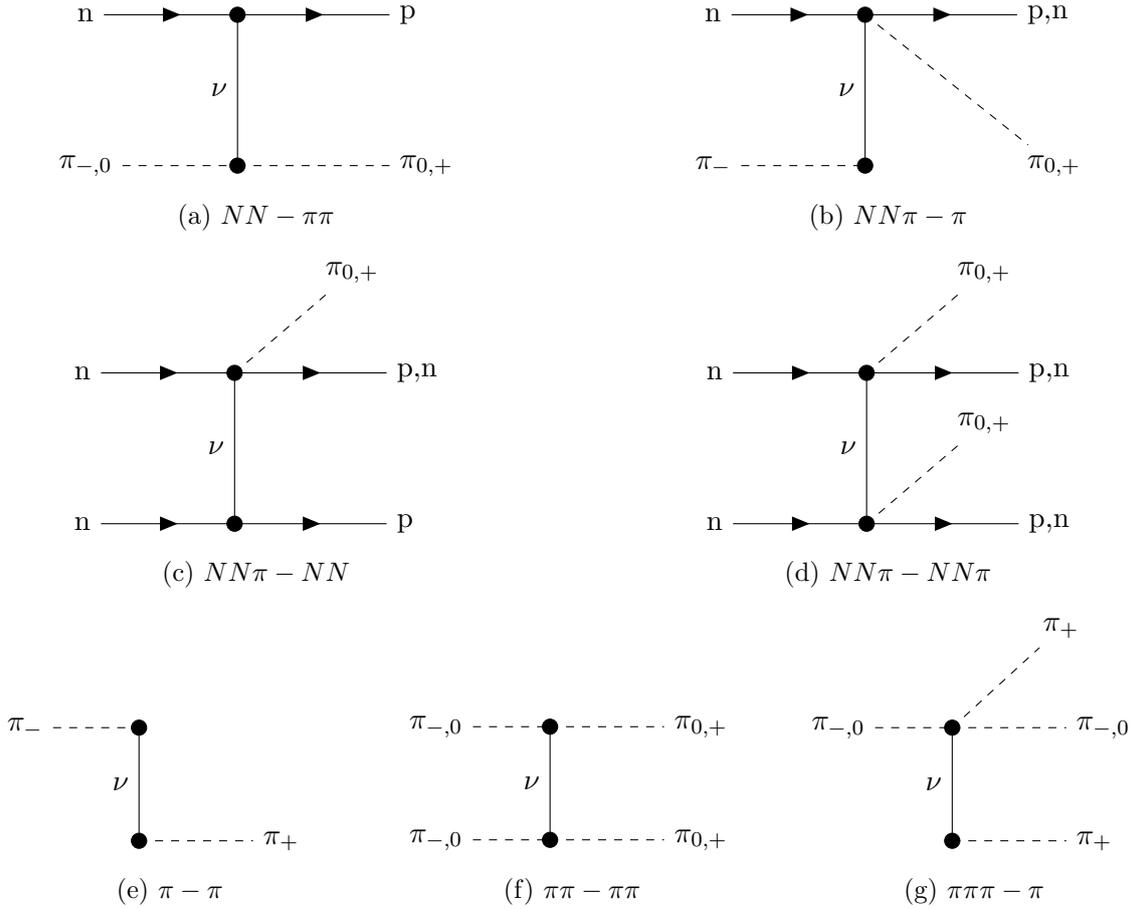
\begin{figure}[t]
    \centering % <-- added
\begin{subfigure}{0.45\textwidth}
    \centering
         \begin{tikzpicture}
          \begin{feynman}
            \vertex (a) {n};
            \vertex [right=2cm of a] (b);
            \vertex [right=2cm of b] (c) {p};
            \vertex [below=2cm of b] (e);
            \vertex [below=2cm of a] (d) {\(\pi_{-,0}\)};
            \vertex [right=2cm of e] (f) {\(\pi_{0,+}\)};
        
            \diagram* {
              (a) -- [fermion] (b) -- [fermion] (c),
              (d) -- [dashed] (e) -- [dashed] (f),
              (b) -- [solid, edge label'=\(\nu\)] (e),
            };
            
            \draw[fill=black] (b) circle(1mm);
            \draw[fill=black] (e) circle(1mm);
          \end{feynman}
        \end{tikzpicture}
  \caption{$NN-\pi\pi$}
  \label{fig:potentialnnpipi1}
\end{subfigure}\hfill % <-- added
\begin{subfigure}{0.45\textwidth}
    \centering
          \begin{tikzpicture}
          \begin{feynman}
            \vertex (a) {n};
            \vertex [right=2cm of a] (b);
            \vertex [right=2cm of b] (c) {p,n};
            \vertex [below=2cm of b] (e);
            \vertex [below=2cm of a] (d) {\(\pi_{-}\)};
            \vertex [right=2cm of e] (f) {\(\pi_{0,+}\)};
        
            \diagram* {
              (a) -- [fermion] (b) -- [fermion] (c),
              (d) -- [dashed] (e),
              (b) -- [dashed] (f),
              (b) -- [solid, edge label'=\(\nu\)] (e),
            };
        
            \draw[fill=black] (b) circle(1mm);
            \draw[fill=black] (e) circle(1mm);
          \end{feynman}
        \end{tikzpicture}
  \caption{$NN\pi-\pi$}
  \label{fig:potentialnnpipi2}
\end{subfigure}

\medskip
\begin{subfigure}{0.45\textwidth}
    \centering
        \begin{tikzpicture}
          \begin{feynman}
            \vertex (a) {n};
            \vertex [right=2cm of a] (b);
            \vertex [right=2cm of b] (c) {p,n};
            \vertex [above right=of b] (g) {\(\pi_{0,+}\)};;
            \vertex [below=2cm of b] (e);
            \vertex [below=2cm of a] (d) {n};
            \vertex [right=2cm of e] (f) {p};
        
            \diagram* {
              (a) -- [fermion] (b) -- [fermion] (c),
              (d) -- [fermion] (e) -- [fermion] (f),
              (b) -- [dashed] (g),
              (b) -- [solid, edge label'=\(\nu\)] (e),
            };
        
            \draw[fill=black] (b) circle(1mm);
            \draw[fill=black] (e) circle(1mm);
          \end{feynman}
        \end{tikzpicture}
  \caption{$NN\pi-NN$}
  \label{fig:potentialnnpinn}
\end{subfigure}\hfill % <-- added
\begin{subfigure}{0.45\textwidth}
    \centering
        \begin{tikzpicture}
          \begin{feynman}
            \vertex (a) {n};
            \vertex [right=2cm of a] (b);
            \vertex [right=2cm of b] (c) {p,n};
            \vertex [above right=of b] (g) {\(\pi_{0,+}\)};;
            \vertex [below=2cm of b] (e);
            \vertex [below=2cm of a] (d) {n};
            \vertex [right=2cm of e] (f) {p,n};
            \vertex [above right=of e] (h) {\(\pi_{0,+}\)};;
        
            \diagram* {
              (a) -- [fermion] (b) -- [fermion] (c),
              (d) -- [fermion] (e) -- [fermion] (f),
              (b) -- [dashed] (g),
              (e) -- [dashed] (h),
              (b) -- [solid, edge label'=\(\nu\)] (e),
            };
        
            \draw[fill=black] (b) circle(1mm);
            \draw[fill=black] (e) circle(1mm);
          \end{feynman}
        \end{tikzpicture}
  \caption{$NN\pi-NN\pi$}
  \label{fig:potentialnnpinnpi}
\end{subfigure}

\medskip
\begin{subfigure}{0.3\textwidth}
          \begin{tikzpicture}
          \begin{feynman}
            \vertex (a) {\(\pi_{-}\)};
            \vertex [right=of a] (b);
            \vertex [right=of b] (c);
            \vertex [below=of b] (e);
            \vertex [below=of a] (d);
            \vertex [right=of e] (f) {\(\pi_{+}\)};
        
            \diagram* {
              (a) -- [dashed] (b),
              (e) -- [dashed] (f),
              (b) -- [solid, edge label'=\(\nu\)] (e),
            };
            
            \draw[fill=black] (b) circle(1mm);
            \draw[fill=black] (e) circle(1mm);
          \end{feynman}
        \end{tikzpicture}
  \caption{$\pi-\pi$}
  \label{fig:potential1pi}
\end{subfigure}\hfill % <-- added
\begin{subfigure}{0.3\textwidth}
         \begin{tikzpicture}
          \begin{feynman}
            \vertex (a) {\(\pi_{-,0}\)};
            \vertex [right=of a] (b);
            \vertex [right=of b] (c) {\(\pi_{0,+}\)};
            \vertex [below=of b] (e);
            \vertex [below=of a] (d) {\(\pi_{-,0}\)};
            \vertex [right=of e] (f) {\(\pi_{0,+}\)};
        
            \diagram* {
              (a) -- [dashed] (b) -- [dashed] (c),
              (d) -- [dashed] (e) -- [dashed] (f),
              (b) -- [solid, edge label'=\(\nu\)] (e),
            };
            
            \draw[fill=black] (b) circle(1mm);
            \draw[fill=black] (e) circle(1mm);
          \end{feynman}
        \end{tikzpicture}
  \caption{$\pi\pi-\pi\pi$}
  \label{fig:potential2pi}
\end{subfigure}\hfill % <-- added
\begin{subfigure}{0.3\textwidth} 
        \begin{tikzpicture}
          \begin{feynman}
            \vertex (a) {\(\pi_{-,0}\)};
            \vertex [right=of a] (b);
            \vertex [right=of b] (c) {\(\pi_{-,0}\)};
            \vertex [above right=of b] (g) {\(\pi_{+}\)};
            \vertex [below=of b] (e);
            \vertex [below=of a] (d);
            \vertex [right=of e] (f) {\(\pi_{+}\)};
        
            \diagram* {
              (a) -- [dashed] (b) -- [dashed] (c),
              (b) -- [dashed] (g),
              (e) -- [dashed] (f),
              (b) -- [solid, edge label'=\(\nu\)] (e),
            };
            
            \draw[fill=black] (b) circle(1mm);
            \draw[fill=black] (e) circle(1mm);
          \end{feynman}
        \end{tikzpicture}
  \caption{$\pi\pi\pi-\pi$}
  \label{fig:potential3pi}
\end{subfigure}
\caption{Diagrams for the neutrino interaction potentials relevant for $NN\pi$ intermediate states. The solid dots denote interaction vertices from the chiral Lagrangian.}
\label{fig:potentialdiagrams}
\end{figure}

Now that we have determined the $T^{(S)}$ physics relevant to the $NN\pi$ intermediate state contribution, we must also compute the relevant neutrino potential contributions corresponding to $\braket{NN\pi | T^{(\nu)} | NN}$. Intuitively, one expects that $\braket{NN\pi | T^{(\nu)} | NN\pi}$ will be dominated by three channels: the usual neutrino insertion between two nucleon lines (but in the presence of a pion state); an insertion between a pion line and a nucleon line; and a double-insertion on the same pion line. The leading irreducible contribution to $\braket{NN\pi | T^{(\nu)} | NN}$ also includes a neutrino insertion between one nucleon line and a $\pi\nu NN$ vertex; all these diagrams are displayed in Figure \ref{fig:potentialdiagrams}. We already discussed the Lorentz decomposition of neutrino couplings to a single nucleon line in Section 4.2, in particular Eq.~\eqref{piNNcurrents} and the following $\chi EFT$-derived form factors, but in order to quantify these channels we must first perform a similar analysis of $\Delta L = 2$ couplings to one, two, and three pions, and to a $\pi\nu NN$ vertex.

All pion couplings are derivable from the generic chiral Lagrangian, which with external vector-like couplings can be parameterised as follows~\cite{Cirigliano2017}:
\begin{equation}
    \mathcal{L}_\pi = \frac{F_\pi^2}{4} Tr \left[ (D_\mu U)^\dagger D^\mu U \right] = \frac{F_\pi^2}{4} Tr \left[ \left( \partial_\mu U^\dagger + i U^\dagger l_\mu - i r_\mu U \right) \left( \partial^\mu U - i l^\mu U + i U r^u \right) \right],
\end{equation}
where $l_\mu$ and $r_\mu$ will facilitate a coupling to the leptonic states via a $W$-propagator, and therefore have an associated isospin operator $\tau^+$. Here we show the extraction of the $\pi\nu$ vertex Feynman rule as an example. To compute the coupling to an axial-vector leptonic current $2 a_\mu \equiv l_\mu - r_\mu$, we set $- l_\mu = r_\mu = a_\mu$, and find:
\begin{align}
    \mathcal{L}_\pi &= \frac{i F_\pi^2}{4} Tr \left[ \frac{-i}{F_\pi} \partial_\mu \vec{\phi} \cdot \vec{\tau} (2 a^\mu \tau^+) + (-2 a_\mu \tau^+) \frac{i}{F_\pi} \partial^\mu \vec{\phi} \cdot \vec{\tau} \right] \nonumber\\
    &= \frac{F_\pi}{2} Tr \left[ \left\{ \vec{\partial_\mu \phi} \cdot \vec{\tau}, \tau^+ \right\} \right] a^\mu \nonumber\\
    &= F_\pi \sqrt{2} \partial_\mu \left[ \pi^- \right] a^\mu
\end{align}
where the quantity in square brackets indicates the isospin constraints on the Feynman vertex, amounting to charge conservation (i.e. only a $\pi_-$ may decay to an $e_-$). Strictly speaking this pion field is factored away from the coupled current at the vertex, so this notation serves only to indicate which diagrams are permitted. If instead, one takes $l_\mu = r_\mu = v_\mu$, a trivial cancellation reveals that the $\pi$ coupling to a vector leptonic current vanishes. Similar computations reveal the patterned structure of $N \pi \nu$ vertices, where we display only those relevant for our $NN\pi$ diagrams:

\begin{align}
    J^\mu_{V,\pi} &= 0 \nonumber\\
    J^\mu_{A,\pi} &= -i F_\pi \sqrt{2} q^\mu [\pi^-] \nonumber\\
    J^\mu_{V,\pi\pi} &= - \sqrt{2} q^\mu [\pi^- \pi^0] \nonumber\\
    J^\mu_{A,\pi\pi} &= 0 \nonumber\\
    J^\mu_{V,\pi\pi\pi} &= 0, \nonumber\\
    J^\mu_{A,\pi\pi\pi} &= i \frac{1}{F_\pi} \sqrt{2} q^\mu [\pi^- \pi^0 \pi^0 + 2 \pi^- \pi^+ \pi^-] .
\end{align}
Also derivable from the above Lagrangian is the Feynman rule for the four-pion vertex, which we will later need to connect the $\pi\pi\pi-\pi$ neutrino potential into a loop diagram. In our case this vertex will connect two momenta $\pm k$ and two momenta $\pm l$, and has amplitude:
\begin{equation}
    i A_{\pi\pi\pi\pi} = \frac{-2}{3 F_\pi^2} \left( k^2 + l^2 - 2M^2 \right),
\end{equation}
where $M^2$ is a mass parameter generated by the explicit chiral symmetry breaking of the quark masses, which at $LO$ is simply $m_\pi^2$~\cite{Scherer2011}.

To study the $\pi\nu NN$ vertex, we also follow~\cite{Cirigliano2017} in introducing an NLO nucleonic chiral Lagrangian:
\begin{align}
    \mathcal{L}_{\pi N, LO} &= i \overline{N} V \cdot D N + g_A \overline{N} S \cdot u N \nonumber\\
    \mathcal{L}_{\pi N, NLO} &= \frac{1}{2 M_N} \left( V^\mu V^\nu - \eta^{\mu\nu} \right) \left( \overline{N} D_\mu D_\nu N \right) + \frac{i g_A}{2 M_N} \overline{N} \left\{ S \cdot D, V \cdot u \right\} N,
\end{align}
where a static nucleon approximation has been made with velocity $V^\mu = (1,\vec{0})$ and spin $S^\mu = (0,\vec{\sigma})$. $D_\mu$ is the chiral covariant derivative derived in terms of covariant vector $\Gamma_\mu$, and $u_\mu$ is the contravariant vector counterpart:
\begin{align}
    \Gamma_\mu &= D_\mu - \partial_\mu = \frac{1}{2} \left[ u^\dagger \left( \partial_\mu - i l_\mu \right) u + u \left( \partial_\mu - i r_\mu \right) u^\dagger \right] \nonumber\\
    u_\mu &= i \left[ u^\dagger \left( \partial_\mu - i l_\mu \right) u - u \left( \partial_\mu - i r_\mu \right) u^\dagger \right].
\end{align}
Note that this is the same chiral order to which the neutrino potential for $NN$ states has been presented here as well as in~\cite{Cirigliano2021long}. We will see that those terms in the Lagrangian which contributed to the vector-coupling for an $NN$ state will contribute only to the axial-vector-coupling here, and vice versa. Again setting $- l_\mu = r_\mu = a_\mu$ for the axial-vector contribution, we find at LO:
\begin{align}
    i \overline{N} V^\mu \vec{D}_\mu N &= i \overline{N} V^\mu \Gamma_\mu N = \frac{i}{2} \overline{N} V^\mu \left( u^\dagger \left( i A_\mu \tau^+ \right) u + u \left( -i A^\mu \tau^+ \right) u^\dagger \right) \nonumber\\
    &= \frac{-1}{2} \overline{N} V^\mu \frac{i}{F_\pi} A_\mu \left[ \tau^+, \vec{\phi} \cdot \vec{\tau} \right] N \nonumber\\
    &= \frac{-i}{2 F_\pi} V^\mu A_\mu [\sqrt{2} \overline{p} \pi^- p - \sqrt{2} \overline{n} \pi^- n - 2 \overline{p} \pi^0 n],
\end{align}
and at NLO,
\begin{align}
    i \overline{N} \overleftarrow{D_\mu} \overrightarrow{D_\nu} N &= \overline{N} \left( \overleftarrow{\partial_\mu} \Gamma_\nu + \Gamma_\mu \overrightarrow{\partial_\nu} \right) N \nonumber\\
    &= \frac{-i}{2 F_\pi} \frac{p^\mu + p'^\mu}{2 M_N} A_\mu [\sqrt{2} \overline{p} \pi^- p - \sqrt{2} \overline{n} \pi^- n - 2 \overline{p} \pi^0 n].
\end{align}
Setting $l_\mu = r_\mu = v_\mu$ for the vector contribution, we find at LO
\begin{align}
    g_A \overline{N} S \cdot u N &= g_A \overline{N} S^\mu \left( u^\dagger v_\mu \tau^+ u - u v_\mu \tau^+ u^\dagger \right) \nonumber\\
    &=  \frac{i g_A}{F_\pi} \overline{N} S^\mu, \left[ \tau^+, \vec{\phi} \cdot \vec{\tau} \right] v_\mu N \nonumber\\
    &= \frac{i g_A}{F_\pi} S^\mu v_\mu, [\sqrt{2} \overline{p} \pi^- p - \sqrt{2} \overline{n} \pi^- n - 2 \overline{p} \pi^0 n],
\end{align}
and at NLO,
\begin{align}
    - \frac{i g_A}{2 M_N} \left\{ S \cdot D, V \cdot u \right\} N &= \frac{g_A}{2 F_\pi M_N} \overline{N} \left\{ S^\mu \partial_\mu, V^\nu v_\nu \left[ \tau^+, \vec{\phi} \cdot \vec{\tau} \right] \right\} N \nonumber\\
    &= \frac{- i g_A}{F_\pi} \frac{V^\mu}{2 M_N} S \cdot (p + p') v_\mu, [\sqrt{2} \overline{p} \pi^- p - \sqrt{2} \overline{n} \pi^- n - 2 \overline{p} \pi^0 n],
\end{align}
as well as the vector current induced by the pseudoscalar $NN\pi$ vertex and the $\pi-\nu$ coupling derived above:
\begin{equation}
    J^\mu_{V,induced} = \frac{i g_A}{F_\pi} \left( \frac{1}{q^2 + m_\pi^2} q^\mu S \cdot q \right), [\sqrt{2} \overline{p} \pi^- p - \sqrt{2} \overline{n} \pi^- n - 2 \overline{p} \pi^0 n].
\end{equation}
Thus in total we have for the currents at the $\pi\nu NN$ vertex:
\begin{align}
    J^\mu_{V,NN\pi} = &\frac{i g_A}{F_\pi} \left( S^\mu - \frac{V^\mu}{2 M_N} S \cdot (p + p') + \frac{1}{q^2 + m_\pi^2} q^\mu S \cdot q \right), \nonumber\\
        & \cdot \left[ \sqrt{2} \overline{p} \pi^- p - \sqrt{2} \overline{n} \pi^- n - 2 \overline{p} \pi^0 n \right] \nonumber\\
    J^\mu_{A,NN\pi} = &\frac{-i}{2 F_\pi} \left( v^\mu + \frac{p^\mu + p'^\mu}{2 M_N} \right), [\sqrt{2} \overline{p} \pi^- p - \sqrt{2} \overline{n} \pi^- n - 2 \overline{p} \pi^0 n].
\end{align}
With all vertex-level building blocks defined, we can now contract them to form complete $\Delta L = 2$ potentials for $NN\pi$ states. Let us first consider the neutrino interaction between a nucleon line and a pion line, shown diagrammatically in Figure \ref{fig:potentialnnpipi1}. Since we have seen that the latter only couples via a vector current, the $NN-\pi\pi$ potential is given by:
\begin{align}
    J^\mu_{V,NN} \cdot J_{\mu V,\pi\pi} &= \left( V^\mu + \frac{p^\mu + p'^\mu}{2 M_N} \right) \left( - \sqrt{2} q^\mu \right) \nonumber\\
    &= -\sqrt{2} \left( q_0 + \frac{q \cdot (p + p')}{2 M_N} \right) \nonumber\\
    &= -\sqrt{2} q_0 \quad \left[ n \pi_0 \rightarrow p \pi_+, n \pi_- \rightarrow p \pi_0 \right],
\end{align}
where the final simplification results from observing that $q \cdot (p + p') = (p - p') \cdot (p + p') = p^2 - p'^2 \simeq E_{1,2}^2$. Following~\cite{Cirigliano2017}, we count the electron energies at $NNLO$ in the chiral power-counting, specifically $E_{1,2} \sim m_\pi \left( \frac{Q}{\Lambda_\chi} \right)^2$, and therefore such terms can be safely neglected from our $NLO$ potentials.

\newpage
The neutrino interaction between a $\pi\nu NN$ vertex and a single $\pi$, shown in Figure \ref{fig:potentialnnpipi2}, proceeds only via an axial-vector current, but delivers a very similar potential to the above (this is less surprising comparing the diagram to that of Figure \ref{fig:potentialnnpipi1}):
\begin{align}
    J^\mu_{A,NN\pi} \cdot J_{\mu A,\pi} = &-\frac{\sqrt{2}}{2} q_0, \nonumber\\
    &\left[\sqrt{2} p \pi_- \rightarrow p \pi_+, - \sqrt{2} n \pi_- \rightarrow n \pi_+, -2 n \pi_- \rightarrow p \pi_0 \right].
\end{align}
In fact this pair of potentials very clearly exhibits the importance of carrying through all isospin algebra; although their kinematic dependence differs only by a constant factor, they share only one isospin channel and discriminate between three others.
The neutrino interaction between a $\pi\nu NN$ vertex and a nucleon line $NN$, shown in Figure \ref{fig:potentialnnpinn}, exhibits more complexity, as both vector $V-V$ and axial-vector $A-A$ channels may contribute. Fortunately, these two components turn out to be identical, and the static nucleon approximation leads to significant simplification:
\begin{align}
    J^\mu_{V,NN\pi} \cdot J_{\mu V,NN} = &J^\mu_{A,NN\pi} \cdot J_{\mu A,NN} \nonumber\\
    = &\frac{i g_A}{F_\pi} \left( S^\mu - \frac{v^\mu}{2 M_N} S \cdot (p + p') + \frac{1}{q^2 + m_\pi^2} q^\mu S \cdot q \right) \left( v_\mu + \frac{p_\mu + p'_\mu}{2 M_N} \right) \nonumber\\
    = &\frac{i g_A}{F_\pi} \Big( \frac{1}{q^2 + m_\pi^2} q_0 S \cdot q - \frac{1}{4 M_N^2} (p^0 + p'^0) S \cdot (p + p') \nonumber\\
    &+ \frac{1}{q^2 + m_\pi^2} \frac{1}{2 M_N} q \cdot (p + p') S \cdot q \Big) \nonumber\\
    = &\frac{i g_A}{F_\pi} \left( \frac{1}{q^2 + m_\pi^2} q_0 \frac{1}{2} \vec{\sigma} \cdot \vec{q} - \frac{\vec{\sigma} \cdot (\vec{p} + \vec{p'})}{4 M_N} \right) \nonumber\\
     \quad \quad &\left[ \sqrt{2} np \rightarrow pp \pi_+, - \sqrt{2} nn \rightarrow pn \pi_+, - 2 nn \rightarrow pp \pi_0 \right].
\end{align}
Similarly, the neutrino interaction between two $\pi\nu NN$ vertices, shown in Figure \ref{fig:potentialnnpinnpi}, receives contributions from both vector $V-V$ and axial-vector $A-A$ channels. The current algebra matches that of the $NN$ neutrino-exchange, albeit with vector and axial-vector contributions interchanged, and an additional constant factor $\frac{-1}{F_\pi^2}$:
\begin{align}
    J^\mu_{V,NN\pi} \cdot J_{\mu V,NN\pi} &= \frac{- 4 g_A^2}{F_\pi^2} \left( S^\mu - \frac{v^\mu}{2 M_N} S \cdot (p + p') + \frac{1}{q^2 + m_\pi^2} q^\mu S \cdot q \right)^2 \nonumber\\
    &= \frac{- 4 g_A^2}{F_\pi^2} \left( S \cdot S  + \frac{2}{q^2 + m_\pi^2} (S \cdot q)^2 + \frac{q^2}{\left(q^2 + m_\pi^2\right)^2} (S \cdot q)^2 \right) \nonumber\\
    &= \frac{g_A^2}{F_\pi^2} \left( \vec{\sigma}_1 \cdot \vec{\sigma}_2  - \frac{q^2 + 2 m_\pi^2}{\left(q^2 + m_\pi^2\right)^2} \vec{\sigma}_1 \cdot \vec{q} \vec{\sigma}_2 \cdot \vec{q} \right) \nonumber\\
     &= \frac{g_A^2}{F_\pi^2} \left( \frac{2}{3} \vec{\sigma}_1 \cdot \vec{\sigma}_2 \left( 1 + \frac{m_\pi^4}{2 \left( q^2 + m_\pi^2 \right)^2} \right)  + \frac{1}{3} S_{12} \left( 1 - \frac{m_\pi^4}{\left( q^2 + m_\pi^2 \right)^2} \right) \right), \nonumber\\
    & \quad \quad \left[ nn \rightarrow pp \pi_0 \pi_0 \right],
\end{align}
and
\begin{equation}
    J^\mu_{A,NN\pi} \cdot J_{\mu A,NN\pi} = \frac{- 1}{F_\pi^2} \left( V_\mu + \frac{p_\mu + p'_\mu}{2 M_N} \right)^2 = \frac{-1}{F_\pi^2} + \mathcal{O} \left(\frac{Q^2}{\Lambda_\chi^2} \right), \quad \left[ nn \rightarrow pp \pi_0 \pi_0 \right],
\end{equation}
Here we have only included the isospin factors for the double-neutral-pion case, as this is the only one relevant for our later analysis. The charged-pion channels are slightly suppressed by factors of either $\pm \sqrt{2}$ or $\pm 2$.

Finally, we have three pion-only potentials, of the forms $\pi-\pi$, $\pi\pi-\pi\pi$, and $\pi\pi\pi-\pi$, which are shown in Figures \ref{fig:potential1pi}, \ref{fig:potential2pi} and \ref{fig:potential3pi}, respectively. The first and third of these involve only axial-vector couplings because both vertices connect an odd number of pions; the second conversely involves only vector couplings. Contracting our previously calculated currents, we have:
\begin{align}
    J^\mu_{A,\pi} \cdot J_{\mu A,\pi} &= \left( -i F_\pi \sqrt{2} q^\mu \right) \left( -i F_\pi \sqrt{2} q_\mu \right) \nonumber\\
    &= -2 F_\pi^2 q^2, \quad \left[ \pi_- \rightarrow \pi_+ \right], \\
    J^\mu_{V,\pi\pi} \cdot J_{\mu V,\pi\pi} &= \left( - \sqrt{2} q^\mu \right) \left( - \sqrt{2} q_\mu \right) \nonumber\\
    &= 2 q^2, \quad \left[ \pi_- \pi_- \rightarrow \pi_0 \pi_0, \pi_- \pi_0 \rightarrow \pi_0 \pi_+, \pi_0 \pi_0 \rightarrow \pi_+ \pi_+ \right], \\
    J^\mu_{A,\pi\pi\pi} \cdot J_{\mu A,\pi} &= \left( \frac{-i}{F_\pi} \sqrt{2} q^\mu \right) \left( -i F_\pi \sqrt{2} q_\mu \right) \nonumber\\
    &= 2 q^2, \quad \left[ \pi_0 \rightarrow \pi_0 \pi_+ \pi_+, 2 \pi_- \rightarrow \pi_- \pi_+ \pi_+ \right].
\end{align}

To express these products of currents as bona fide two-nucleon potentials, we we must include the neutrino propagator $\frac{1}{q^2}$ as well as the leptonic structure (and dimensionful constants) which apply equally to all topologies. The resultant potentials are collected in Table \ref{tab:pinnpotentials}.

\renewcommand{\arraystretch}{1.3}
\begin{table}[t!]
\centering
\begin{tabular}{cll}
    \hline
    Potential $V^\nu_{...}$ & $G_F^2 V_{ud}^2 m_{\beta\beta}$ ( \quad ... \quad ) $\overline{u}(k_1) \hat{P}_R C \overline{u}^T(k_2)$ \\
    \hline
    $(NN,\pi\pi)$   &  $-\sqrt{2} \frac{q_0}{q^2}$ \\
    $(NN\pi,\pi)$  &  $-\frac{\sqrt{2}}{2} \frac{q_0}{q^2}$ \\
    $(NN\pi,NN)$   &  $\frac{i g_A}{F_\pi} \frac{1}{q^2} \left( \frac{1}{q^2 + m_\pi^2} q_0 \frac{1}{2} \vec{\sigma} \cdot \vec{q} - \frac{\vec{\sigma} \cdot (\vec{p} + \vec{p'})}{4 M_N} \right)$ \\
    $(NN\pi,NN\pi)$   &  $\frac{-1}{F_\pi^2} \frac{1}{q^2} \left(1 - \vec{\sigma}_1 \cdot \vec{\sigma}_2  - \frac{q^2 + 2 m_\pi^2}{\left(q^2 + m_\pi^2\right)^2} \vec{\sigma}_1 \cdot \vec{q} \vec{\sigma}_2 \cdot \vec{q} \right)$ \\
    $(\pi,\pi)$   &  $-2 F_\pi^2$ \\
    $(\pi\pi,\pi\pi)$   &  $2$ \\
    $(\pi\pi\pi,\pi)$   &   $2$ \\
    \hline
\end{tabular}
\caption{Neutrino interaction potentials $V^\nu_{(X,Y)} = J^\mu_{V,X} \cdot J_{\mu_V,Y} + J^\mu_{A,X} \cdot J_{\mu_A,Y}$ as functions of exchange momentum $q$. These potentials correspond to the diagrams of Figure~\ref{fig:potentialdiagrams}.}
\label{tab:pinnpotentials}
\end{table}
In computing these $NLO$ potentials, we experience the full benefits of the $\chi EFT$ formulation of nuclear forces. The highly symmetric interactions permitted for pions deliver simple kinematics for all permitted $\nu$-interactions. Indeed, only the $J^\mu_{V,NN\pi}\cdot J_{\mu V,NN}$ potential has any spin- or pion mass-dependence up to $NLO$, a fact which will dramatically ease the estimation of our leading $0\nu\beta\beta$ diagrams.

\subsection{\texorpdfstring{Diagrammatics and power-counting of the leading $NN\pi$ divergences}{Diagrammatics and power-counting of the leading NNpi divergences}}

With both off-shell pion production/absorption $NN \rightarrow NN\pi$ and $NN\pi \rightarrow NN$ and neutrino potentials connecting $NN\pi$ and $NN$ states characterised, we can now present the complete set of diagrams whose divergences will contribute to the size of the contact $0\nu\beta\beta$ counterterm. These are shown in Figure \ref{fig:pinndimanal}, alongside the workings which demonstrate their (order-of-magnitude) $NLO$ status in the chiral power-counting; the diagrams with only $NN$ intermediate states are also presented in Figure \ref{fig:nndimanal}.

We largely follow the power-counting formalised in~\cite{VanKolck2020}. As is standard in chiral perturbation theory, pion propagators (being light compared to the cutoff scale) contribute $1/Q^2$, while nucleon propagators (being heavy compared to the cutoff scale, and therefore static) contribute $1/Q$. Loop integrals generically contribute $Q^4/(4\pi)^2$, where the factor $(4\pi)^{-2}$ arises from angular integrations of $\frac{dq^4}{(2\pi)^4}$. From the exponential form of $U$ in the chiral Lagrangian, a vertex with $d$ derivatives, $p$ pion fields, and $n$ nucleon fields can be expected to contribute $Q^d F_\pi^{2-p-b} \Lambda_\chi^{2-d-b/2}$. At this point two large scales enter into the power-counting: $F_\pi$ and $\Lambda_\chi$; Manohar and Georgi's ``naive dimensional analysis" procedure shows that in order for the leading non-renormalizable four-pion vertex to be sufficiently and naturally suppressed, $\Lambda_\chi$ can be at most $4\pi F_\pi$, and in the chiral power-counting we take this bound to be saturated.

For multi-nucleon interactions,~\cite{Weinberg1991} showed that one must treat nucleon propagators and loops differently when they appear as reducible subdiagrams, i.e. elastic intermediate states. When integrating over these internal nucleon lines, pole contributions effectively restrict the nucleon from a static treatment, with the result~\cite{Bedaque2002,Hammer2020,VanKolck2020} that the nucleon propagator is enhanced to $\sim M_N/Q^2$. The loop integral contribution is adjusted to $Q^5/(4\pi M_N)$, but since in the power-counting $M_N \sim \Lambda_\chi \sim 4\pi Q$, is of the same order as the irreducible loop integral.

An additional feature of the chiral power-counting specific to our application is that it distinguishes between $p$- and $s$-wave pion production from / absorption to $NN$ states. As discussed in Section 4.2 and in~\cite{Hanhart2004}, the leading $NN\pi$ vertex in the chiral Lagrangian only gives nonzero contributions for $p$-wave pion production, and is assigned size $\frac{g_A}{2} \frac{Q}{F_\pi}$ in agreement with the generic vertex expression from~\cite{VanKolck2020} (we choose to retain simple constant factors since they prove substantial for certain $NN\pi$ diagrams). The first vertex to contribute to $s$-wave pion production is suppressed by one chiral order, specifically by a ratio $\frac{\omega_\pi}{2 M_N}$ where $\omega_\pi$ is the pion energy. We recall that neutral pion production receives no contribution from the $p$-wave.

We conduct our power-counting in the manner appropriate for comparison to the integrand $a_\chi(k)$ in the matching expression~\cite{Cirigliano2021long}:
\begin{equation}
    2 g_{\nu}^{NN}(\mu) = \frac{1 + 2 g_A^2}{2} - \int_0^\mu dk a_\chi(k) + \int_0^\Lambda dk a_<(k) + \int_\Lambda^\infty dk a_<(k).
    \label{integrandmatching2}
\end{equation}
Thus the factor $\frac{M_N^2}{(4\pi)^2}$ arising from the two $NN$ contact interactions on either side of each bubble diagram is neglected, and we implicitly include the neutrino-momentum measure $d|k|$ such that each estimate is formally dimensionless. As a proof of concept, Figure \ref{fig:nndimanal} shows the application of our chosen power-counting to the elastic diagrams $\mathcal{A}_C$ containing divergences; we separate these diagrams to make explicit the contributions from the pseudoscalar-induced axial $NN\nu$ coupling, whose presence will play a more direct role in the $NN\pi$-inelastic diagrams to follow. As expected, all three diagrams enter at leading chiral order $\left(\frac{Q}{\Lambda_\chi} \right)^0$.

\renewcommand{\arraystretch}{1.3}
\begin{figure}
\makebox[1 \textwidth][c]{       %centering table
\resizebox{1.3 \textwidth}{!}{   %resize table
\centering
\begin{tblr}{c l c}

    \hline
    Diagram & Dimensional Analysis & Chiral Order \\
    \hline
    \raisebox{-0.5\totalheight}{\begin{tikzpicture}
      \begin{feynman}
        \vertex (l);
        \vertex [right=of l] (ma);
        \vertex [right=of ma] (mb);
        \vertex [right=of mb] (mc);
        \vertex [right=of mc] (r);

        \vertex [above=0.8cm of ma] (na);
        \vertex [above=0.8cm of mb] (nb);
        \vertex [above=0.8cm of mc] (nc);
        \vertex [above=0.8cm of na] (a);
        \vertex [above=0.8cm of nb] (b);
        \vertex [above=0.8cm of nc] (c);
        \vertex [below=0.8cm of ma] (ka);
        \vertex [below=0.8cm of mb] (kb);
        \vertex [below=0.8cm of mc] (kc);
        \vertex [below=0.8cm of ka] (d);
        \vertex [below=0.8cm of kb] (e);
        \vertex [below=0.8cm of kc] (f);
    
        \diagram* {
          (l) -- [fermion, bend left, edge label'=n] (b) -- [fermion, bend left, edge label'=p] (r),
          (l) -- [fermion, bend right, edge label'=n] (e) -- [fermion, bend right, edge label'=p] (r),
          (b) -- [solid, edge label'=\(\nu\)] (e),
        };

        \draw[fill=black] (l) circle(1mm);
        \draw[fill=black] (r) circle(1mm);
        \draw[fill=black] (b) circle(1mm);
        \draw[fill=black] (e) circle(1mm);
      \end{feynman}
\end{tikzpicture}} 
    &  $\begin{aligned} g_V \cdot g_V + g_A \cdot g_A &= 1 + g_A^2 \\ &\simeq 2.6 \end{aligned}$ & LO\\

    \raisebox{-0.5\totalheight}{\begin{tikzpicture}
      \begin{feynman}
        \vertex (l);
        \vertex [right=of l] (ma);
        \vertex [right=of ma] (mb);
        \vertex [right=of mb] (mc);
        \vertex [right=of mc] (r);

        \vertex [above=0.8cm of ma] (na);
        \vertex [above=0.8cm of mb] (nb);
        \vertex [above=0.8cm of mc] (nc);
        \vertex [above=0.8cm of na] (a);
        \vertex [above=0.8cm of nb] (b);
        \vertex [above=0.8cm of nc] (c);
        \vertex [below=0.8cm of ma] (ka);
        \vertex [below=0.8cm of mb] (kb);
        \vertex [below=0.8cm of mc] (kc);
        \vertex [below=0.8cm of ka] (d);
        \vertex [below=0.8cm of kb] (e);
        \vertex [below=0.8cm of kc] (f);
    
        \diagram* {
          (l) -- [fermion, bend left, edge label'=n] (b) -- [fermion, bend left, edge label'=p] (r),
          (l) -- [fermion, bend right, edge label'=n] (e) -- [fermion, bend right, edge label'=p] (r),
          (b) -- [dashed, edge label'=\(\pi_-\)] (nb) -- [solid, edge label'=\(\nu\)] (e),
        };

        \draw[fill=black] (l) circle(1mm);
        \draw[fill=black] (r) circle(1mm);
        \draw[fill=black] (b) circle(1mm);
        \draw[fill=black] (e) circle(1mm);
        \draw[fill=black] (nb) circle(1mm);
      \end{feynman}
\end{tikzpicture}} 
    &  $\begin{aligned} \frac{g_A}{2} \frac{Q}{F_\pi} \cdot \frac{1}{Q^2} \cdot \sqrt{2} F_\pi Q \cdot g_A &= g_a^2 \frac{\sqrt{2}}{2} \\ &\simeq 1.1 \end{aligned}$ & LO\\

    \raisebox{-0.5\totalheight}{\begin{tikzpicture}
      \begin{feynman}
        \vertex (l);
        \vertex [right=of l] (ma);
        \vertex [right=of ma] (mb);
        \vertex [right=of mb] (mc);
        \vertex [right=of mc] (r);

        \vertex [above=0.8cm of ma] (na);
        \vertex [above=0.8cm of mb] (nb);
        \vertex [above=0.8cm of mc] (nc);
        \vertex [above=0.8cm of na] (a);
        \vertex [above=0.8cm of nb] (b);
        \vertex [above=0.8cm of nc] (c);
        \vertex [below=0.8cm of ma] (ka);
        \vertex [below=0.8cm of mb] (kb);
        \vertex [below=0.8cm of mc] (kc);
        \vertex [below=0.8cm of ka] (d);
        \vertex [below=0.8cm of kb] (e);
        \vertex [below=0.8cm of kc] (f);
    
        \diagram* {
          (l) -- [fermion, bend left, edge label'=n] (b) -- [fermion, bend left, edge label'=p] (r),
          (l) -- [fermion, bend right, edge label'=n] (e) -- [fermion, bend right, edge label'=p] (r),
          (b) -- [dashed, edge label'=\(\pi_-\)] (nb) -- [solid, edge label'=\(\nu\)] (kb)  -- [dashed, edge label'=\(\pi_-\)] (e),
        };

        \draw[fill=black] (l) circle(1mm);
        \draw[fill=black] (r) circle(1mm);
        \draw[fill=black] (b) circle(1mm);
        \draw[fill=black] (e) circle(1mm);
        \draw[fill=black] (nb) circle(1mm);
        \draw[fill=black] (kb) circle(1mm);
      \end{feynman}
\end{tikzpicture}} 
    &  $\begin{aligned} \left( \frac{g_A}{2} \frac{Q}{F_\pi} \cdot \frac{1}{q^2} \cdot \sqrt{2} F_\pi Q \right) \cdot \left( \frac{g_A}{2} \frac{Q}{F_\pi} \cdot \frac{1}{Q^2} \cdot \sqrt{2} F_\pi Q \right) &= \frac{g_a^2}{2} \\ &\simeq 0.8 \end{aligned}$ & LO\\
    \hline
\end{tblr}
}
}
\caption{Dimensional analysis of leading $NN$ neutrino-exchange diagrams. They are all at LO in the chiral power-counting.}
\label{fig:nndimanal}
\end{figure}
\renewcommand{\arraystretch}{1.2}
\begin{figure}
\vspace{-15pt}
\makebox[1 \textwidth][c]{       %centering table
\resizebox{1.2 \textwidth}{!}{   %resize table
\centering
\begin{tblr}{c c l c}
    \hline
    & Diagram & Dimensional Analysis & Chiral Order \\
    \hline
    (a) &
         \raisebox{-0.5\totalheight}{\begin{tikzpicture}
          \begin{feynman}
            \vertex (l);
            \vertex [right=of l] (ma);
            \vertex [right=of ma] (mb);
            \vertex [right=of mb] (mc);
            \vertex [right=of mc] (r);
    
            \vertex [above=0.8cm of ma] (na);
            \vertex [above=0.8cm of mb] (nb);
            \vertex [above=0.8cm of mc] (nc);
            \vertex [above=0.8cm of na] (a);
            \vertex [above=0.8cm of nb] (b);
            \vertex [above=0.8cm of nc] (c);
            \vertex [below=0.8cm of ma] (ka);
            \vertex [below=0.8cm of mb] (kb);
            \vertex [below=0.8cm of mc] (kc);
            \vertex [below=0.8cm of ka] (d);
            \vertex [below=0.8cm of kb] (e);
            \vertex [below=0.8cm of kc] (f);
        
            \diagram* {
              (l) -- [fermion, bend left, edge label'=n] (b) -- [fermion, bend left, edge label'=p] (r),
              (l) -- [fermion, bend right, edge label'=n] (e) -- [fermion, bend right, edge label'=p] (r),
              (b) -- [solid, edge label'=\(\nu\)] (e),
              (b) -- [dashed, bend left, edge label=\(\pi_0\)] (e) 
            };
    
            \draw[fill=black] (l) circle(1mm);
            \draw[fill=black] (r) circle(1mm);
            \draw[fill=black] (b) circle(1mm);
            \draw[fill=black] (e) circle(1mm);
          \end{feynman}
    \end{tikzpicture}} 
    &  $\begin{aligned} & \frac{1}{F_\pi^2} \left( 1 + 2 g_A^2 \right) \cdot \frac{1}{Q^2} \cdot \int \frac{dQ^4}{(4 \pi)^2} \\ &= \left( 1 + 2 g_A^2 \right) \frac{1}{(4\pi)^2} \simeq \left( 1 + 2 g_A^2 \right) \left( \frac{Q}{\Lambda_\chi} \right)^2 \quad\quad \simeq 4.2 \left( \frac{Q}{\Lambda_\chi} \right)^2 \end{aligned}$ & $NNLO$\\
    (b) &
    \raisebox{-0.5\totalheight}{\begin{tikzpicture}
      \begin{feynman}
        \vertex (l);
        \vertex [right=1.0cm of l] (ma);
        \vertex [right=1.0cm of ma] (mb);
        \vertex [right=1.0cm of mb] (mc);
        \vertex [right=1.0cm of mc] (md);
        \vertex [right=1.0cm of md] (me);
        \vertex [right=1.0cm of me] (r);

        \vertex [above=0.8cm of ma] (na);
        \vertex [above=0.8cm of mb] (nb);
        \vertex [above=0.8cm of mc] (nc);
        \vertex [above=0.8cm of md] (nd);
        \vertex [above=0.8cm of me] (ne);
        \vertex [above=0.8cm of na] (a);
        \vertex [above=0.8cm of nb] (b);
        \vertex [above=0.8cm of nc] (c);
        \vertex [above=0.8cm of nd] (d);
        \vertex [above=0.8cm of ne] (e);
        \vertex [below=0.8cm of ma] (ka);
        \vertex [below=0.8cm of mb] (kb);
        \vertex [below=0.8cm of mc] (kc);
        \vertex [below=0.8cm of md] (kd);
        \vertex [below=0.8cm of me] (ke);
        \vertex [below=0.8cm of ka] (f);
        \vertex [below=0.8cm of kb] (g);
        \vertex [below=0.8cm of kc] (h);
        \vertex [below=0.8cm of kd] (i);
        \vertex [below=0.8cm of ke] (j);
    
        \diagram* {
          (l) -- [fermion, bend left, edge label'=n] (a) -- [fermion] (e) -- [fermion, bend left, edge label'=p] (r),
          (l) -- [fermion, bend right, edge label'=n] (f) -- [fermion] (j) -- [fermion, bend right, edge label'=p] (r),
          (nb) -- [solid, edge label=\(\nu\)] (nd),
          (f) -- [dashed, bend left, edge label'=\(\pi_-\)] (mc) -- [dashed, edge label=\(\pi_-\)] (nb),
          (nd) -- [dashed, edge label=\(\pi_+\)] (mc) -- [dashed, bend right, edge label'=\(\pi_+\)]  (e)
        };

        \draw[fill=black] (l) circle(1mm);
        \draw[fill=black] (r) circle(1mm);
        \draw[fill=black] (e) circle(1mm);
        \draw[fill=black] (f) circle(1mm);
        \draw[fill=black] (nb) circle(1mm);
        \draw[fill=black] (nd) circle(1mm);
        \draw[fill=black] (mc) circle(1mm);
      \end{feynman}
\end{tikzpicture}}
    &  $\begin{aligned} &\left( 2 Q^2 \right) \cdot \left( \frac{1}{Q^2} \right)^3 \cdot \int \frac{dQ^4}{(4 \pi)^2} = 2 \frac{1}{(4\pi)^2} \quad\quad\quad\quad \simeq 2 \left( \frac{Q}{\Lambda_\chi} \right)^2 \end{aligned}$ & $NNLO$\\
    (c) &
    \raisebox{-0.5\totalheight}{\begin{tikzpicture}
      \begin{feynman}
        \vertex (l);
        \vertex [right=1.0cm of l] (ma);
        \vertex [right=1.0cm of ma] (mb);
        \vertex [right=1.0cm of mb] (mc);
        \vertex [right=1.0cm of mc] (md);
        \vertex [right=1.0cm of md] (me);
        \vertex [right=1.0cm of me] (r);

        \vertex [above=0.8cm of ma] (na);
        \vertex [above=0.8cm of mb] (nb);
        \vertex [above=0.8cm of mc] (nc);
        \vertex [above=0.8cm of md] (nd);
        \vertex [above=0.8cm of me] (ne);
        \vertex [above=0.8cm of na] (a);
        \vertex [above=0.8cm of nb] (b);
        \vertex [above=0.8cm of nc] (c);
        \vertex [above=0.8cm of nd] (d);
        \vertex [above=0.8cm of ne] (e);
        \vertex [below=0.8cm of ma] (ka);
        \vertex [below=0.8cm of mb] (kb);
        \vertex [below=0.8cm of mc] (kc);
        \vertex [below=0.8cm of md] (kd);
        \vertex [below=0.8cm of me] (ke);
        \vertex [below=0.8cm of ka] (f);
        \vertex [below=0.8cm of kb] (g);
        \vertex [below=0.8cm of kc] (h);
        \vertex [below=0.8cm of kd] (i);
        \vertex [below=0.8cm of ke] (j);
    
        \diagram* {
          (l) -- [fermion, bend left, edge label'=n] (a) -- [fermion] (e) -- [fermion, bend left, edge label'=p] (r),
          (l) -- [fermion, bend right, edge label'=n] (f) -- [fermion] (j) -- [fermion, bend right, edge label'=p] (r),
          (mb) -- [solid, bend left, edge label=\(\nu\)] (nc) -- [solid, bend left] (md),
          (f) -- [dashed, bend left, edge label'=\(\pi_-\)] (mb) -- [dashed, edge label'=\(\pi_0\)] (md) -- [dashed, bend right, edge label'=\(\pi_+\)] (e) 
        };

        \draw[fill=black] (l) circle(1mm);
        \draw[fill=black] (r) circle(1mm);
        \draw[fill=black] (e) circle(1mm);
        \draw[fill=black] (f) circle(1mm);
        \draw[fill=black] (mb) circle(1mm);
        \draw[fill=black] (md) circle(1mm);
      \end{feynman}
\end{tikzpicture}}
    &  $\begin{aligned} &\left( 2 F_\pi^2 Q^2 \right) \cdot \left( \frac{1}{Q^2} \right)^4 \cdot \int \frac{dQ^4}{(4 \pi)^2} = 2 \frac{F_\pi}{Q} \frac{1}{(4\pi)^2} \quad\quad \simeq 2 \left( \frac{Q}{\Lambda_\chi} \right)^2 \end{aligned}$ & $NNLO$\\

    (d) &
    \raisebox{-0.5\totalheight}{\begin{tikzpicture}
      \begin{feynman}
        \vertex (l);
        \vertex [right=1.0cm of l] (ma);
        \vertex [right=1.0cm of ma] (mb);
        \vertex [right=1.0cm of mb] (mc);
        \vertex [right=1.0cm of mc] (md);
        \vertex [right=1.0cm of md] (me);
        \vertex [right=1.0cm of me] (r);

        \vertex [above=0.8cm of ma] (na);
        \vertex [above=0.8cm of mb] (nb);
        \vertex [above=0.8cm of mc] (nc);
        \vertex [above=0.8cm of md] (nd);
        \vertex [above=0.8cm of me] (ne);
        \vertex [above=0.8cm of na] (a);
        \vertex [above=0.8cm of nb] (b);
        \vertex [above=0.8cm of nc] (c);
        \vertex [above=0.8cm of nd] (d);
        \vertex [above=0.8cm of ne] (e);
        \vertex [below=0.8cm of ma] (ka);
        \vertex [below=0.8cm of mb] (kb);
        \vertex [below=0.8cm of mc] (kc);
        \vertex [below=0.8cm of md] (kd);
        \vertex [below=0.8cm of me] (ke);
        \vertex [below=0.8cm of ka] (f);
        \vertex [below=0.8cm of kb] (g);
        \vertex [below=0.8cm of kc] (h);
        \vertex [below=0.8cm of kd] (i);
        \vertex [below=0.8cm of ke] (j);
    
        \diagram* {
          (l) -- [fermion, bend left, edge label'=n] (a) -- [fermion] (e) -- [fermion, bend left, edge label'=p] (r),
          (l) -- [fermion, bend right, edge label'=n] (f) -- [fermion] (j) -- [fermion, bend right, edge label'=p] (r),
          (kc) -- [solid, edge label'=\(\nu\)] (nc),
          (f) -- [dashed, bend left, edge label'=\(\pi_-\)] (kc),
          (nc) -- [dashed, bend right, edge label'=\(\pi_-\)] (kc) -- [dashed, bend right, edge label'=\(\pi_+\)] (e) 
        };

        \draw[fill=black] (l) circle(1mm);
        \draw[fill=black] (r) circle(1mm);
        \draw[fill=black] (e) circle(1mm);
        \draw[fill=black] (f) circle(1mm);
        \draw[fill=black] (kc) circle(1mm);
        \draw[fill=black] (nc) circle(1mm);
      \end{feynman}
\end{tikzpicture}}
    &  $\begin{aligned} &\left( 2 Q^2 \right) \cdot \left( \frac{1}{Q^2} \right)^3 \cdot \int \frac{dQ^4}{(4 \pi)^2} = 2 \frac{1}{(4\pi)^2} \quad\quad\quad\quad \simeq 2 \left( \frac{Q}{\Lambda_\chi} \right)^2 \end{aligned}$ & $NNLO$\\

   (e) &
    \raisebox{-0.5\totalheight}{\begin{tikzpicture}
      \begin{feynman}
        \vertex (l);
        \vertex [right=of l] (ma);
        \vertex [right=of ma] (mb);
        \vertex [right=of mb] (mc);
        \vertex [right=of mc] (r);

        \vertex [above=0.8cm of ma] (na);
        \vertex [above=0.8cm of mb] (nb);
        \vertex [above=0.8cm of mc] (nc);
        \vertex [above=0.8cm of na] (a);
        \vertex [above=0.8cm of nb] (b);
        \vertex [above=0.8cm of nc] (c);
        \vertex [below=0.8cm of ma] (ka);
        \vertex [below=0.8cm of mb] (kb);
        \vertex [below=0.8cm of mc] (kc);
        \vertex [below=0.8cm of ka] (d);
        \vertex [below=0.8cm of kb] (e);
        \vertex [below=0.8cm of kc] (f);
    
        \diagram* {
          (l) -- [fermion, bend left, edge label'=n] (b) -- [fermion, bend left, edge label'=p] (r),
          (l) -- [fermion, bend right, edge label'=n] (d) -- [fermion, edge label'=p] (e) -- [fermion, edge label'=p] (f) -- [fermion, bend right, edge label'=p] (r),
          (b) -- [solid, edge label'=\(\nu\)] (mb),
          (d) -- [dashed, bend left, edge label'=\(\pi_-\)] (mb) -- [dashed, bend left, edge label'=\(\pi_0\)] (f) 
        };

        \draw[fill=black] (l) circle(1mm);
        \draw[fill=black] (r) circle(1mm);
        \draw[fill=black] (b) circle(1mm);
        \draw[fill=black] (d) circle(1mm);
        \draw[fill=white] (f) circle(1mm);
        \draw[fill=black] (mb) circle(1mm);
      \end{feynman}
\end{tikzpicture}} 
    &  $\begin{aligned} &\left( \sqrt{2} Q \right) \cdot \left( \frac{1}{Q^2} \right)^2 \cdot \frac{1}{Q} \cdot \left( \frac{g_a}{2} \frac{Q}{F_\pi} \right) \cdot \left( \frac{g_A}{4} \frac{Q \omega_\pi}{F_\pi M_N} \right) \cdot \int \frac{dQ^4}{(4 \pi)^2} \\ &= \frac{1}{4\sqrt{2}} \frac{1}{(4\pi)^2} g_A^2 \frac{Q^3}{F_\pi^2 M_N} \simeq \frac{g_A^2}{4\sqrt{2}} \left( \frac{Q}{\Lambda_\chi} \right)^3 \quad\quad \simeq 0.3 \left( \frac{Q}{\Lambda_\chi} \right)^3 \end{aligned}$ & $N^3LO$\\

    (f) &
    \raisebox{-0.5\totalheight}{\begin{tikzpicture}
      \begin{feynman}
        \vertex (l);
        \vertex [right=of l] (ma);
        \vertex [right=of ma] (mb);
        \vertex [right=of mb] (mc);
        \vertex [right=of mc] (r);

        \vertex [above=0.8cm of ma] (na);
        \vertex [above=0.8cm of mb] (nb);
        \vertex [above=0.8cm of mc] (nc);
        \vertex [above=0.8cm of na] (a);
        \vertex [above=0.8cm of nb] (b);
        \vertex [above=0.8cm of nc] (c);
        \vertex [below=0.8cm of ma] (ka);
        \vertex [below=0.8cm of mb] (kb);
        \vertex [below=0.8cm of mc] (kc);
        \vertex [below=0.8cm of ka] (d);
        \vertex [below=0.8cm of kb] (e);
        \vertex [below=0.8cm of kc] (f);
    
        \diagram* {
          (l) -- [fermion, bend left, edge label'=n] (b) -- [fermion, bend left, edge label'=p] (r),
          (l) -- [fermion, bend right, edge label'=n] (d) -- [fermion, edge label'=p] (f) -- [fermion, bend right, edge label'=p] (r),
          (b) -- [solid, edge label'=\(\nu\)] (mb),
          (d) -- [dashed, bend left, edge label'=\(\pi_-\)] (mb),
          (b) -- [dashed, bend left, edge label'=\(\pi_0\)] (f) 
        };

        \draw[fill=black] (l) circle(1mm);
        \draw[fill=black] (r) circle(1mm);
        \draw[fill=black] (b) circle(1mm);
        \draw[fill=black] (d) circle(1mm);
        \draw[fill=white] (f) circle(1mm);
        \draw[fill=black] (mb) circle(1mm);
      \end{feynman}
\end{tikzpicture}} 
    &  $\begin{aligned} &\left( \frac{\sqrt{2}}{2} Q \right) \cdot \left( \frac{1}{Q^2} \right)^2 \cdot \frac{1}{Q} \cdot \left( \frac{g_a}{2} \frac{Q}{F_\pi} \right) \cdot \left( \frac{g_A}{4} \frac{Q \omega_\pi}{F_\pi M_N} \right) \cdot \int \frac{dQ^4}{(4 \pi)^2} \\ &= \frac{1}{8\sqrt{2}} \frac{1}{(4\pi)^2} g_A^2 \frac{Q^3}{F_\pi^2 M_N} \simeq \frac{g_A^2}{8\sqrt{2}} \left( \frac{Q}{\Lambda_\chi} \right)^3 \quad\quad \simeq 0.15 \left( \frac{Q}{\Lambda_\chi} \right)^3 \end{aligned}$ & $N^3LO$\\

    (g) &
    \raisebox{-0.5\totalheight}{\begin{tikzpicture}
      \begin{feynman}
        \vertex (l);
        \vertex [right=of l] (ma);
        \vertex [right=of ma] (mb);
        \vertex [right=of mb] (mc);
        \vertex [right=of mc] (r);

        \vertex [above=0.8cm of ma] (na);
        \vertex [above=0.8cm of mb] (nb);
        \vertex [above=0.8cm of mc] (nc);
        \vertex [above=0.8cm of na] (a);
        \vertex [above=0.8cm of nb] (b);
        \vertex [above=0.8cm of nc] (c);
        \vertex [below=0.8cm of ma] (ka);
        \vertex [below=0.8cm of mb] (kb);
        \vertex [below=0.8cm of mc] (kc);
        \vertex [below=0.8cm of ka] (d);
        \vertex [below=0.8cm of kb] (e);
        \vertex [below=0.8cm of kc] (f);
    
        \diagram* {
          (l) -- [fermion, bend left, edge label'=n] (a) -- [fermion] (c) -- [fermion, bend left, edge label'=p] (r),
          (l) -- [fermion, bend right, edge label'=n] (d) -- [fermion, edge label'=n] (f) -- [fermion, bend right, edge label'=p] (r),
          (c) -- [solid, edge label'=\(\nu\)] (f),
          (d) -- [dashed, edge label'=\(\pi_0\)] (c) 
        };

        \draw[fill=black] (l) circle(1mm);
        \draw[fill=black] (r) circle(1mm);
        \draw[fill=black] (c) circle(1mm);
        \draw[fill=black] (f) circle(1mm);
        \draw[fill=white] (d) circle(1mm);
        
      \end{feynman}
\end{tikzpicture}} 
    &  $\begin{aligned} &\left( \frac{g_A}{F_\pi} \right) \cdot \frac{1}{Q^2} \cdot \frac{1}{Q} \cdot \left( \frac{g_A}{4} \frac{Q \omega_\pi}{F_\pi M_N} \right) \cdot \int \frac{dQ^4}{(4 \pi)^2} \\ &= \frac{g_A^2}{4} \left( \frac{Q}{\Lambda_\chi} \right)^3 \quad\quad\quad\quad\quad\quad\quad\quad\quad\quad \simeq 0.4 \left( \frac{Q}{\Lambda_\chi} \right)^3 \end{aligned}$ & $N^3LO$\\

    (h) &
     \raisebox{-0.5\totalheight}{\begin{tikzpicture}
      \begin{feynman}
        \vertex (l);
        \vertex [right=of l] (ma);
        \vertex [right=of ma] (mb);
        \vertex [right=of mb] (mc);
        \vertex [right=of mc] (r);

        \vertex [above=0.8cm of ma] (na);
        \vertex [above=0.8cm of mb] (nb);
        \vertex [above=0.8cm of mc] (nc);
        \vertex [above=0.8cm of na] (a);
        \vertex [above=0.8cm of nb] (b);
        \vertex [above=0.8cm of nc] (c);
        \vertex [below=0.8cm of ma] (ka);
        \vertex [below=0.8cm of mb] (kb);
        \vertex [below=0.8cm of mc] (kc);
        \vertex [below=0.8cm of ka] (d);
        \vertex [below=0.8cm of kb] (e);
        \vertex [below=0.8cm of kc] (f);
    
        \diagram* {
          (l) -- [fermion, bend left, edge label'=n] (a) -- [fermion] (b) -- [fermion, edge label'=p] (c) -- [fermion, bend left, edge label'=p] (r),
          (l) -- [fermion, bend right, edge label'=n] (d) -- [fermion, edge label'=n] (e) -- [fermion, edge label'=p] (f) -- [fermion, bend right, edge label'=p] (r),
          (b) -- [solid, edge label'=\(\nu\)] (e),
          (d) -- [dashed, edge label'=\(\pi_0\)] (c) 
        };

        \draw[fill=black] (l) circle(1mm);
        \draw[fill=black] (r) circle(1mm);
        \draw[fill=black] (b) circle(1mm);
        \draw[fill=black] (e) circle(1mm);
        \draw[fill=white] (d) circle(1mm);
        \draw[fill=white] (c) circle(1mm);
      \end{feynman}
\end{tikzpicture}} 
    &  $\begin{aligned} &\left( 1 + 2 g_A^2 \right) \cdot \left( \frac{g_A}{4} \frac{Q \omega_\pi}{F_\pi M_N} \right)^2 \cdot \frac{1}{Q^2} \cdot \left( \frac{1}{Q} \right)^2 \cdot \int \frac{dQ^4}{(4 \pi)^2} \\ &= \left( 1 + 2 g_A^2 \right) \frac{g_A^2}{16} \frac{1}{(4\pi)^2} \simeq \left( 1 + 2 g_A^2 \right) \frac{g_A^2}{16} \left( \frac{Q}{\Lambda_\chi} \right)^4 \simeq 0.4 \left( \frac{Q}{\Lambda_\chi} \right)^4 \end{aligned}$ & $N^4LO$\\
    \hline
\end{tblr}
}
}
    \caption{Dimensional analysis of leading $NN\pi$ neutrino-exchange diagrams.}
    \label{fig:pinndimanal}
\end{figure}

In Figure \ref{fig:pinndimanal}, a complete classification is presented of $NN$ $0\nu\beta\beta$ diagrams whose neutrino exchange occurs in an $NN\pi$ hadronic intermediate state. From the six neutrino potentials exhibited and two competing pion production/absorption channels (charged and neutral), one might naively have expected as many as $24$ distinct topologies. However, isospin symmetry places severe restrictions on these combinatorics, leaving only the seven families of diagrams illustrated in Figure \ref{fig:pinndimanal}. Note that we show only one representative of each family, whose pion production and absorption occur via the impulse diagram defined in Section 4.2. As described there, each $p$-wave charged-pion production / absorption can occur either via an impulse or a seagull (pion-rescattering) diagram, both at leading order. In our power-counting estimate, we will neglect the seagull diagrams, as simple power-counting shows that the presence of an additional loop integral suppresses such diagrams by two chiral orders. Each $s$-wave neutral-pion production likewise can occur either via an impulse diagram or a $\Delta$-excitation diagram, both at sub-leading order. We elect not to include these $\Delta$-excitation diagrams in our estimate for the reason of consistency with our truncation of the inelastic hadronic states to $NN\pi$. Including $\Delta$ degrees of freedom in one stage of our calculation would necessitate their inclusion throughout, in particular in the intermediate hadronic state present for neutrino exchange, which lies beyond the scope of this work. Furthermore, each charged-pion production / absorption diagram also includes a subleading $s$-wave component; we do not explicitly consider these, but they serve as a reminder that generically, diagrams which enter the chiral power-counting at some particular order may also experience contributions at any or all higher orders.

This classification reveals four broader types of $NN\pi$ diagrams according to the isospin properties of their virtual pion; all labels in the following few paragraphs refer to Figure~\ref{fig:pinndimanal}. First, lone ``eye" diagram (a) simply inserts a pion-exchange at the same vertices of the neutrino-exchange, as captured by our $NN\pi-NN\pi$ neutrino potential. The pion production vertices contribute near unity factors of $\frac{Q}{F_\pi}$, but the integration of the pion loop suppresses diagram (a) to $NNLO$ in the chiral counting, two orders below the $NN$ contributions. The eye diagram topology was fully evaluated in Appendix C of~\cite{Cirigliano2021long}, and forms the basis of the $NN\pi$ uncertainty estimate quoted in that work. Second, three ``surfer" diagrams (b-d) are distinguished by both charged-pion production and absorption, and by neutrino exchange via a purely pionic potential; these diagrams are also suppressed to $NNLO$. However, all the other families of diagrams will experience additional suppressions. Next, three ``mountain" diagrams (e-g) involve a lone neutral-pion interaction (some with an additional charged-pion interaction), with the neutrino exchanged via a mixed pion-nucleon potential; all these diagrams are suppressed to $N^3LO$ due to the $s$-wave interaction of the neutral pion. Note that each of these diagrams has a further multiplicity of $2$ since the neutral pion could interact on either nucleon line, not impacting the power-counting estimate but giving a distinct momentum-routing. Finally, lone ``bridge" diagram (h) is present, whose virtual pion experiences one order of suppression at both the production and absorption ends, while neutrino exchange occurs via a purely nucleonic potential, just as in the $NN$ contribution of~\cite{Cirigliano2021long}. This diagram is thus suppressed to $N^4LO$, along with a family of variations depending on the placement of the pion production vertices on the nucleon lines. Overall, one expects from the power-counting that the surfer diagrams will dominate with the mountain diagrams subleading.

Before proceeding to the integrals, we can now give a power-counting-driven estimate of the error bar which $0\nu\beta\beta$ diagrams with an $NN\pi$ intermediate state place on the elastic contact term size. Summing all diagrams constructively alongside the aforementioned multiplicities for each family, approximating $\frac{Q}{\Lambda\chi} \sim \frac{1}{7.5}$, and comparing with the estimate $g_{NN}(\mu_\chi = m_\pi) \simeq 1.32$ from~\cite{Cirigliano2021long}, we find:
\begin{align}
    \Delta g_{NN\pi} \sim &\left( 4.2 + 2 \cdot 2 + 2 \cdot 2 + 2 \cdot 2 \right) \left( \frac{Q}{\Lambda\chi} \right)^2 \nonumber\\
    &+ \left( 8 \cdot 0.3 + 8 \cdot 0.15 + 8 \cdot 0.4 \right) \left( \frac{Q}{\Lambda\chi} \right)^3 \nonumber\\
    &+ \left(4 \cdot 0.4 \right) \left( \frac{Q}{\Lambda\chi} \right)^4 \nonumber\\
    &= 0.288_{NNLO} + 0.016_{N^3LO} + 0.0005_{N^4L0}, \nonumber\\
    \vspace{5mm}
    \frac{\Delta g_{NN\pi}}{g_{NN}} &= 21.8\%_{NNLO} + 1.2\%_{N^3LO} + 0.04\%_{N^4LO}.
\end{align}
This estimate is in agreement with the error bars allocated by~\cite{Cirigliano2021long}, suggesting that we can expect as much as a $22\%$ modification to the size of contact counterterm $g_\nu^{NN}$ through the inclusion of our $NNLO$ corrections (the eye and surfer diagrams). Even further, our estimate shows that if all four described NNLO diagrams (a-d) are taken into account in a complete regularisation analysis, the remaining $NN\pi$ contributions only induce uncertainties on the order of $1\%$.

\subsection{\texorpdfstring{Dimensional regularisation of the NNLO $NN\pi$ divergences}{Dimensional regularisation of the NNLO NNpi divergences}}

In the preceding subsection, we identified a collection of four neutrino-exchange diagrams, the ``eye" diagram (a) and ``surfer" diagrams (b-d), as encoding the leading $NN\pi$ contribution to $NN$ $0\nu\beta\beta$-decay. These diagrams and their power-counting are shown in the top four rows of Figure \ref{fig:pinndimanal}, and were expected to enter at $NNLO$ in the chiral power-counting, with as much as a $8\%$ modification to the elastic contact coefficient anticipated. Here we will test that hypothesis by evaluating the divergent part of all four diagrams in the dimensional regularisation framework, and follow the matching strategy of~\cite{Cirigliano2021long} in order to compute the corrected contact term.

We first consider the surfer diagrams (b-d), which require novel calculation. All three give a generalised amplitude with the structure:
\begin{equation}
    \delta A_{S,i} = -i K \int \frac{d^4l}{(2\pi)^4} \frac{i}{l^2 - m_\pi^2 + i\epsilon} \frac{i}{l^2 - m_\pi^2 + i\epsilon} I_C(l^2,p^2,p'^2) I_{S,i}(l^2), \quad i = 1,2,3,
\end{equation}
where $I_C(l^2,p^2,p'^2)$ is the two-nucleon bubble integral, shown in~\cite{Cirigliano2021long} to have asymptotic ($p \sim p'$) real component $ \frac{1}{8 l} \theta(|l| - 2|p|)$, the partial amplitudes $I_{S,i}(l^2)$ encode the kinematics of the distinct pion-neutrino subdiagrams, and $K$ encodes constant factors. We will see momentarily that the leading divergences of $I_{S,i}(l^2)$ in  which we are interested will resolve to the form $A l^2 + B$, and so very fortunately we can factorise the internal one- and external two-loop integrals.

But first, one further factorisation can be performed, as all of the $I_{S,i}$ enjoy the same spin-structure at the $NN\pi$ vertices. Following the normalisation of~\cite{Bernard1995,Hanhart2004}, this $p$-wave charged-pion vertex is determined by the following term of the $LO$ chiral Lagrangian:
\begin{equation}
    \mathcal{L}_{\pi N} \ni \frac{g_A}{2 F_\pi} \overline{N} \tau_a \sigma^j \nabla^j \pi_a N.
\end{equation}
The isospin algebra (index $a$) is implicitly accounted for in our enforcement of charge conservation at each $NN\pi$ vertex; however the spin algebra requires more thought. Each $I_{S,i}$ will include an operator $\sigma_1 \cdot l \sigma_2 \cdot l$, where $\sigma_i$ indicates the vector of Pauli spin operators on nucleon $i$. This operator may be decomposed into a spherical component $\frac{1}{3} l \cdot l \sigma_1 \cdot \sigma_2$, which we will need, and a tensor component $- \frac{1}{3} S_12$, which vanishes since we are only considering $^1S_0$ $NN$ partial waves on the external legs. In this partial wave, $\sigma_1 \cdot \sigma_2 = -3$, so the total spin algebra factor is simply $- l^2$.

%Complex-valued $l_0$ has four poles: two at $\sqrt{\vec{l}^2 - m_\pi^2} - i\epsilon$, and two at $- \sqrt{\vec{l}^2 - m_\pi^2} + i\epsilon$. Thus we can perform a Wick rotation of the form $i l_0 \mapsto l_0$ and $f(i l_0) \mapsto -i f(l_0)$ without picking up any pole contributions. Now taking $\epsilon \rightarrow 0$, the Euclidean integral is performed directly:
%
%\begin{align}
%     \frac{\delta T_{S,i}}{I_{S,i}} =& -\frac{i}{8} \int \frac{d^4l_E}{(2\pi)^4} \frac{1}{\left( l_E^2 + m_\pi^2 \right)^2} \frac{1}{|l_E|} \theta(|l_E| - 2|p|) \nonumber\\
%     =& -\frac{i}{8} \frac{1}{8 \pi^2} \int_{2|p|}^\infty dl \frac{l^2}{\left( l^2 + m_\pi^2 \right)^2} \nonumber\\
%     =& -\frac{i}{8} \frac{1}{8 \pi^2} \left[ \frac{\pi}{4 m_\pi} + \frac{p^2}{m_\pi^2 + 4 p^2} - \frac{\tan^{-1} \left( \frac{2p}{m_\pi} \right)}{2 m_\pi} \right] \nonumber\\
%     &\underset{p \rightarrow 0}{=}& \frac{-i}{256 \pi m_\pi}.
%\end{align}
%
%where in the intermediate steps we have defined $l = |l_E|$ and $p = |\vec{p}|$. It is not surprising that this result is finite, because we expect these diagrams to contain only logarithmic divergences, and from dimensional analysis the pion-neutrino subdiagrams must each be divergent.

Now we perform the external integration separately for both $I_{S,i} = 1$ and $I_{S,i} = l^2$, a complete basis for our solution. Because we are working within a static nucleon approximation, pion-exchange momentum $l$ needs only be treated within $D=3$, averting the usual Wick rotation step. For $I_{S,i} = 1$, the resultant Euclidean integral is logarithmically solvable and divergent, and following~\cite{Cirigliano2021long} we regulate it with a cutoff scale $\mu_\pi \sim \Lambda_\chi$:
\begin{align}
     \delta A_{S,i,1} =& -\frac{1}{8} \left(\frac{g_A}{2 F_\pi}\right)^2 \int \frac{d^3l}{(2\pi)^3} \frac{l^2}{\left( l^2 + m_\pi^2 \right)^2} \frac{1}{|l|} \theta(|l| - 2|p|) \nonumber\\
     =& -\frac{1}{8} \frac{1}{8 \pi^2} \left(\frac{g_A}{2 F_\pi}\right)^2 \int_{2|p|}^{\mu_\pi} dl \frac{l^3}{\left( l^2 + m_\pi^2 \right)^2} \nonumber\\
     =& -\frac{1}{8} \frac{1}{2 \pi^2} \frac{1}{2} \left(\frac{g_A}{2 F_\pi}\right)^2 \left[ -\frac{m_\pi^2}{m_\pi^2 + 4 p^2} + \frac{m_\pi^2}{m_\pi^2 + {\mu_\pi}^2} + \log\left( \frac{m_\pi^2 + {\mu_\pi}^2}{m_\pi^2 + 4 p^2} \right) \right] \nonumber\\
     \underset{p \rightarrow 0}{=}& \frac{- g_A^2}{128 \pi^2 F_\pi^2}\left[ -1 + \frac{m_\pi^2}{m_\pi^2 + {\mu_\pi}^2} + \log\left( 1 + \frac{{\mu_\pi}^2}{m_\pi^2} \right) \right].
\end{align}
For $I_{S,i} = l^2$, the Euclidean integral is solvable but naively appears quadratically divergent. The situation is analogous to the $NN$ one-pion-exchange potential as treated in~\cite{Kaplan1996}, whose momentum-space integral is proportional to:
\begin{equation}
    \frac{g_A^2}{4 F_\pi^2} \int \frac{d^3l}{(2\pi)^3} \frac{l^2}{l^2 + m_\pi^2} \frac{1}{|l|} = \frac{g_A^2}{4 F_\pi^2} \int \frac{d^3l}{(2\pi)^3} \left(1 - \frac{m_\pi^2}{l^2 + m_\pi^2} \right) \frac{1}{|l|}.
\end{equation}
In the above, the latter term gives a convergent contribution to the one-pion exchange bubble diagram (seen in Figure \ref{fig:GEtwodiagrams}). The former term is linearly divergent, but has the same form as the four-nucleon contact interaction $C$; in spatial coordinates, they are both $\delta^{(3)}(r)$ potentials. Thus by setting the contact interaction $C \mapsto \tilde{C} = C + \frac{g_A^2}{4 F_\pi^2}$, the linear divergence is absorbed into an existing, observable constant encoding short-range physics.

We assume, perhaps liberally, that our leading quadratic divergence can be absorbed into short-range parameters in a similar fashion; we isolate and subtract the linear term from the integrand, assuming that it can absorbed into the coefficient of the $\chi EFT$ $NNNN\pi\pi$ vertex, which will only contribute to our process at a much lower order. This leaves only a logarithmic divergence, regularised as follows:
\newpage
\begin{align}
     \delta A_{S,i,l^2} =& -\frac{1}{8} \left(\frac{g_A}{2 F_\pi}\right)^2 \int \frac{d^3l}{(2\pi)^3} \frac{l^4}{\left( l^2 + m_\pi^2 \right)^2} \frac{1}{|l|} \theta(|l| - 2|p|) \nonumber\\
     =& -\frac{1}{8} \frac{1}{8 \pi^2} \left(\frac{g_A}{2 F_\pi}\right)^2 \int_{2|p|}^{\mu_\pi} dl \frac{l^5}{\left( l^2 + m_\pi^2 \right)^2} \nonumber\\
     \underset{subtraction}{\rightarrow}& -\frac{i}{8} \frac{1}{8 \pi^2} \left(\frac{g_A}{2 F_\pi}\right)^2 \int_{2|p|}^{\mu_\pi} dl \frac{m_\pi^4}{l \left( l^2 + m_\pi^2 \right)^2} - \frac{2 m_\pi^2}{l \left( l^2 + m_\pi^2 \right)} \nonumber\\
     =& -\frac{1}{8} \frac{1}{2 \pi^2} \frac{1}{2} \left(\frac{g_A}{2 F_\pi}\right)^2 m_\pi^2 \left[ - \frac{m_\pi^2 }{m_\pi^2 + {\mu_\pi}^2} + \frac{m_\pi^2}{m_\pi^2 + 4 p^2} + 2\log \left( \frac{m_\pi^2 + 4 p^2}{m_\pi^2 + {\mu_\pi}^2} \right) \right] \nonumber\\
     \underset{p \rightarrow 0}{=}& \frac{-g_A^2 m_\pi^2}{128 \pi^2 F_\pi^2}\left[ 1 - \frac{m_\pi^2}{m_\pi^2 + {\mu_\pi}^2} - 2\log \left( 1 + \frac{{\mu_\pi}^2}{m_\pi^2} \right) \right].
     \label{l2result}
\end{align}
Now we apply dimensional regularisation to each of the $I_{S,i}$, making use of standard integration results similar to those shown by~\cite{Scherer2011} in computing the one-loop chiral correction to the nucleon mass. $I_{S,1}$, corresponding to diagram (b) in Figure \ref{fig:pinndimanal}, is regularised as follows:
\begin{align}
    I_{S,1} &= 2 \int \frac{d^4k}{2\pi^4} \frac{1}{(k-l)^2 - m_\pi^2 + i\epsilon} \frac{k^2}{k^2 - m_\nu^2 + i\epsilon} \nonumber\\
    &=  2 \int \frac{d^4k}{2\pi^4} \frac{1}{(k-l)^2 - m_\pi^2 + i\epsilon} +  2 \frac{1}{(k-l)^2 - m_\pi^2 + i\epsilon} \cdot \frac{m_\nu^2}{k^2 - m_\nu^2 + i\epsilon} \nonumber\\
    &= \frac{ 2 m_\pi^2}{16\pi^2} \left( \frac{2}{\varepsilon} + \gamma_E - 1 - \log(4\pi) - \log\left( \frac{{\mu_\nu}^2}{m_\pi^2}\right)  \right) + \mathcal{O}\left(m_\nu^2\right),
\end{align}
where ${\mu_\nu}$ is the scale of dimensional regularisation, and where we employ a finite $m_\nu$ as a regulator but find that such contributions are suppressed by $\frac{m_\nu^2}{m_\pi^2}$. Thus, applying the $\overline{MS}$ scheme such that $\frac{2}{\varepsilon} + \gamma_E - \log(4\pi) - 1$ is absorbed into the coupling constants of the Lagrangian (i.e. $g_A \mapsto g_A^{(r)}$, $F_\pi \mapsto F_\pi^{(r)}$) overall we find:
\begin{equation}
    \delta A_{S,1} = \frac{M_N^2}{(4\pi)^2} \frac{g_A^2 m_\pi^2}{64 \pi^2 F_\pi^2}\left[ -1 + \frac{m_\pi^2}{m_\pi^2 + \mu_\nu^2} + \log\left( 1 + \frac{\mu_\nu^2}{m_\pi^2} \right) \right] \left[ \log\left( \frac{\mu_\pi^2}{m_\pi^2}\right) \right].
\end{equation}
Examining diagram (c) in Figure \ref{fig:pinndimanal}, we find that $I_{S,2} = I_{S,1}$ up to corrections of $\mathcal{O}\left( \frac{m_\nu^2}{m_\pi^2} \right)$, since
\begin{equation}
    \int \frac{d^4k}{2\pi^4} \frac{1}{k^2 - m_\pi^2 + i\epsilon} \frac{k^2}{k^2 - m_\nu^2 + i\epsilon} = \int \frac{d^4k}{2\pi^4} \frac{1}{k^2 - m_\pi^2 + i\epsilon} + \frac{1}{k^2 - m_\pi^2 + i\epsilon} \frac{m_\nu^2}{k^2 - m_\nu^2 + i\epsilon}
\end{equation}
is equivalent to the integrand of $I_{S,1}$ after a shift in $k$ (sans corrections of $\mathcal{O}\left( \frac{m_\nu^2}{m_\pi^2} \right)$). This equivalence is facilitated by chiral symmetry, via the equivalence of the $\pi\pi-\pi\pi$ and $\pi\pi\pi-\pi$ potentials seen in Table \ref{tab:nupotterms}.

Now considering diagram (d) in Figure \ref{fig:pinndimanal}, we will see that the four-pion vertex $\sim \frac{-2}{3 F_\pi^2} \left( k^2 + l^2 - 2m_\pi^2 \right)$ will facilitate a demarcation into terms which are proportional either to $l^2$ or to a constant. Regularising,
\begin{align}
    I_{S,3} =& \frac{-4 F_\pi^2}{3 F_\pi^2} \int \frac{d^4k}{2\pi^4} \frac{k^2 + l^2 - 2 m_\pi^2}{ \left( k^2 - m_\pi^2 + i\epsilon \right)^2} \frac{k^2}{k^2 - m_\nu^2 + i\epsilon} +  \mathcal{O}\left(m_\nu^2\right) \nonumber\\
    =& - \frac{4}{3} \int \frac{d^4k}{2\pi^4} \frac{k^2}{\left(k^2 - m_\pi^2 + i\epsilon\right)^2} \frac{k^2}{ k^2 - m_\nu^2 + i\epsilon  } \nonumber\\
    & \quad\quad - \frac{4}{3} (l^2 - 2 m_\pi^2) \int \frac{d^4k}{2\pi^4} \frac{1}{ \left( k^2 - m_\pi^2 + i\epsilon \right)^2} \frac{k^2}{k^2 - m_\nu^2 + i\epsilon} +  \mathcal{O}\left(m_\nu^2\right) \nonumber\\
    =& - \frac{1}{12\pi^2} \left( 2 m_\pi^2 \log \left( \frac{\mu^2}{m_\pi^2} \right) + \left( l^2 - 2 m_\pi^2 \right) \log \left( \frac{{\mu_\nu}^2}{m_\pi^2} \right) \right) +  \mathcal{O}\left(m_\nu^2\right) \nonumber\\
    =& - \frac{l^2}{12 \pi^2}  \log \left( \frac{{\mu_\nu}^2}{m_\pi^2} \right) +  \mathcal{O}\left(m_\nu^2\right),
\end{align}
where in the final steps we have enforced the $\overline{MS}$ scheme and cancelled all but the divergent term proportional to $l^2$. Then combining with the result of the $l^2$ external integration Eq.~\eqref{l2result}, the regularised amplitude is:
\begin{equation}
    \delta A_{S,3} = \frac{M_N^2}{(4\pi)^2} \frac{g_A^2 m_\pi^2}{96 \pi^2 F_\pi^2}\left[ 1 - \frac{m_\pi^2}{m_\pi^2 + {\mu_\pi}^2} - 2\log \left( 1 + \frac{{\mu_\pi}^2}{m_\pi^2} \right) \right] \left[ \log\left( \frac{{\mu_\nu}^2}{m_\pi^2}\right) \right]
\end{equation}
\begin{figure}[t!]
    \centering
    \includegraphics[width=.49\textwidth]{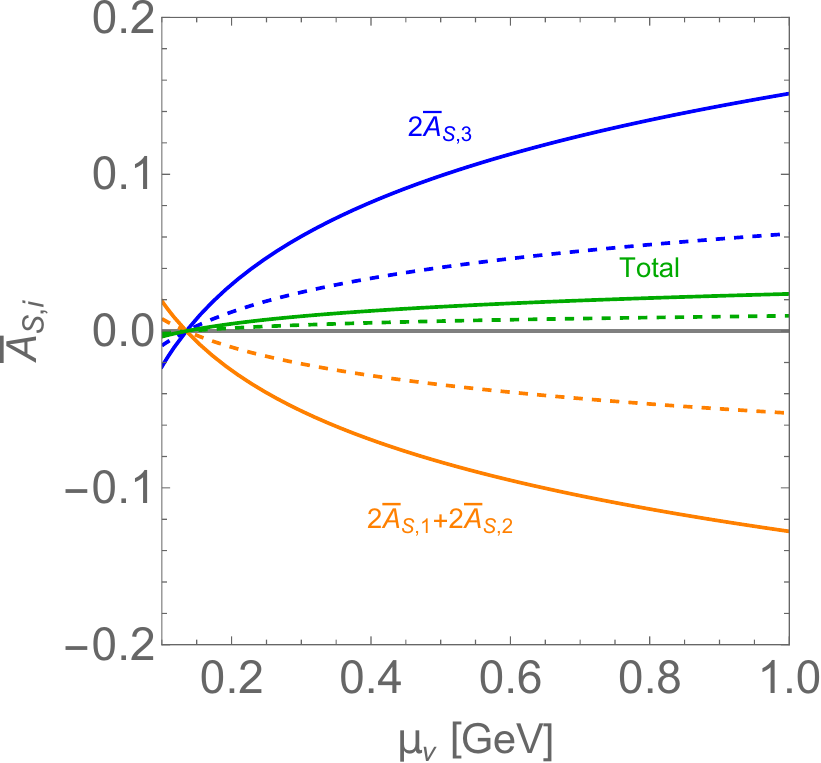}
    \hspace{2pt}
    \includegraphics[width=.49\textwidth]{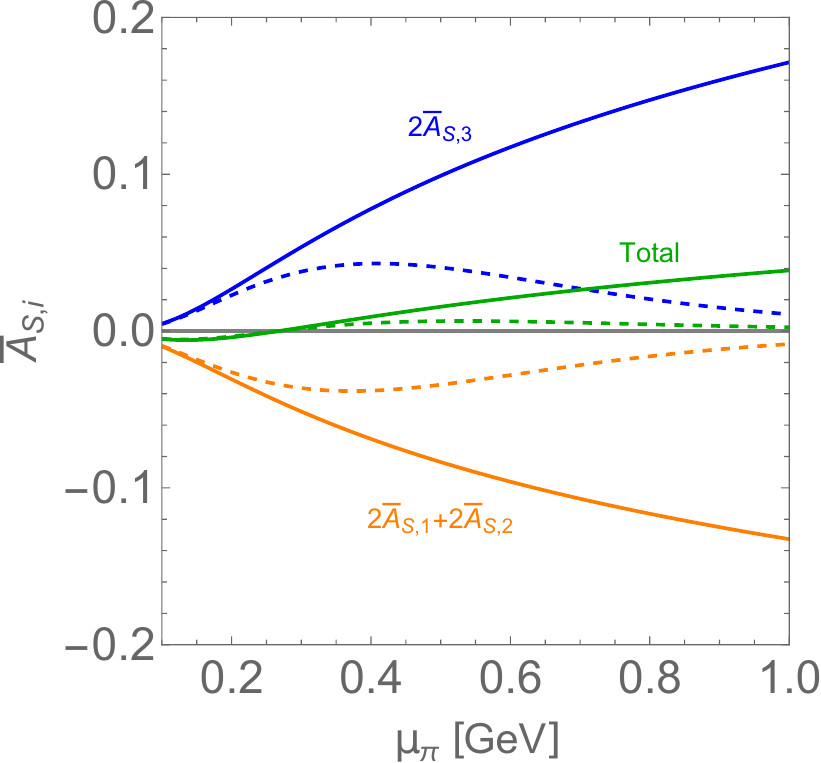}
    \caption{Dimensionless amplitudes for surfer diagrams $1$ and $2$ (orange), $3$ (blue), and their sum (green), as functions of regularisation scales $\mu_\nu$ (left) and $\mu_\pi$ (right), with the respective other scale set to $\mu_\pi = 0.5$~GeV. In both plots, the solid lines show the direct $\chi EFT$ result, while the dashed lines show the enhanced model defined in the text, where $g_A$ is replaced with its dipole form factor.}
    \label{fig:surferamps_both}
\end{figure}
All of the calculated $A_{S,i}$ are displayed as functions of $\mu_\nu$ and $\mu_\pi$ in Figure \ref{fig:surferamps_both}, where the remarkable cancellation between the logarithmic divergences of $2 A_{S,1} + 2 A_{S,2}$ and $2 A_{S,3}$ is apparent. Shown alongside are enhanced models, where $g_A$ is replaced with the dipole form factor defined in Eq.~\eqref{eq:dipolega} with scale $\Lambda_A = 1.0$~GeV, with the intention of making the amplitudes more valid for the intermediate momenta between $m_\pi$ and $\Lambda_\chi$. These enhanced models will prove essential for a matching analysis analogous to that performed by~\cite{Cirigliano2021long} in the $NN$ case. In addition to these contributions, we quote the amplitude for the eye diagram (top row in Figure \ref{fig:nndimanal}) as computed in Appendix C of~\cite{Cirigliano2021long}:
\begin{align}
    \delta A_E &= \frac{M_N^2}{(4\pi)^2} \int_0^{\mu_\nu} dk \delta a_E(k) \nonumber\\
    &\text{where} \quad \delta a_E(k) = \frac{1+2 g_A^2}{2 \left( 4\pi F_\pi \right)^2} \left[ k \left( 1 + \log \frac{{\mu_\pi} m_\pi}{\left( k + \sqrt{k^2 + m_\pi^2} \right)^2} \right) + m_\pi F\left( \frac{k}{m_\pi} \right) \right], \nonumber\\
    &\quad\quad F(x) = \sqrt{1-x^2} \left[ \tan^{-1} \left( \frac{x^2}{\sqrt{1-x^4}}\right) - \tan^{-1} \left( \frac{x}{\sqrt{1-x^2}}\right) \right]
\end{align}
and where we have made the substitution $1 + 3 g_A^2 \mapsto 1 + 2 g_A^2$ in accordance with the induced pseudoscalar component of the $NN$ neutrino potential in $\chi EFT$. Note that the final integration in neutrino momentum $k$ for this diagram has not been analytically solved; in order to compare apples-to-apples with our analytically-regulated amplitudes for the surfer diagrams, we define partial amplitudes:
\begin{equation}
    a_{s,i}({\mu_\nu}) \equiv \frac{d}{d{\mu_\nu}} A_{s,i}({\mu_\nu}).
\end{equation}
Strictly speaking, these distributions are not physically meaningful; they encode the density of an amplitude which one obtains while increasing the dimensional regularisation scale, rather than amplitude densities over loop momentum $k$ defined before regularisation. However we will only use them for visualisation purposes and to verify our power-counting, instead applying $A_{s,i}({\mu_\nu})$ directly when we compute the contact counterterm coefficient.

Recall that three adjustable scales appear in this procedure: the neutrino loop regularisation scale ${\mu_\nu}$ appearing above, which is the running renormalisation scale of the counterterm coefficient $g^\nu_{NN}$; the pion loop regularisation scale ${\mu_\pi}$, which we treat with a simple momentum cutoff as does Appendix C of~\cite{Cirigliano2021long}; and the matching scale $\Lambda$, which was seen in~\cite{Cirigliano2021long} to introduce negligible error into the counterterm estimate for $1 < \Lambda < 4$~GeV. We will follow~\cite{Cirigliano2021long} in choosing the renormalisation point ${\mu_\nu} = m_\pi$ for quoted values of $g^\nu_{NN}$, but emphasise that the entire function $g^\nu_{NN}({\mu_\nu})$ is the physical quantity which this analysis aims to predict.

\begin{figure}[t!]
    \centering
    \includegraphics[width=1\textwidth]{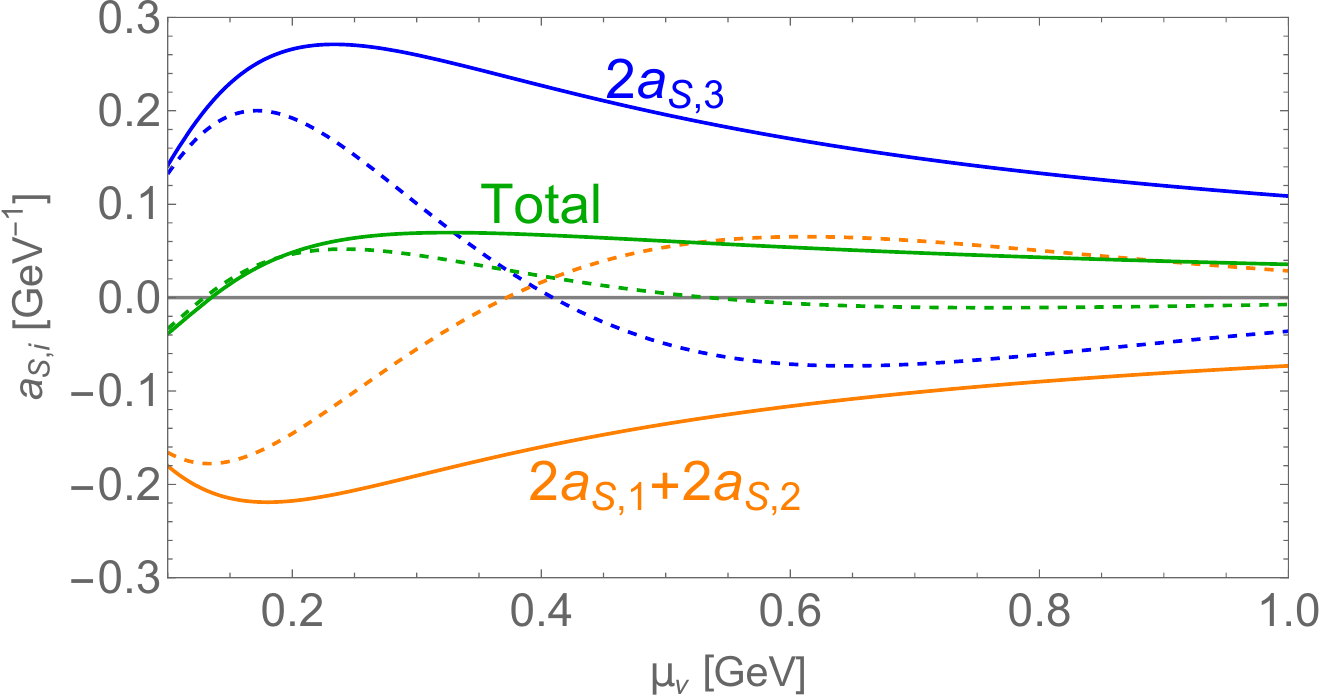}
    \caption{Partial amplitudes (defined in text) for surfer diagrams $1$ and $2$ (orange), $3$ (blue), and their sum (green), as functions of regularisation scale $\mu_\nu$ with $\mu_\pi = 0.5$~GeV. The solid lines show the direct $\chi EFT$ result, while the dashed lines include a dipole form factor for $g_A$.}
    \label{fig:surfersamps_partial}
\end{figure}
\begin{figure}[t!]
    \centering
    \includegraphics[width=1\textwidth]{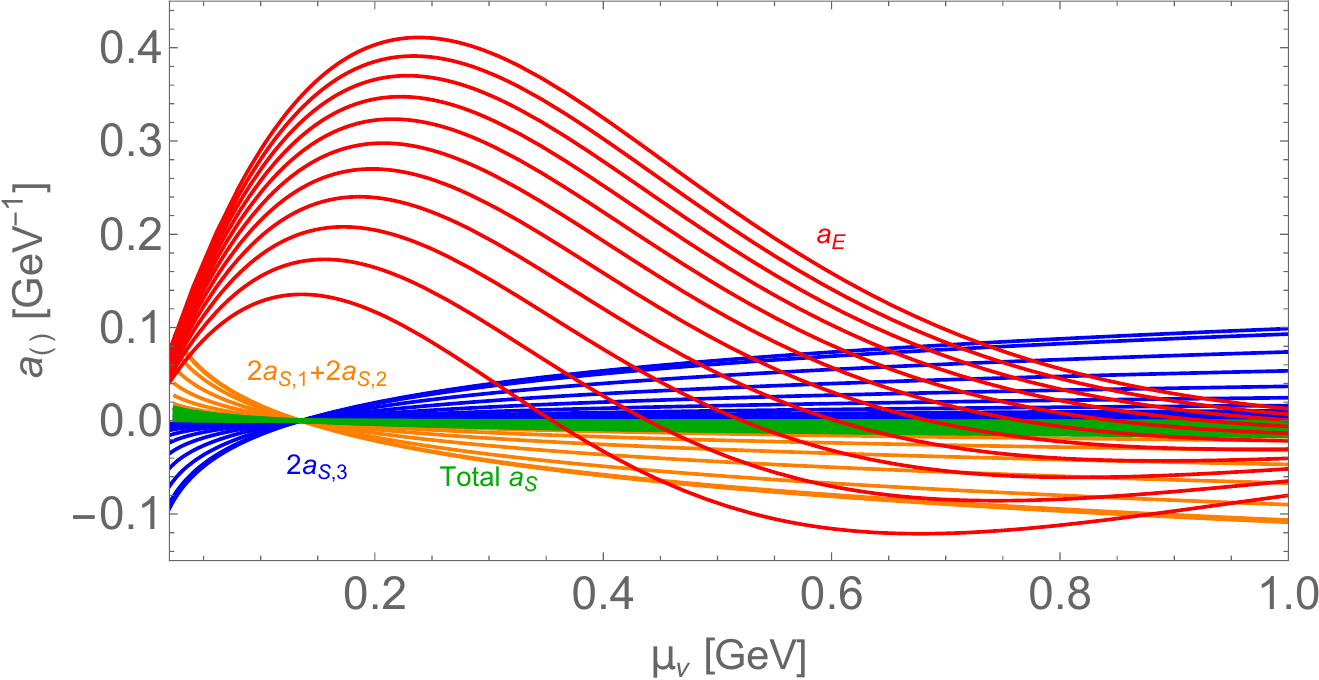}
    \\
    \includegraphics[width=1\textwidth]{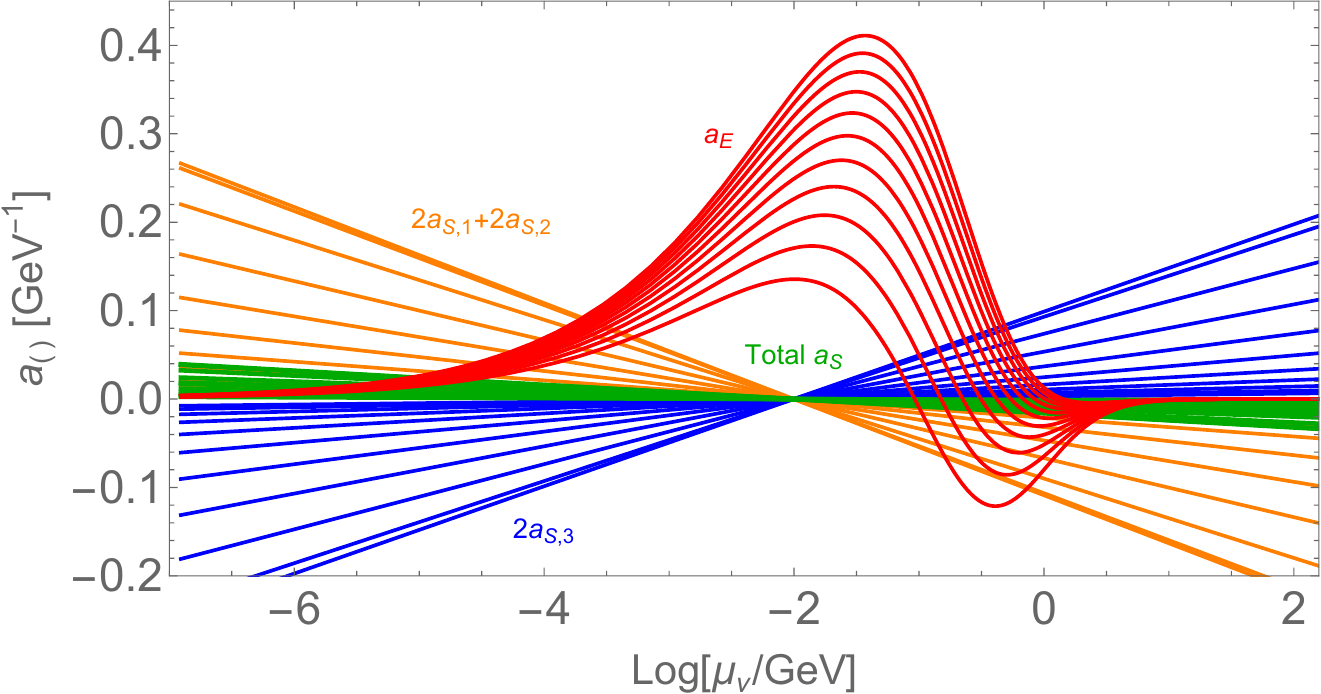}
    \caption{Partial amplitudes for the eye diagram (red), surfer diagrams $1$ and $2$ (orange), $3$ (blue), their sum (green), as functions of regularisation scale $\mu_\nu$ shown linearly (top) and logarithmically (bottom). $\mu_\pi$ is varied between $0.5 \Lambda_\chi$ and $1.5 \Lambda_\chi$, and the dipole $g_A$ form factor is included throughout.}
    \label{fig:allamps_partial}
\end{figure}

The partial amplitudes for all three surfer diagrams (b-d) are shown in Figure \ref{fig:surfersamps_partial} for example pion regularisation point $\mu_\pi = 0.5$~GeV, where again the significant cancellation is apparent, as well as some indication that virtual momenta above nuclear scales contribute diminishing amounts to these diagrams. To illustrate sensitivity of our results to the pion regularisation point, Figure \ref{fig:allamps_partial} again shows the partial amplitudes for the surfer diagrams as well as for the eye diagram treated in~\cite{Cirigliano2021long}, for a range of $\mu_\pi$ values between $0.5 \Lambda_\chi$ and $1.5 \Lambda_\chi$. One notices immediately that although the partial amplitude for the eye diagram suffers from some significant pion regularisation scale-dependence, the sum of partial amplitudes for the surfer diagrams is nearly vanishing across a broad range of both $\mu_\pi$ and $\mu_\nu$; in fact the total surfer diagram contribution exactly vanishes at $\mu_\nu = m_\pi$ (the representative nuclear scale chosen here and by~\cite{Cirigliano2021long}), regardless of $\mu_\pi$. We can say roughly that the chiral symmetry inherent in our treatment of this nuclear system has enforced a natural smallness of pion-loops through both the permitted neutrino currents and the combinatorics. This exact cancellation will be broken by electromagnetic effects which separate $m_{\pi^0}$ and $m_{\pi^\pm}$, but these should only be relevant one chiral order lower than our diagrams, at $N^3LO$.

To illustrate the size of these contributions relative the partial amplitudes $a_\chi$, $a_<$, and $a_>$ needed for the matching analysis of~\cite{Cirigliano2021long}, we interpret our $\chi EFT$ result as a correction on $a_\chi$, while we interpret its enhanced $g_A$ form-factor counterpart as a correction on $a_<$. In a more complete analysis, phenomenological pion-nucleon form factors would be included in a corrected $a_<$ alongside an appropriate extension of the half-off-shell form factors used in~\cite{Cirigliano2021long} to improve the robustness of the $NN$ contact interactions. We will treat these possible extensions as falling within the uncertainties accorded to the choice of the short-range potential, and see that this simple parameterisation of intermediate-momentum behavior is enough to deliver a corrected matching result with expected scale-dependence properties. There is certainly no need to adjust $a_>$ within the precision desired in this analysis, since as discussed in~\cite{Cirigliano2021long} the role of this partial amplitude is primarily to enforce finiteness in the UV, and at any rate the quark-level diagrams used in its calculation already include at least some inelastic intermediate nuclear states. 

\begin{figure}[t!]
    \centering
    \includegraphics[width=.99\textwidth]{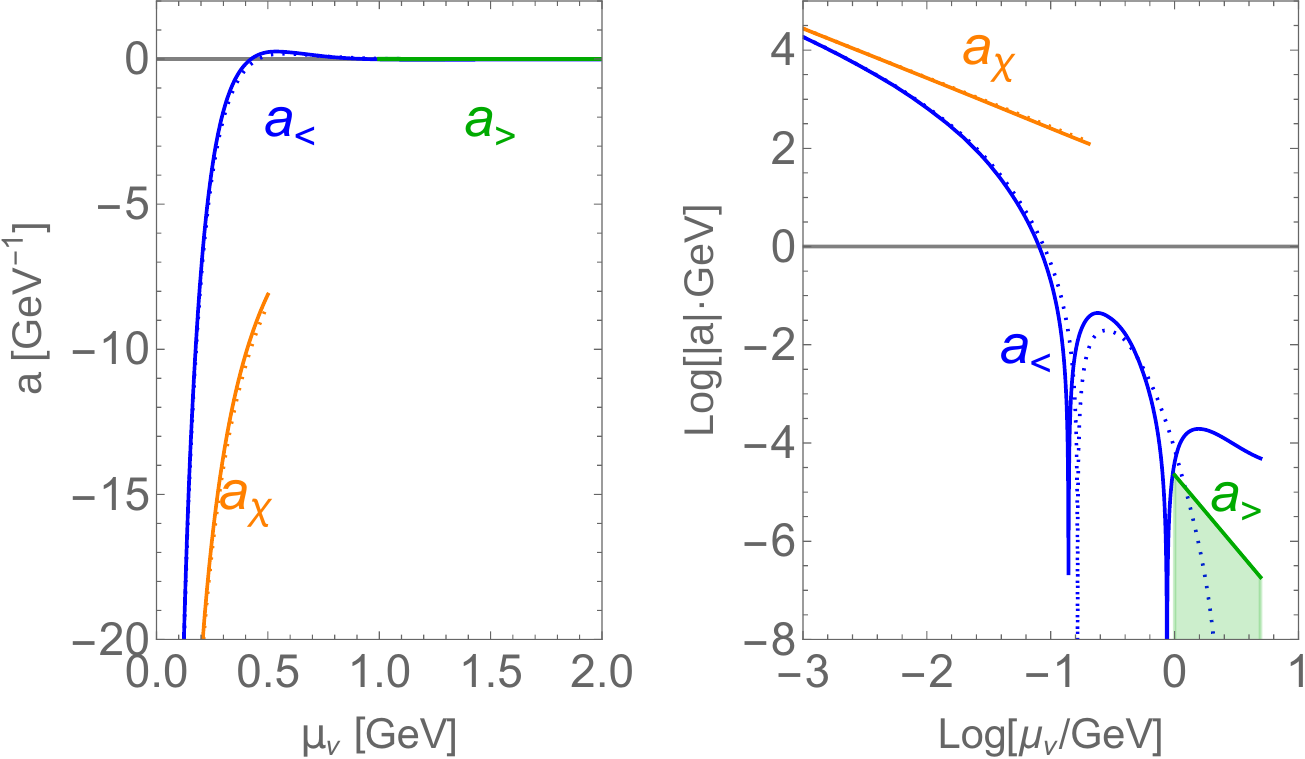}
    \caption[Total partial amplitudes for the matching analysis, where $a_\chi$ is accurate in the low-momentum exchange region up to $\sim m_\pi$, $a_<$ is the enhanced model for the intermediate-momentum exchange region, $a_>$ is the high-momentum exchange OPE from perturbative QCD, and the pion regularisation scale is fixed at $\mu_\pi = \Lambda_\chi$. The solid lines include all four $NN\pi$ partial amplitudes, while the dotted lines include only the $NN$ results.]{Total partial amplitudes for the matching analysis following Figure 6 of~\cite{Cirigliano2021long}, where $a_\chi$ is accurate in the low-momentum exchange region up to $\sim m_\pi$, $a_<$ is the enhanced model for the intermediate-momentum exchange region, $a_>$ is the high-momentum exchange OPE from perturbative QCD, and the pion regularisation scale is fixed at $\mu_\pi = \Lambda_\chi$. The solid lines include all four $NN\pi$ partial amplitudes, while the dotted lines include only the $NN$ results of~\cite{Cirigliano2021long}. }
    \label{fig:matchingboth}
\end{figure}
In Figure \ref{fig:matchingboth} the corrected partial amplitudes for the matching analysis are plotted (solid) in comparison to the $NN$-only partial amplitudes from~\cite{Cirigliano2021long} (dotted). Viewed globally, these corrections are almost invisible to the eye, certainly smaller than say $5\%$, showing the most discrepancy in the $0.1 < \mu_\nu < 0.5$~GeV range where the eye partial amplitude is strongest.

\begin{figure}[t!]
    \centering
    \includegraphics[width=1\textwidth]{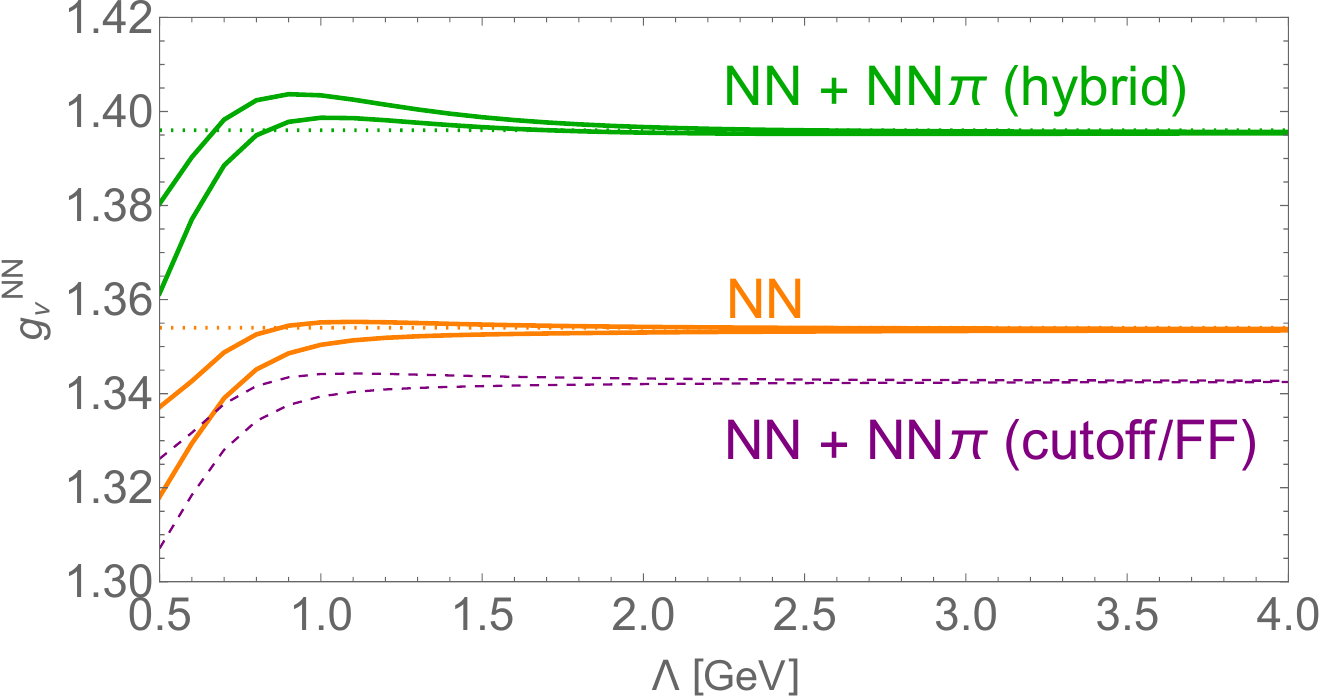}
    \caption[Computed low-energy constant $g_\nu^{NN}$ from combined $NN$ and $NN\pi$ amplitudes using the naive low-momentum schemes described in the text (purple), a more robust low- and intermediate-momentum scheme (green), in comparison with the $NN$-only result (orange). The weak dependence of $g_\nu^{NN}$ on matching scale $\Lambda$ is seen, and all computations are performed at pion regularisation scale $\mu_\pi = \Lambda_\chi$ and neutrino regularisation scale $\mu_\nu = m_\pi$.]{Computed low-energy constant $g_\nu^{NN}$ from combined $NN$ and $NN\pi$ amplitudes using the naive low-momentum schemes described in the text (purple), a more robust low- and intermediate-momentum scheme (green), in comparison with the $NN$-only result following~\cite{Cirigliano2021long} (orange). The weak dependence of $g_\nu^{NN}$ on matching scale $\Lambda$ is seen, and all computations are performed at pion regularisation scale $\mu_\pi = \Lambda_\chi$ and neutrino regularisation scale $\mu_\nu = m_\pi$.}
    \label{fig:allgnn}
\end{figure}

With these partial amplitudes in hand, we can perform a preliminary matching analysis analogous to that of~\cite{Cirigliano2021long} to estimate the corrected size of the contact counterterm $g_\nu^{NN}$. In practice this amounts to implementing Eq.~\eqref{integrandmatching} on our corrected partial amplitudes, and in particular we propose a hybrid matching scheme where the pure $\chi EFT$ corrections and those including a $g_A$ form factor are applied separately to $a_\chi$ and $a_<$, thus preserving the demarcation between low- and intermediate-momentum physics. However, other schemes are plausible: naively one might assume that due to our incomplete knowledge of the intermediate-momentum range, we should only apply our corrections to $a_\chi$, either with a hard cutoff or with the softer inclusion of the $g_A$ dipole form factor. The results of all three approaches, at $\mu_\nu = m_\pi$, $\mu_\pi = \Lambda_\chi$, and as a function of matching scale $\Lambda$, are displayed in Figure \ref{fig:allgnn}. We find that in our hybrid matching scheme, the size of $g_\nu^{NN}(\mu_\nu = m_\pi)$ is enhanced from $1.35$ to $1.40$, an increase of just under $4\%$, while generic independence of sufficiently large matching scale $\Lambda$ is preserved. The suppression of $1\%$ in the naive cutoff and form-factor schemes, in our view, neglects relevant intermediate-momentum scale physics which is present in our analysis. However, at the very least it provides an additional confirmation that the $NN\pi$ correction $g_\nu^{NN}$ is small, well below the $~38\%$ uncertainty allocated by~\cite{Cirigliano2021long}. Note also that our value of $1.35$ for the $NN$-only counterterm size differs from the value of $1.34$ quoted by~\cite{Cirigliano2021long}; this difference appears to be entirely attributable to updated measurement of $m_\pi$, and so does not represent a correction but rather an update on the $NN$-only result.

\begin{figure}[t!]
    \centering
    \includegraphics[width=.49\textwidth]{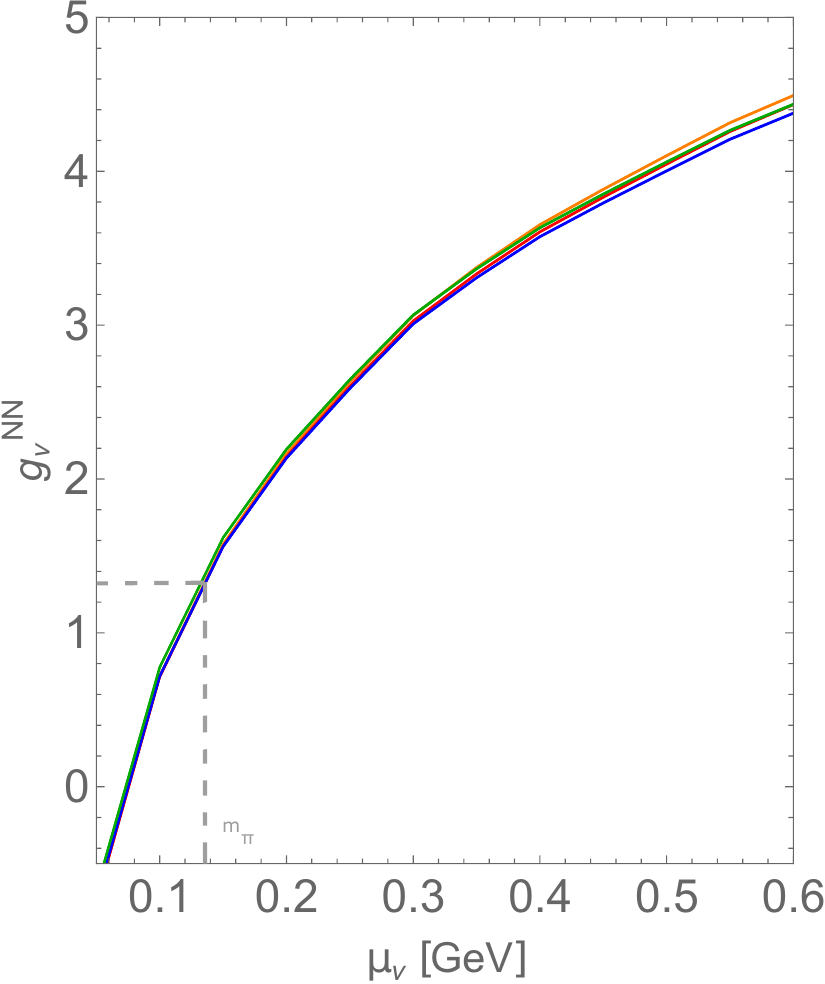}
    \hspace{2pt}
    \includegraphics[width=.49\textwidth]{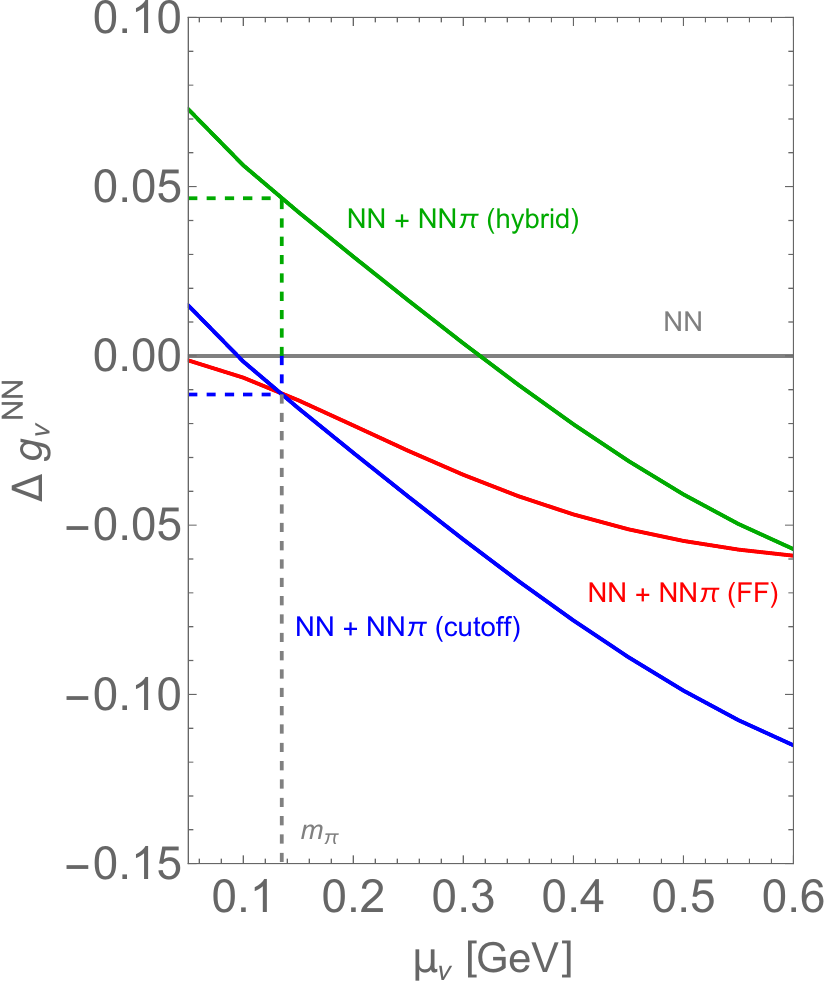}
    \caption{Computed low-energy constant $g_\nu^{NN}$ from combined $NN$ and $NN\pi$ amplitudes using the naive low-momentum schemes described in the text (purple), a more robust low- and intermediate-momentum scheme (green), in comparison with the $NN$-only result (orange). The logarithmic dependence of $g_\nu^{NN}$ on neutrino regularisation scale $\mu_\nu$ is seen, and all computations are performed at pion regularisation scale $\mu_\pi = \Lambda_\chi$ and matching scale $\Lambda = \Lambda_\chi$. Global similarity of the running coupling is demonstrated on the left, while the right shows the $NN_\pi$ corrections $\Delta g_\nu^{NN}$ by scheme, where the values at chosen scale $\mu_\nu = m_\pi$ correspond to Figure \ref{fig:allgnn}.}
    \label{fig:allgnn_nu_both}
\end{figure}

In Figure \ref{fig:allgnn_nu_both}, we illustrate the dependence of these results on the neutrino regularisation scale $\mu_\nu$. As expected, the size of the correction remains small across all $\mu_\nu$, and globally we can state that the shape of the renormalisation group of $g_\nu^{NN}$ is adjusted very little. However, the total corrections to $g_\nu^{NN}$ are scale-dependent; in the hybrid scheme they vary between $-0.06 < \Delta g_\nu^{NN} < 0.07$ for scales $0.05 < \mu_\nu < 0.6$~GeV. The results according to the naive cutoff and form factor schemes diverge from one another as one moves away from $\mu_\nu = m_\pi$ and the surfer diagram contributions become larger; however all three schemes exhibit decreasing scale dependence of $\Delta g_\nu^{NN}$. In other words, accounting for $NN\pi$ diagrams has served to improve the scale-dependence of $g_\nu^{NN}$, albeit very slightly.

In total, we quote the following $NN\pi$-corrected result for the size of the $\Delta L=2$ contact counterterm first estimated by~\cite{Cirigliano2021long}, representing a $\sim 4\%$ increase in its size:
\begin{equation}
    g_\nu^{NN} \simeq 1.40(20)_{V_S}(5)_{\text{parameters}}(3)_{\text{inelastic}} = 1.4(3)
\end{equation}
Regarding our uncertainties, we retain the short-range potential and parameter uncertainties claimed by~\cite{Cirigliano2021long}, since we have not sought to make improvements on the calculation in either of these areas, and since we have not introduced any dramatically different short-range physics or physical parameters into the picture which would be more prone to uncertainty. In addition we have checked that reasonable variation of our model parameters (namely $\Lambda_A$ in the $g_A$ dipole form factor) does not exceed $\delta = 0.05$.

Our remaining inelastic uncertainty is derived from the power-counting performed in Figure \ref{fig:pinndimanal}, from which all unaccounted for subleading $NN\pi$ diagrams contribute are expected to contribute not more than $0.016$ uncertainty to the total correction amplitude. $NN\pi\pi$ and more-pion intermediate states only begin to contribute at $N^4LO$ in our power-counting, and can be expected to be safely sub-percentile. We suggest that the inclusion of $\Delta$ and heavier resonance intermediate states in our analysis, because it would require a full extension of the underlying $\chi EFT$ framework to include explicit heavier degrees of freedom, is already accounted for in the short-range potential uncertainty $\delta_{V_S} = 0.20$. To account for any $N^3LO$ $\Delta$-resonance diagrams which might contribute at $N^3LO$ to our power-counting of diagrams, we conservatively double our remaining inelastic uncertainty to give $\delta_{inelastic} = 0.03$.

Therefore in addition to identifying a $\sim 5\%$ increase in the size of the contact counterterm owing to $NN\pi$ intermediate states, our $\chi EFT$ framework demonstrates a reduction of inelastic uncertainties from the $\sim 38\%$ of~\cite{Cirigliano2021long} to $\sim 3\%$, corresponding to a total uncertainty improvement from $\sim 46\%$ to $\sim 21\%$. Thus the dominant source of contact counterterm uncertainty, rather than inelastic states, is the selection of a short-range internucleon potential for intermediate momenta.

\section{Discussion}

The complexity of the nuclear many-body problem and resultant wide spread in estimates of NMEs remains the primary obstacle to improving the precision of our theoretical understanding of $0\nu\beta\beta$-decay. While development continues for a variety of computational methods based on phenomenological nuclear theory, a growing proportion of the nuclear theory community has come to view \textit{ab initio}, first principles nuclear theory as the most promising way forwards. Uncertainty quantification in \textit{ab initio} nuclear theory is more rigorous than in phenomenological models, granting insights into disagreements between calculations, and ultimately the systematic improvement of many-body methods as well as the NMEs they predict \cite{Cirigliano2022b}. 

In this work, we have attempted to expand upon the calculation of the contact counterterm contribution to $0\nu\beta\beta$-decay, as first noted in Ref.~\cite{Cirigliano2018} and quantified for elastic intermediate-states in Refs.~\cite{Cirigliano2021short,Cirigliano2021long}. At renormalisation point $\mu = m_\pi$, this estimate was:
\begin{equation}
    g_\nu^{NN}|_{NN} \simeq 1.32(50)_{\text{inelastic}}(20)_{\text{V}_\text{S}}(5)_{\text{parameters}} = 1.3(6),
\end{equation}
where a total of $\sim 46\%$ uncertainty is present, dominated by the $\sim 38\%$ uncertainty arising from the choice to only account for elastic intermediate hadronic states. By accounting for $NN\pi$ intermediate states, we have generated an improved estimate of the contact coefficient:
\begin{equation}
    g_\nu^{NN}|_{NN + NN\pi} \simeq 1.40(20)_{V_S}(5)_{\text{parameters}}(3)_{\text{inelastic}} = 1.4(3)
\end{equation}
where a total of $\sim 21\%$ uncertainty is present, only $\sim 3\%$ now stemming from the intermediate state truncation (an order of magnitude reduction). As indicated in Section 4.5, additional work will be required to ensure the robustness of this intermediate-state correction through intermediate-momentum scales. Further improvements in precision could mainly be obtained through careful characterisation of the short-range nuclear potentials employed in modelling that intermediate-momentum scale.

A distinct verification of the size of the contact-counterterm, justifying in part the relation to charge-independence breaking low-energy constants made in \cite{Cirigliano2018}, was recently performed by \cite{Richardson2021,Richardson2024} using a large-$N_c$ analysis. We note that their determination, like ours, favours a slightly larger $g_\nu^{NN}$ than the elastic truncation, and that our results agree within uncertainties.

After the completion of this work, the author became aware of a new and alternative approach by \cite{Yang2024}, in which the $0\nu\beta\beta$-decay amplitude is computed using the manifestly Lorentz-invariant chiral Lagrangian of \cite{Epelbaum2012}. Just as Ref.~\cite{Epelbaum2012} shows how to absorb all UV divergences for S-wave $NN$ scattering, Ref.~\cite{Yang2024} claims that this relativistic Lagrangian leads to an $0\nu\beta\beta$-decay amplitude which requires no counterterm at LO. The resultant LO amplitude is numerically compatible with the non-relativistic approach with a counterterm, and therefore our improved-precision estimate as well; it would be interesting to consider whether the subleading corrections presented here have an analogous effect in the relativistic setting.

To obtain an \textit{ab initio} estimate of the NME for $0\nu\beta\beta$-decay in a given isotope, input from the two- and few-nucleon scale must be provided to an appropriate many-body method; popular contenders include the self-consistent Green's function (SCGF) approach \cite{Soma2020}, the in-medium similarity renormalisation group (IMSRG) \cite{Hergert2016}, coupled-cluster (CC) methods \cite{Hagen2014}, and the in-medium generator coordinate method (IM-GCM) \cite{Yao2018}. NME estimates from these approaches, built upon long-range few-nucleon inputs only, skew smaller than estimates from most phenomenological nuclear models, with the consequence that a given $0\nu\beta\beta$-decay search will probe less of new-physics parameter space. The most recent calculation based on IMSRG \cite{Belley2023} suggested as much as an order-of-magnitude decrease in experimental reach; or rather, that biased estimates based on phenomenological nuclear models have overinflated experimental reach by this amount.

Ref.~\cite{Wirth2021} was the first to produce NME estimates which include the contact-counterterm as input, demonstrating (within a no-core shell model) a $\sim 15\%$ enhancement in $^6$~He compared to the long-range only estimate. Refs.~\cite{Belley2023} and \cite{Holt2022} similarly applied IMSRG and IM-GCM incorporating the contact-counterterm to experimentally-relevant isotopes such as $^{76}$~Ge; compared to the long-range only NME estimates, these results show $40-90\%$ enhancements depending on the isotope. As a result, a significant portion of the experimental reach which was lost in proceeding from a phenomenological to an \text{ab initio} nuclear model is restored. Since our improved-precision estimate of the contact term has resulted in only a very small adjustment of the central value, within even the improved uncertainties, we should therefore anticipate a tightening of the \text{ab initio} NME estimates, without much movement.

Clearly it is essential both to the interpretation and design of future $0\nu\beta\beta$-decay searches that the remaining theoretical uncertainties embedded in these \textit{ab initio} estimates be suitably resolved. As \cite{Belley2023} explains, many of these theoretical uncertainties will arise from neglected many-body effects e.g. in IMSRG; these are beyond the scope of this work and represent significant future endeavours of the nuclear theory community. However, other theoretical uncertainties arise from the two-nucleon scale, and it is precisely those uncertainties on the contact counterterm which we have here reduced by a factor of two.

%\appendix
%\section{Some title}

\acknowledgments

The author wishes to thank, first and foremost, their doctoral supervisor Frank Deppisch for essential guidance and feedback in pursuing this line of research -- and also to thank Matteo Agostini, Patrick Bolton, James Canning, Vincenzo Cirigliano, Jonathan Engel, Jason Holt, Emanuelle Mereghetti, and Fedor Šimkovic for illuminating conversations which helped to bring this work to completion. G.V.G. in particular acknowledges support from the UCL Centre for Doctoral Training in Data Intensive Science funded by STFC, and from the UCL Overseas Research Scholarship / Graduate Research Scholarship, and the Science and Technology Facilities Council, part of U.K. Research and Innovation, Grant No. ST/P00072X/1, ST/T000880/1, ST/T004169/1 and ST/W00058X/1, as well as support from the University of Southampton and Grant No. ST/W006251/1 during the writing up of this work.

%This is the most common positions for acknowledgments. A macro is available to maintain the same layout and spelling of the heading.

%\paragraph{Note added.} This is also a good position for notes added after the paper has been written.

% Bibliography

%% [A] Recommended: using JHEP.bst file
%% \bibliographystyle{JHEP}
%% \bibliography{biblio.bib}

%% or
%% [B] Manual formatting (see below)
%% (i) We suggest to always provide author, title and journal data or doi:
%% in short all the informations that clearly identify a document.
%% (ii) please avoid comments such as "For a review'', "For some examples",
%% "and references therein" or move them in the text. In general, please leave only references in the bibliography and move all
%% accessory text in footnotes.
%% (iii) Also, please have only one work for each \bibitem.

%\addcontentsline{toc}{chapter}{Bibliography}

% Actually generates your bibliography.
\bibliographystyle{JHEP}
\bibliography{biblio.bib,background.bib}

\end{document}